\documentclass[12pt,onecolumn]{article}
\usepackage{color}
\usepackage{amsmath}
\usepackage{amssymb}
\usepackage{amsfonts}
\usepackage{geometry}
\usepackage{setspace}
\usepackage{amsmath,bm}
\usepackage{graphicx}
\usepackage[section]{placeins}
\usepackage{subfigure}
\usepackage[sort&compress]{natbib}
\usepackage{booktabs}  
\usepackage{makecell}
\usepackage{lineno} 
\usepackage{caption}
\usepackage[colorlinks, citecolor=blue]{hyperref}

\author{Xiaoying Zhuang$^{2}$, Shuwei Zhou$^{1,2*}$, Mao Sheng$^{3}$, Gengsheng Li$^{3}$}
\title {On the hydraulic fracturing in naturally-layered porous media using the phase field method}
\begin{document}

\captionsetup[figure]{labelfont={bf},name={Fig.},labelsep=period}
\bibliographystyle{apa}
\setcitestyle{authoryear,round,aysep={,},yysep={,}}
\date{}
\maketitle

\spacing {2}
\noindent
1 Department of Geotechnical Engineering, College of Civil Engineering, Tongji University, Shanghai 200092, P.R. China\\
2 Institute of Continuum Mechanics, Leibniz University Hannover, Hannover 30167, Germany\\
3 State Key Laboratory of Petroleum Resources and Prospecting, China University of Petroleum, Beijing, China

* Corresponding author: Shuwei Zhou (shuwei.zhou@ikm.uni-hannover.de; zhoushuwei1016@126.com)

\begin{abstract}
\noindent In the hydraulic fracturing of natural rocks, understanding and predicting crack penetrations into the neighboring layers is crucial and relevant in terms of cost-efficiency in engineering and environmental protection. This study constitutes a phase field framework to examine hydraulic fracture propagation in naturally-layered porous media. Biot's poroelasticity theory is used to couple the displacement and flow field, while a phase field method helps characterize fracture growth behavior. Additional fracture criteria are not required and fracture propagation is governed by the equation of phase field evolution. Thus, penetration criteria are not required when hydraulic fractures reach the material interfaces. The phase field method is implemented within a staggered scheme that sequentially solves the displacement, phase field, and fluid pressure. We consider the soft-to-stiff and the stiff-to-soft configurations, where the layer interface exhibits different inclination angles $\theta$. Penetration, singly-deflected, and doubly-deflected fracture scenarios can be predicted by our simulations. In the soft-to-stiff configuration, $\theta=0^\circ$ exhibits penetration or symmetrical doubly-deflected scenarios, and $\theta=15^\circ$ exhibits singly-deflected or asymmetric doubly-deflected scenarios. Only the singly-deflected scenario is obtained for $\theta=30^\circ$. In the stiff-to-soft configuration, only the penetration scenario is obtained with widening fractures when hydraulic fractures penetrate into the soft layer.
\end{abstract}


\section {Introduction}\label{Introduction}

Hydraulic fracturing (HF) in porous media is a challenging area for mechanical, environmental, energy, and geological engineering \citep{cui2013resistance, mikelic2013phase, li2014hydraulic, lei2017role, guo2018modelling, zhou2018interpretation, liu2019experimental, jeanpert2019fracture}. On the one hand, HF applies pressurized fluid \citep{liu2014situ} to form highly permeable fractures into the rock strata and the fracture network facilitate the linking of wellbores with the expected natural resources. On the other hand, HF is also relatively controversial because of its potential impact on the engineering geological environment. Unexpected fractures may be stimulated and propagate into neighboring rock strata, thereby resulting in a risk of water contamination. Moreover, uncontrolled hydraulic fractures along dominant or previously unknown faults may cause an increase in seismic activity \citep{kim2013induced}, which is detrimental to the stability of the entire geological system. Therefore, better prediction of hydraulic fracture propagation is one of the most critical issues in recent years \citep{figueiredo2017effects, ren2017equivalent, zhou2018phase2, liu2018methodology}, especially in engineering geology.

\begin{figure}[htbp]
	\centering
	\subfigure[Penetration]{\includegraphics[width = 5cm]{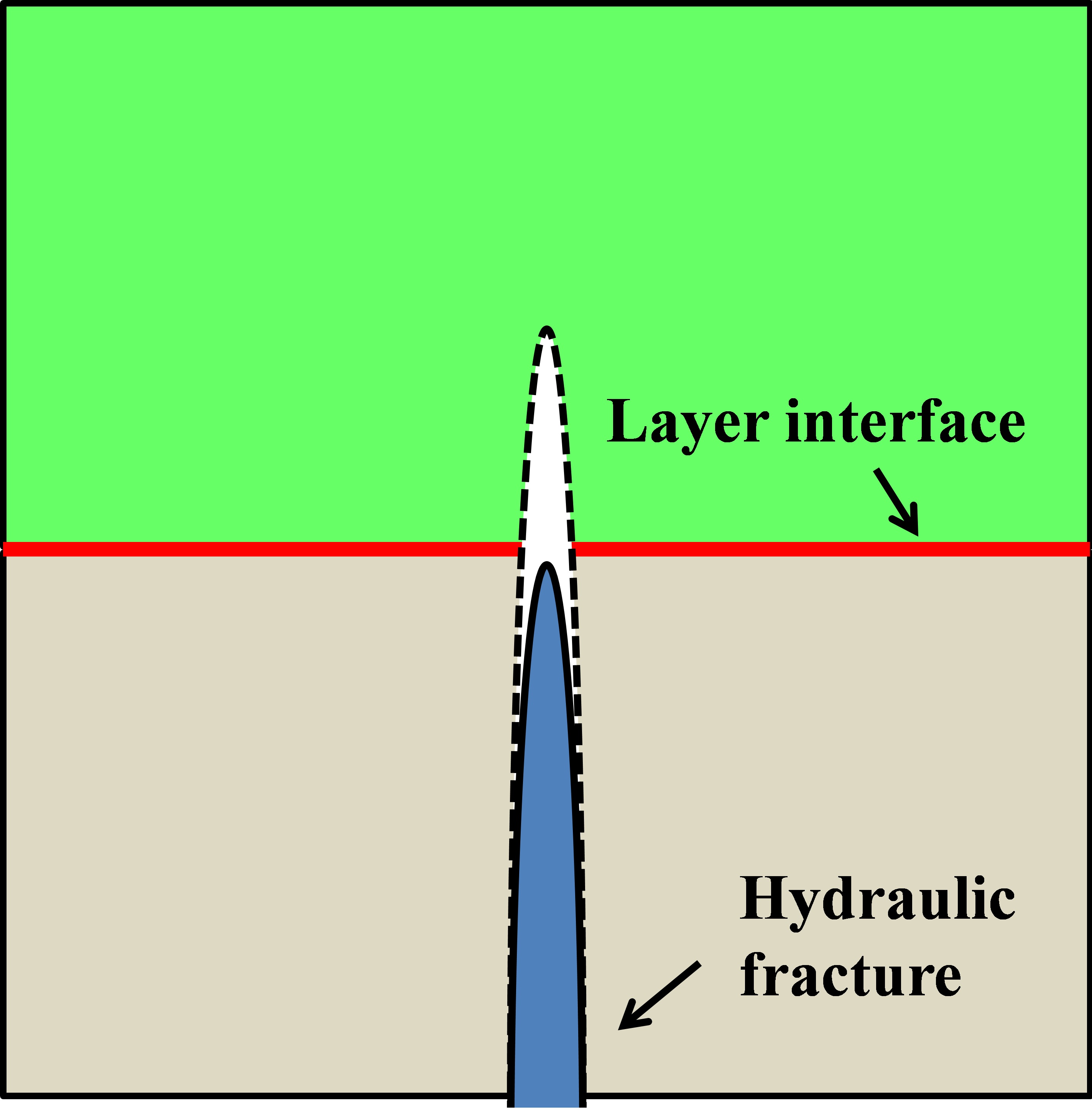}}
	\subfigure[Singly-deflected]{\includegraphics[width = 5cm]{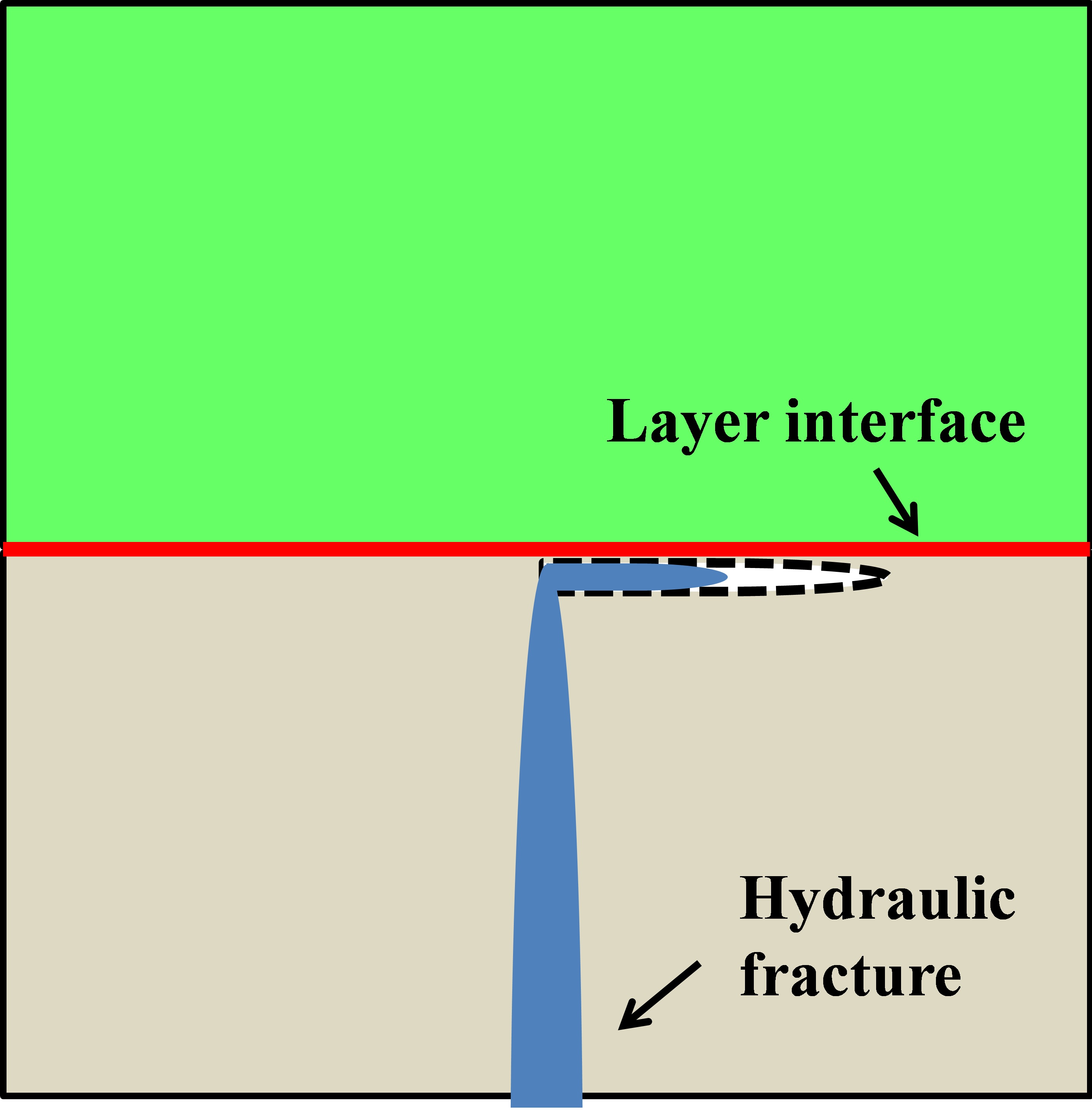}}
	\subfigure[Doubly-deflected]{\includegraphics[width = 5cm]{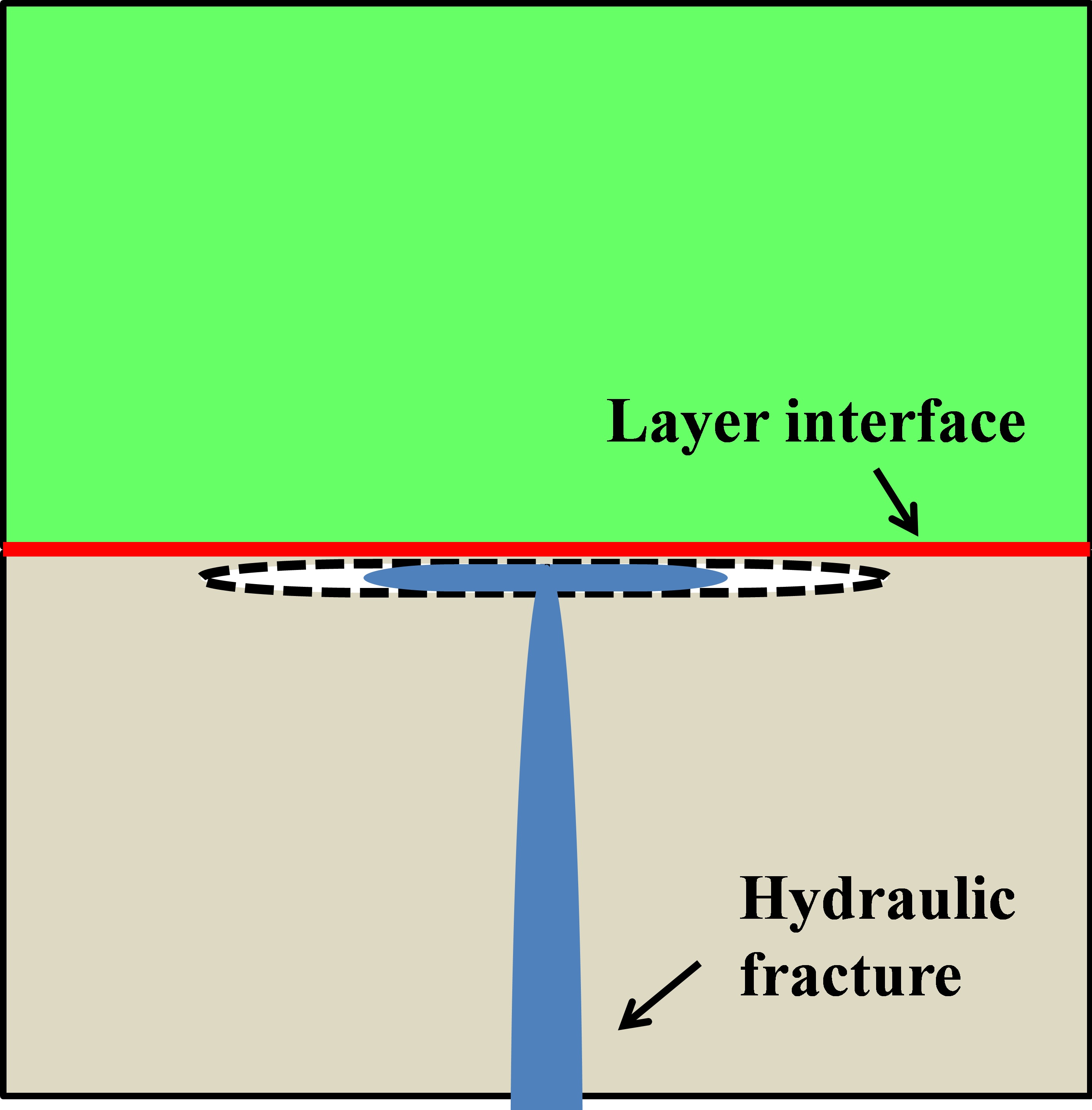}}\\
	\caption{Three fracture patterns when hydraulic fractures reach the layer interface \citep{vahab2018x}}
	\label{Three fracture patterns when hydraulic fractures reach the layer interface}
	\end{figure}

Naturally-layered rock strata are an important and representative part of the engineering geological environment. In general, numerous natural or artificial discontinuities are contained in naturally-layered reservoirs, such as fractures or material interfaces. Many studies have examined how fluid-driven fractures interact with natural discontinuities. Some contributions can be referred to \citet{van1982hydraulic, biot1983fracture, gudmundsson2001hydrofractures, dyskin2009orthogonal, dahi2011numerical, zhang2011simulation, behnia2014numerical, khoei2015enriched, wang2015numerical}. These studies have indicated that three fracture patterns exist in layered domains, particular when hydraulic fractures reach the layer interfaces: penetration, singly-deflected, and doubly-deflected scenarios (see Fig. \ref{Three fracture patterns when hydraulic fractures reach the layer interface}). However, these studies have yet to achieve a consensus on mechanisms for different fracture penetration patterns. Hence, predicting hydraulic fractures in naturally-layered media remains an open research topic and new insights should be provided to conduct further research.

For example, advanced numerical approaches can be applied in naturally-layered media, although hydraulic fractures can be also investigated either analytically \citep{santillan2017phase} or experimentally \citep{cooke2001fracture, xing2018laboratory}. The numerical models for fracture can be classified into discrete and continuous approaches, including the extended finite element method (XFEM) \citep{moes2002extended}, cohesive element method \citep{nguyen2001cohesive}, element-erosion method \citep{belytschko1987three}, phantom-node method \citep{rabczuk2008new}, mesh-free methods \citep{zhuang2012fracture, zhuang2014improved}, cracking particle methods \citep{ rabczuk2010simple}, peridynamics \citep{ren2016dual,ren2017dual}, gradient damage models \citep{peerlings1996some}, screened-Poisson models \citep{areias2016damage}, and phase field models (PFMs) \citep{miehe2010thermodynamically, miehe2010phase, borden2012phase, zhou2019phase2}. Although some continuous and discrete approaches have been used in HF, predicting fracturing networks across different rock formations, such as reservoir and cap layers, remains a challenging topic. For example, an XFEM framework \citep{vahab2018x} was recently established to investigate hydraulic fracture propagation in a layered domain. Nevertheless, mandatory penetration criteria are applied when the fractures reach the layer interfaces. Although singly-deflected and penetration scenarios can be observed, the hydraulic fractures are ``man-made", and not the automatically predicted ones with respect to physical principles.

The current study investigates the propagation of hydraulic fractures in naturally layered geological formations by applying the phase field model \citep{zhou2018phase2} to solve the fracture penetration issue at the layer interface and provide a new perspective. Note that the used PFM is based on the quasi-static formulation of \citet{zhou2018phase2} for fluid-driven fracture, which is different from the PFMs for single-phase solid \citep{zhou2018phase, zhou2018phase3} and for dynamic fluid-driven fracture \citep{zhou2019phase}. In the numerical investigation, we consider two configurations, namely, the soft-to-stiff and the stiff-to-soft, and also different inclination angles of the layer interface. The phase field characteristics of fluid-driven fracture in layered geological formations are firstly and systematically explored. In addition, the displacement and stress characteristics related to geological system stability are investigated. The relationship between the phase field (fracture pattern) and the stiffness contrast and inclination angle of the geological formations, is also further revealed. Based on minimization of the energy functional, our simulations automatically reveal the three potential fracture scenarios in naturally-layered geological formations, i.e., the penetration, singly-deflected, and doubly-deflected scenarios, which must be achieved by other numerical methods such as XFEM \citep{vahab2018x} by imposing ``man-made" penetration criteria.

The phase field simulation in this study can deepen our understanding of crack patterns and their governing factors in layered geological formations. The successful application of PFM also shows its immense potential in modeling the interface crack between two layers and how cracks appear into neighboring layers, which are not easily captured with conventional methods. The remainder of this paper is organized as follows: Sections \ref{Mathematical models for fracture propagation}, \ref{Numerical treatment}, and \ref{Further validation of PFM} present the theoretical model, numerical implementation, and validation examples, respectively. Sections \ref{The soft to stiff configuration} and \ref{The stiff to soft configuration} provide the numerical investigations on the soft-to-stiff and stiff-to-soft configurations, respectively. Lastly, Section \ref{Conclusions} concludes this study.

\section {Mathematical models for fracture propagation}\label{Mathematical models for fracture propagation}
\subsection {Energy functional}\label{Problem definition}

\begin{figure}[htbp]
	\centering
	\includegraphics[width = 6cm]{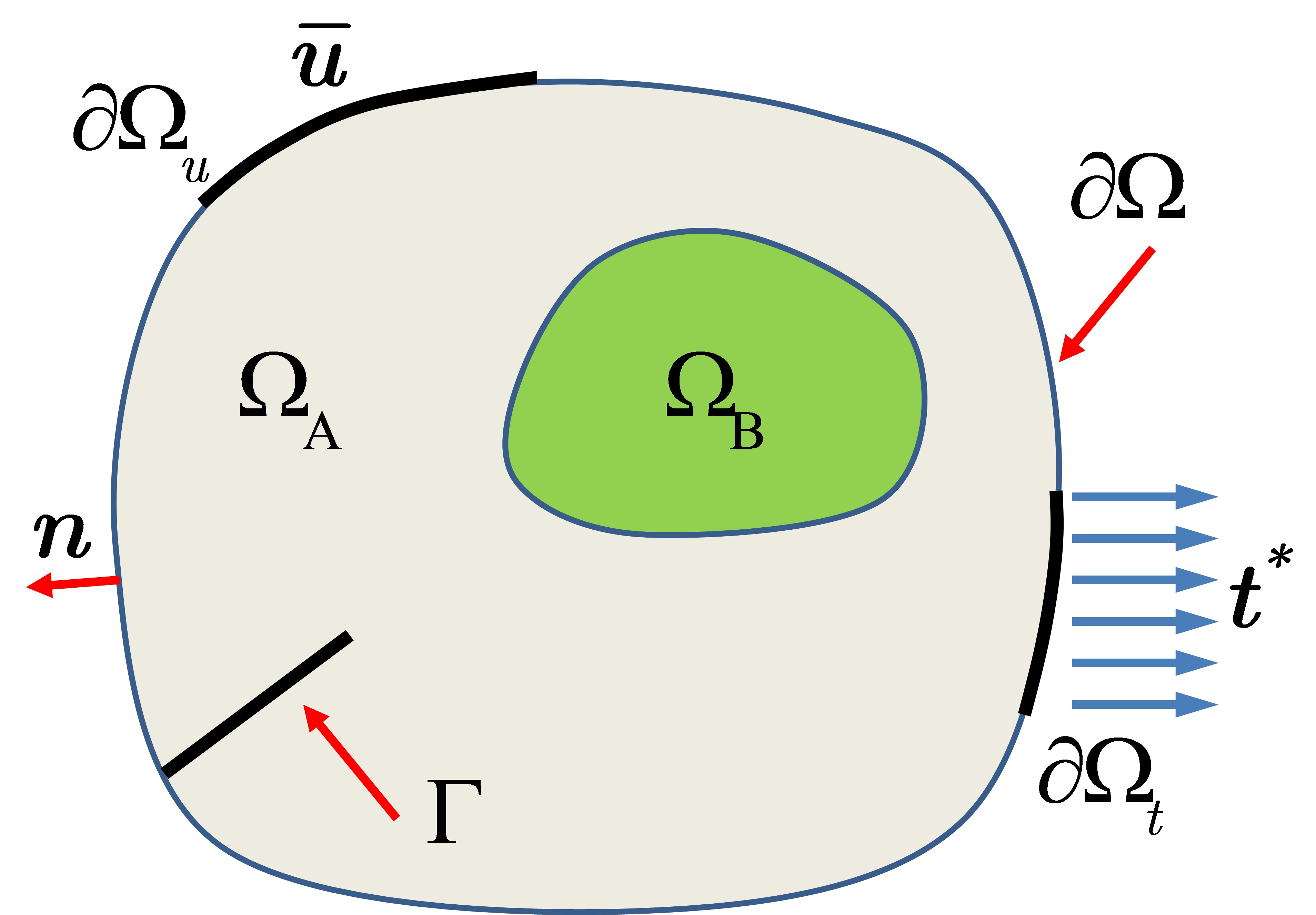}
	\caption{Layered porous media}
	\label{Layered porous media}
	\end{figure}

Let $\Omega$ be a cracked two-dimensional permeable porous solid  and $\bm x$ be the position vector in Fig. \ref{Layered porous media}. Two different layers, $\Omega_A$ and $\Omega_B$, compose the calculation domain $\Omega$ ($\Omega_A \cup \Omega_B= \Omega$). The full bond between the two layers is assumed and continuity conditions are naturally fulfilled. The boundary of the domain $\Omega$ is $\partial \Omega$. Two disjointed parts, $\partial \Omega_u$ and $\partial \Omega_t$, are defined with the prescribed displacement $\bar{\bm {u}}(\bm x,t)$ and traction $\bm {t}^*(\bm x,t)$. Moreover, a body force $\bm b(\bm x,t)$ is applied on $\Omega$. We also define the outward unit normal vector $\bm n$ and the internal fracture in the domain as $\Gamma$ in Fig. \ref{Layered porous media}. The main purpose of this study is to investigate fracture propagation in the domains $\Omega_A$ and $\Omega_B$, particularly the fracture behavior around the interface of the two layers. Furthermore, this research assumes that the pore size is considerably smaller than the fracture length scale; the porous media are elastic and homogeneous with compressible and viscous fluids.

We use the variational approach of Griffith's theory \citep{francfort1998revisiting} to study fracture propagation. For the purely single phase problem, the energy functional $\Psi(\bm u,\Gamma)$ for the entire calculation domain is composed of only the elastic energy $\Psi_{\varepsilon}(\bm \varepsilon)$, dissipation energy $\Psi_f$ and external work $W_{ext}$ \citep{zhou2018phase}. By contrast, the influence of fluid pressure $p$ must be considered in a porous medium and the energy functional involves an additional pressure-related term \citep{bourdin2012variational, wheeler2014augmented, mikelic2015quasi, mikelic2015phase, lee2016pressure, miehe2016phase, zhou2018phase2}:
	\begin{equation}
	\Psi(\bm u,p,\Gamma) = \underbrace{\int_{\Omega}\psi_{\varepsilon}(\bm \varepsilon) \mathrm{d}{\Omega}}_{\Psi_{\varepsilon}}\underbrace{-\int_{\Omega}\alpha p \cdot (\nabla \cdot \bm u) \mathrm{d}{\Omega}}_{\text{pressure-related term}}+\underbrace{\int_{\Gamma}G_c \mathrm{d}S}_{\Psi_f}\underbrace{-\int_{\Omega} \bm b\cdot{\bm u}\mathrm{d}{\Omega} - \int_{\partial\Omega_{t}} \bm {t^*}\cdot{\bm u}\mathrm{d}S}_{W_{ext}}
	\label{functional2}
	\end{equation}

\noindent where $\alpha$ and $G_c$ represent the Biot coefficient and critical energy release rate, respectively, and $\bm \varepsilon$ represents the linear strain tensor.
	
In Eq. \eqref{functional2}, $\psi_{\varepsilon}(\bm \varepsilon)$ denotes the elastic energy density, which can be expressed as follows in an intact isotropic and linear elastic solid \citep{miehe2010phase}:
	\begin{equation}
	\psi_{\varepsilon}(\bm \varepsilon) = \frac{1}{2}\lambda\varepsilon_{ii}\varepsilon_{jj}+\mu\varepsilon_{ij}\varepsilon_{ij}
	\end{equation}

\noindent where $\lambda,\mu>0$ are the Lam\'e constants.

\subsection{Phase field description}\label{Phase field description}

The phase field model \citep{zhou2018phase2}, which smears the discontinuous fracture $\Gamma$ over the domain $\Omega$, is used to describe fracture propagation in the porous media. In this study, the phase field varies from 0 to 1, $\phi=0$ means that the material is unbroken, and $\phi=1$ is for a ``fully" broken region. Note that the value of the phase field can be used in identifying a fracture, while a damaged region bounded by a threshold value of phase field is commonly adopted to reflect the fracture shape similar to those in the discrete settings \citep{borden2012phase}. Thereafter, the crack surface density is represented by the phase field and its gradient as follows \citep{bourdin2000numerical}:
	\begin{equation}
	\gamma(\phi,\bigtriangledown\phi)=\frac{\phi^2}{2l_0}+\frac{l_0}2\nabla\phi\cdot \nabla\phi
	\label{phase field approximation}
	\end{equation}

\noindent where $l_0$ is the length scale parameter.

The regularized formulation \eqref{phase field approximation} facilitate the transfer of the  dissipation energy integration over a discontinuity into an integration over a continuous domain. This treatment considerably facilitates the numerical implementation in fracture problems. Therefore, substituting Eq. \eqref{phase field approximation} into Eq. \eqref{functional2} yields the dissipation energy $\Psi_f$:
	\begin{equation}
	\int_{\Gamma}G_c \mathrm{d}S=G_c\int_{\Gamma} \mathrm{d}S\approx G_c\int_{\Omega} \gamma \mathrm{d}{\Omega} =\int_{\Omega}G_c\left[\frac{\phi^2}{2l_0}+\frac{l_0}2\nabla\phi\cdot \nabla\phi\right]\mathrm{d}{\Omega}
	\label{phase field approximation for the fracture energy}
	\end{equation}
\indent{In the PFM, the elastic energy appears as a driving term in the evolution equation. Therefore, for the purpose of removing unrealistic fracture patterns in the simulation \citep{miehe2010phase}, the elastic energy must be decomposed. The strain decomposition} is used first where the strain $\bm \varepsilon$ is composed of the tensile strain tensor $\bm\varepsilon^+$ and compressive $\bm\varepsilon^-$:
	\begin{equation}
	\bm\varepsilon^{\pm}=\sum_{a=1}^d \langle\varepsilon_a\rangle^{\pm}\bm n_a\otimes\bm n_a
	\label{strain decomposition}
	\end{equation}

\noindent where $\varepsilon_a$ denotes the principal strains, and $\bm n_a$ represents the direction vectors. The two operators in Eq. \eqref{strain decomposition} are $\langle\centerdot\rangle^+=\mathrm{max}(\centerdot,0)$ and $\langle\centerdot\rangle^-=\mathrm{min}(\centerdot,0)$. Thereafter, the positive and negative elastic energy densities can be defined in terms of the tensile and compressive strain tensors:
	\begin{equation}
	\psi_{\varepsilon}^{\pm}(\bm \varepsilon) = \frac{\lambda}{2}\langle tr(\bm\varepsilon)\rangle^{\pm 2}+\mu tr \left(\bm\varepsilon^{\pm 2}\right) 
	\end{equation}

According to \citet{miehe2010phase}, the total elastic energy density is expressed as
	\begin{equation}
	\psi_{\varepsilon}(\bm\varepsilon)=\left[(1-k)(1-\phi)^2+k\right]\psi_{\varepsilon}^+(\bm \varepsilon)+\psi_{\varepsilon}^-(\bm \varepsilon)
	\label{decomposition of the elastic energy}
	\end{equation}

\noindent where $0<k\ll1$ is a stability parameter. In Eq. \eqref{decomposition of the elastic energy}, if the phase field $\phi = 1$, then the stiffness against tension is only $k$ times of its original value, and $k=0$ will lead to a singularity in the stiffness matrix. Therefore, the positive stability parameter can effectively avoid this singularity and its detrimental effect on convergence in the simulation.

\subsection{Governing equations for the phase field evolution}\label{Governing equations for phase field evolution}

The energy functional \eqref{functional2} is renewed according to Eqs. \eqref{phase field approximation for the fracture energy} and \eqref{decomposition of the elastic energy}. For the variational approach \citep{francfort1998revisiting}, fracture initiation and growth at time $t$ is a process that the energy functional $\Psi$ seeks for a minimum value. In a natural manner, setting the first variation of the functional $\Psi$ as 0 yields
	\begin{multline}
	\delta \Psi=\underbrace{\int_{\partial \Omega_t}\left[(\sigma_{ij}-\alpha p\delta_{ij})n_j-t_i^* \right] \delta u_i \mathrm{d}{S}}_{\textcircled{1}}
-\underbrace{\int_{\Omega}\left[(\sigma_{ij}-\alpha p \delta_{ij})_{,j}+b_i\right]\delta  u_i \mathrm{d}{\Omega}}_{\textcircled{2}}
-\\ \underbrace{\int_{\Omega} \left[ 2(\phi-1)(1-k)\psi_{\varepsilon}^+ + \frac {G_c \phi}{l_0}-G_c l_0\frac{\partial^2\phi}{\partial x_i^2} \right]\delta\phi \mathrm{d}{\Omega}}_{\textcircled{3}} + \underbrace{\int_{\partial\Omega}\left( \frac{\partial\phi}{\partial x_i}n_i\right)\delta\phi \mathrm{d}S}_{\textcircled{4}}=0
	\label{first variation of the functional}
	\end{multline}

\noindent where the component of the effective stress tensor $\bm \sigma(\bm\varepsilon)$ is
	\begin{equation}
	\sigma_{ij}=\left [(1-k)(1-\phi)^2+k \right]\frac {\partial{\psi_\varepsilon^+}}{\partial {\varepsilon_{ij}}}+\frac {\partial{\psi_\varepsilon^-}}{\partial {\varepsilon_{ij}}}
	\end{equation}

Thereafter, the Cauchy stress tensor $\bm\sigma^{por}$ \citep{lee2016pressure} is set as follows:
	\begin{equation}
	\bm \sigma^{por}(\bm\varepsilon)=\bm \sigma(\bm\varepsilon)-\alpha p \bm I,\hspace{0.5cm} in \hspace{0.1cm} \Omega
	\label{Cauchy stress}
	\end{equation}

In Eq. \eqref{Cauchy stress}, the dimensionless Biot coefficient $\alpha$ reflects the extent of perturbation in the total stress $\bm\sigma^{por}$ owing to the changes in fluid pressure $p$ \citep{segall1998note}. Given that Eq. \eqref{first variation of the functional} consistently holds for all possible $\delta \bm u$ and $\delta \phi$, in Eq. \eqref{first variation of the functional}$\textcircled{2}$ and $\textcircled{3}$, except the displacement and phase field variations, the main bodies in the integrals must constantly be 0. Therefore, Eq. \eqref{first variation of the functional}$\textcircled{2}$ and $\textcircled{3}$ produce the following governing equations:
	\begin{equation}
	  \left\{
	   \begin{aligned}
	\frac {\partial {\sigma_{ij}^{por}}}{\partial x_j}+b_i=0
	\\ \left[\frac{2l_0(1-k)\psi_{\varepsilon}^+}{G_c}+1\right]\phi-l_0^2\frac{\partial^2 \phi}{\partial {x_i^2}}=\frac{2l_0(1-k)\psi_{\varepsilon}^+}{G_c}
	   \end{aligned}\right.
	\label{governing equations 0}
	\end{equation}

To ensure the availability of PFM, the irreversibility condition must be established which indicates that a fracture cannot be healed. An easy treatment is the introduction of a history field $H(\bm x,t)$, which represents the maximum tensile elastic energy density at the time interval [0,t] \citep{miehe2010phase, miehe2010thermodynamically, borden2012phase}. That is, the history field $H$ can be expressed as follows:
	\begin{equation}
	H(\bm x,t) = \max \limits_{s\in[0,t]}\psi_\varepsilon^+\left(\bm\varepsilon(\bm x,s)\right)
	\end{equation}

History field $H$ satisfies the Kuhn-Tucker conditions \citep{miehe2010phase}. Thereby, a monotonic increase in the phase field is ensured under compression or unloading. Substituting $H(\bm x,t)$ into $\psi_\varepsilon^+$, Eq. \eqref{governing equations 0} yields the following strong form:
	\begin{equation}
	  \left\{
	   \begin{aligned}
	\frac {\partial {\sigma_{ij}^{por}}}{\partial x_i}+b_i=0
		\\ \left[\frac{2l_0(1-k)H}{G_c}+1\right]\phi-l_0^2\frac{\partial^2 \phi}{\partial {x_i^2}}=\frac{2l_0(1-k)H}{G_c}
	\label{governing equation1}
	   \end{aligned}\right.
	\end{equation}

Apart from the Dirichlet boundary condition, Eq. \eqref{governing equation1} is also subjected to the Neumann condition:
	\begin{equation}
	\sigma_{ij}^{por}n_j=t_i, \hspace{1cm} \mathrm{on}\hspace{0.5cm} \partial\Omega_{t}
	\label{boundary condition of the phase field}
	\end{equation}
	
In addition, the entire calculation domain has an initial phase field $\phi_0=0$ and the approach of \citet{borden2012phase} is used to artificially induce a pre-existing crack.

\subsection{Flow field}\label{Governing equations for the flow field}

Three parts of the flow domain are distinguished: unbroken domain (reservoir domain) $\Omega_r(t)$, fracture domain $\Omega_f(t)$, and transition domain $\Omega_t(t)$ \citep{zhou2018phase2}. These three domains are defined by setting two  phase field thresholds, $c_1$ and $c_2$. The subdomain is an unbroken domain $\Omega_r(t)$ if $\phi\le c_1$, but a fracture domain $\Omega_f(t)$ if $\phi\ge c_2$. In the case of $c_1<\phi<c_2$, the subdomain is a transition domain. Given that a sharp fracture is diffused in PFM, determining the hydraulic and solid parameters in the transition domain is evidently a challenge. For simplicity, many studies such as \citet{mikelic2015phase, lee2016pressure, lee2017initialization, lee2017iterative, zhou2018phase2} have adopted linear interpolation between the unbroken and fractured domains for the flow field, which was proven to receive favorable numerical results. The current study uses a linear relationship between the hydraulic and solid parameters of the unbroken and fractured domains. Thereafter, two indicator functions, $\chi_r$ and $\chi_f$, are naturally established by applying the phase field $\phi$ \citep{lee2016pressure}:
	\begin{equation}
	\chi_r(\cdot,\phi)=\left\{
		\begin{aligned}
		&1,\hspace{2 cm}&\phi\le c_1\\ &\frac{c_2-\phi}{c_2-c_1} &c_1<\phi<c_2
\\&0,&\phi\ge c_2
		\end{aligned}\right.,\hspace{0.5cm}	\chi_f(\cdot,\phi)=\left\{
		\begin{aligned}
		&0,\hspace{2 cm}&\phi\le c_1\\ &\frac{\phi-c_1}{c_2-c_1} &c_1<\phi<c_2
		\\&1,&\phi\ge c_2
		\end{aligned}\right.
	\label{function1}
	\end{equation}

Note that the indicator functions $\chi_r$ and $\chi_f$ have also been established in \citet{mikelic2015phase, lee2016pressure, lee2017initialization, lee2017iterative, zhou2018phase2}, indicating that the fracture pattern is relatively insensitive to the indicator functions. In addition, \citet{lee2016pressure} suggested a relation $c_1=0.5-m_x$ and $c_2=0.5+m_x$, with $0<m_x<0.5$. However, this study considers $c_1$ and $c_2$ as two independent values.

Darcy's law is applied to describe the flow field in the porous media. In the entire domain, mass conservation is expressed as follows\citep{zhou2018phase2}:
	\begin{equation}
	\rho S \frac{\partial p}{\partial t}+\nabla\cdot(\rho\bm v)=q_m-\rho\alpha\chi_r\frac{\partial \varepsilon_{vol}}{\partial t}
	\label{mass conservation of the whole domain}
	\end{equation}

\noindent  where $\rho$, $S$, $\bm v$, and $\varepsilon_{vol}$ are the flow density, storage coefficient, flow velocity, volumetric strain of the domain, and source term, respectively. By denoting $\rho_r$ and $\rho_f$ as the fluid densities in $\Omega_r$ and $\Omega_f$, we have $\rho=\rho_r\chi_r+\rho_f\chi_f$; similarly, $\alpha=\alpha_r\chi_r+\alpha_f\chi_f$. Given that the Biot coefficient $\alpha=1$ for the fracture domain, $\alpha=\alpha_{r}\chi_r+\chi_f$, with $\alpha_{r}$ representing the Biot coefficient in $\Omega_r$. In addition, the volumetric strain $\varepsilon_{vol}=\nabla\cdot\bm u$. Note that we provide one feasible model for coupling the fluid flow with the phase field. We mention that more general methods such as the one proposed by \citet{pudasaini2016novel} can be applied to provide improved description of the fluid flow through the porous media.

The storage coefficient $S$ can be expressed as follows \citep{zhou2018phase2}:
 	\begin{equation}
	S=\varepsilon_pc+\frac{(\alpha-\varepsilon_p)(1-\alpha)}{K_{Vr}}
	\end{equation}

\noindent where $\varepsilon_p$, $c$, and $K_{Vr}$ are the porosity, fluid compressibility, and bulk modulus of $\Omega_r$, respectively. Naturally, $c=c_r\chi_r+c_f\chi_f$, with $c_r$ and $c_f$ as the respective fluid compressibility in $\Omega_r$ and $\Omega_f$. Note that we set $\varepsilon_p=1$ for $\Omega_f$. Therefore, $\varepsilon_p=\varepsilon_{pr}\chi_r+\chi_f$, with $\varepsilon_{pr}$ representing the porosity of the reservoir domain. 

Darcy's velocity $\bm v$ is subsequently expressed as
	\begin{equation}
	\bm v=-\frac{K}{\mu_e}(\nabla p+\rho\bm g)
	\label{velocity of the whole domain}
	\end{equation} 

\noindent where $K$ and $\mu_e$ represent the effective permeability and fluid viscosity, respectively. $K=K_{r}\chi_r+K_f\chi_f$, with $K_{r}$ and $K_{f}$ being the permeabilities of $\Omega_r$ and $\Omega_f$, respectively. $\mu_e=\mu_{r}\chi_r+\mu_f \chi_f$, while $\mu_{r}$ and $\mu_{f}$ are the fluid viscosity in $\Omega_r$ and $\Omega_f$, respectively; $\bm g$ denotes the gravity. Moreover, the novel method proposed by \citet{pudasaini2016novel} can extensively describe the fluid flow velocity in porous media. The feasibility and advantages of this method in the phase field modeling will be examined in future research.

Lastly, the following equation that governs fluid flow in $\Omega$ is expressed in terms of $p$:
	\begin{equation}
	\rho S \frac{\partial p}{\partial t}-\nabla\cdot \frac{\rho K}{\mu_e}(\nabla p+\rho\bm g)=q_m-\rho\alpha\chi_r\frac{\partial \varepsilon_{vol}}{\partial t}
	\label{governing equation of the whole domain}
	\end{equation}

\noindent which is subjected to the Dirichlet and Neumann conditions as shown in \citet{zhou2018phase2}.

\section {Numerical implementation}\label {Numerical treatment}

Finite element method (FEM) is used to solve the governing equations of the multi-field fracture problem in the porous media. Therefore, weak-form formulations are required \citep{zhou2018phase2} with a modified stiffness matrix to solve the displacement field. The implicit Generalized-$\alpha$ method \citep{borden2012phase} is used to discretize the time domain and unconditionally enhance numerical stability. In addition, we subsequently solve the three fields in a staggered manner. That is, solving each field is independent in one time step. Note that we couple $\bm u$ and $p$ and solve them in one staggered step. Table \ref{Solution flow-chart for the fracture propagation in naturally-layered porous media} presents the flowchart of the staggered scheme for the hydraulic fracture propagation in naturally-layered porous media. The Newton-Raphson iteration method is used \citep{zhou2018phase2, zhou2019phase} in each segregated step.

	\begin{table}[htbp]
	\caption{Solution flowchart for fracture propagation in the naturally-layered porous media}
	\label{Solution flow-chart for the fracture propagation in naturally-layered porous media}
	\centering
	\begin{tabular}{l}
	\toprule[1pt]
	\textbf{Initiation}\\
	\hspace{1cm} $\bm u_0$, $\bm \phi_0$, $H_0$, $p_0$\\
	\textbf{For} every successive time step $n+1$ \textbf{do}\\
	\hspace{1cm} $\bm u_{n}$, $\phi_{n}$, $H_{n}$, and $p_{n}$ are known\\
	\hspace{1cm}\textbf{Initiation}\\
			\hspace{2cm} Construct the initial guess $(\bm u, p)_{n+1}^{j=0}$, $H_{n+1}^{j=0}$, and $\phi_{n+1}^{j=0}$\\
	\hspace{1cm}\textbf{For} every successive iteration step $j+1$ \textbf{do}\\
			\hspace{2cm} $(\bm u,p)_{n}^j$, $H_{n}^j$, and $\phi_{n}$ are known\\
			\hspace{2cm} 1. Solve $(\bm u, p)_{n+1}^{j+1}$ by using $(\bm u, p)_{n+1}^{j}$, $H_{n+1}^{j}$, and $\phi_{n+1}^{j}$\\
			\hspace{2cm} 2. Update $H_{n+1}^{j+1}$ by using $(\bm u, p)_{n+1}^{j}$ and $H_{n+1}^{j}$\\
			\hspace{2cm} 3. Solve $\phi_{n+1}^{j+1}$ by using $(\bm u, p)_{n+1}^{j+1}$, $H_{n+1}^{j+1}$, and $\phi_{n+1}^{j}$\\
			\hspace{2cm} 4. Evaluate the global relative error of $(\bm u, p, H, \phi)$\\
		\hspace{1cm} \textbf{until} the tolerance $< \varepsilon_t=1\times10^{-4}$\\
	\textbf{End}\\
	\bottomrule[1pt] 
	\end{tabular}
	\end{table}

\section{Validation of PFM for hydraulic fracture}\label{Further validation of PFM}

The fluid-driven fracture in a permeable solid is affected by competing dissipative processes owing to fluid viscosity and medium toughness \citep{detournay2016mechanics}. Therefore, we check the fracture propagation and fluid pressure in the following cases \citep{santillan2017phase} to further validate the used phase field method:
\begin{itemize}
		\item Toughness-dominated regime. The porous medium is impermeable and fracturing expends substantially higher energy than the viscous dissipation.
		\item Viscous-dominated regime. The porous medium is impermeable and energy dissipation in fracture propagation is considerably lower than the viscous dissipation.
		\item Leak-off toughness-dominated regime. The porous medium is permeable and more fluid is stored in the porous medium than in the fracture. In addition, fracturing expends significantly higher energy than the viscous dissipation.
\end{itemize}

A 60 m $\times$ 30 m rectangular domain is considered with an initial injection notch of 1.2 m $\times$ 0.2 m in the domain center. After being modeled through the initial history field, the initial notch has a fluid injection rate of $Q_f$. All the outer boundaries of the domain are permeable with $p=0$ and fixed in the displacement. The parameters listed in Tables \ref{Basic parameters for the validation example} and \ref{Parameters in the toughness dominated, viscous dominated, and leak-off toughness dominated regimes} are used in the simulation. Note that these parameters correspond to those used in \citet{santillan2017phase} for different fluid regimes.

	\begin{table}[htbp]
	\small
	\caption{Basic parameters for the validation example}
	\label{Basic parameters for the validation example}
	\centering
	\begin{tabular}{llllllll}
		\hline
		$\mu$ & 4.722 GPa & $\lambda$ &7.083 GPa & $G_c$ &120 N/m & $k$ &$1\times10^{-9}$ \\ $l_0$ & 0.4 m & $c_1$ & 0.4 & $c_2$ & 1.0 & $\varepsilon_{pr}$ & 0.088 \\ $\rho_{r}$,  $\rho_{f}$& $1.0\times10^{3}$ kg/m$^3$  & $\alpha_r$ &0.088 & $q_r$ & 0 & $K_r$ & $1\times10^{-15}$ m$^2$ \\ $K_f$ & $5\times10^{-8}$ m$^2$ &$c_r$ & $1\times10^{-8}$ 1/Pa & $c_f$ & $1\times10^{-8}$ 1/Pa & $\mu_r$ & $1\times10^{-3}$ Pa$\cdot$s \\
		\hline
		\normalsize
	\end{tabular}
	\end{table}

	\begin{table}[htbp]
	\small
	\caption{Parameters in the toughness-dominated, viscous-dominated, and leak-off toughness-dominated regimes}
	\label{Parameters in the toughness dominated, viscous dominated, and leak-off toughness dominated regimes}
	\centering
	\begin{tabular}{llllll}
		\hline
		\multicolumn{2}{l}{Toughness dominated}&\multicolumn{2}{l}{Viscous dominated}&\multicolumn{2}{l}{Leak-off toughness dominated}\\ \hline
		\multicolumn{2}{l}{Leak-off term is ignored}&\multicolumn{2}{l}{Leak-off term is ignored}&\multicolumn{2}{l}{Leak-off term is included}\\
		$Q_f$&$10^{-3}$ m$^2$/s&$Q_f$&$2\times10^{-4}$ m$^2$/s&$Q_f$&$10^{-2}$ m$^2$/s\\$\mu_f$&$1\times10^{-6}$ Pa$\cdot$s&$\mu_f$&0.1 Pa$\cdot$s&$\mu_f$&$1\times10^{-6}$ Pa$\cdot$s\\
		\hline
		\normalsize
	\end{tabular}
	\end{table}

\begin{figure}[htbp]
	\centering
	\subfigure[Toughness dominated]{\includegraphics[width = 5.5cm]{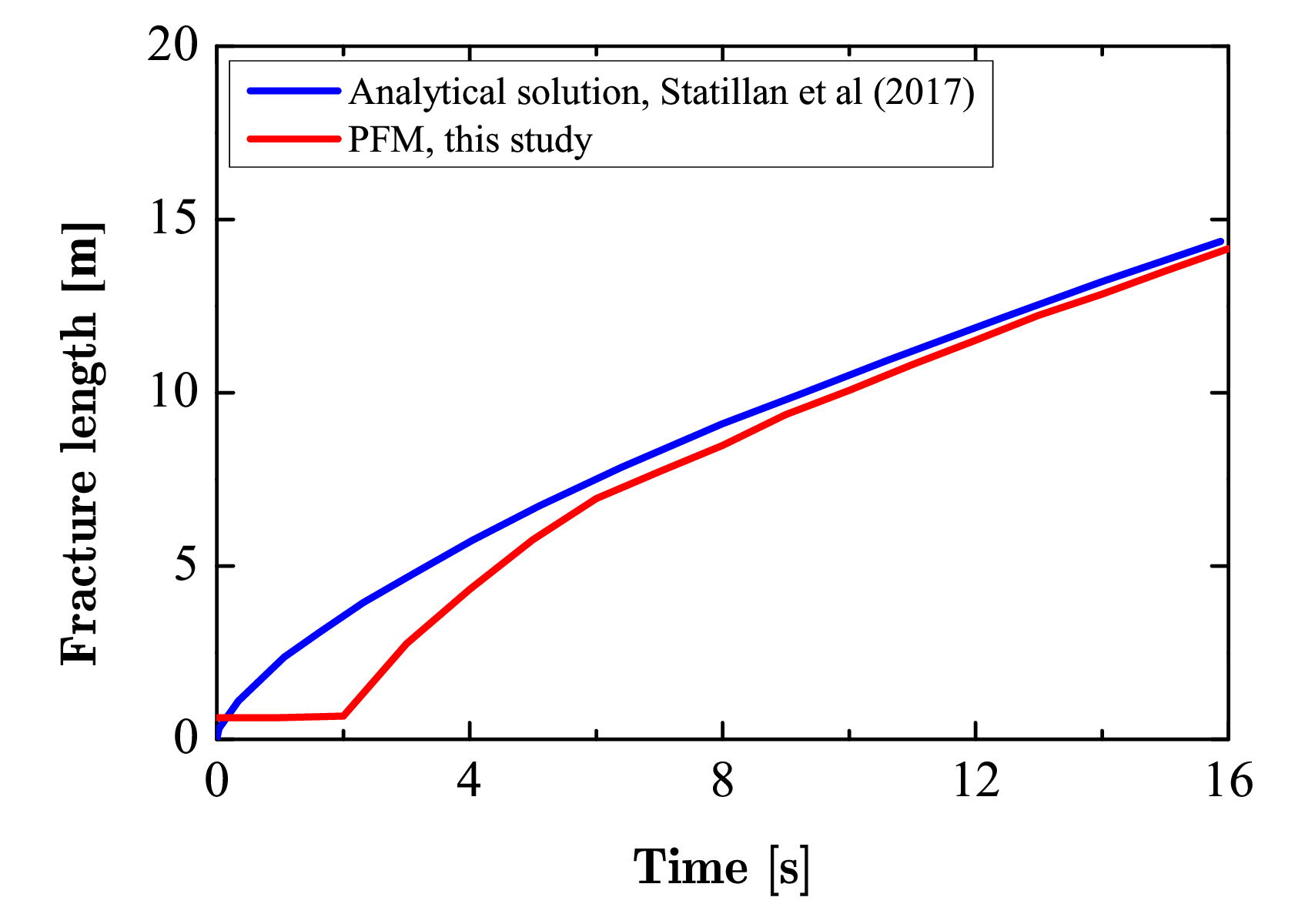}}
	\subfigure[Viscous dominated]{\includegraphics[width = 5.5cm]{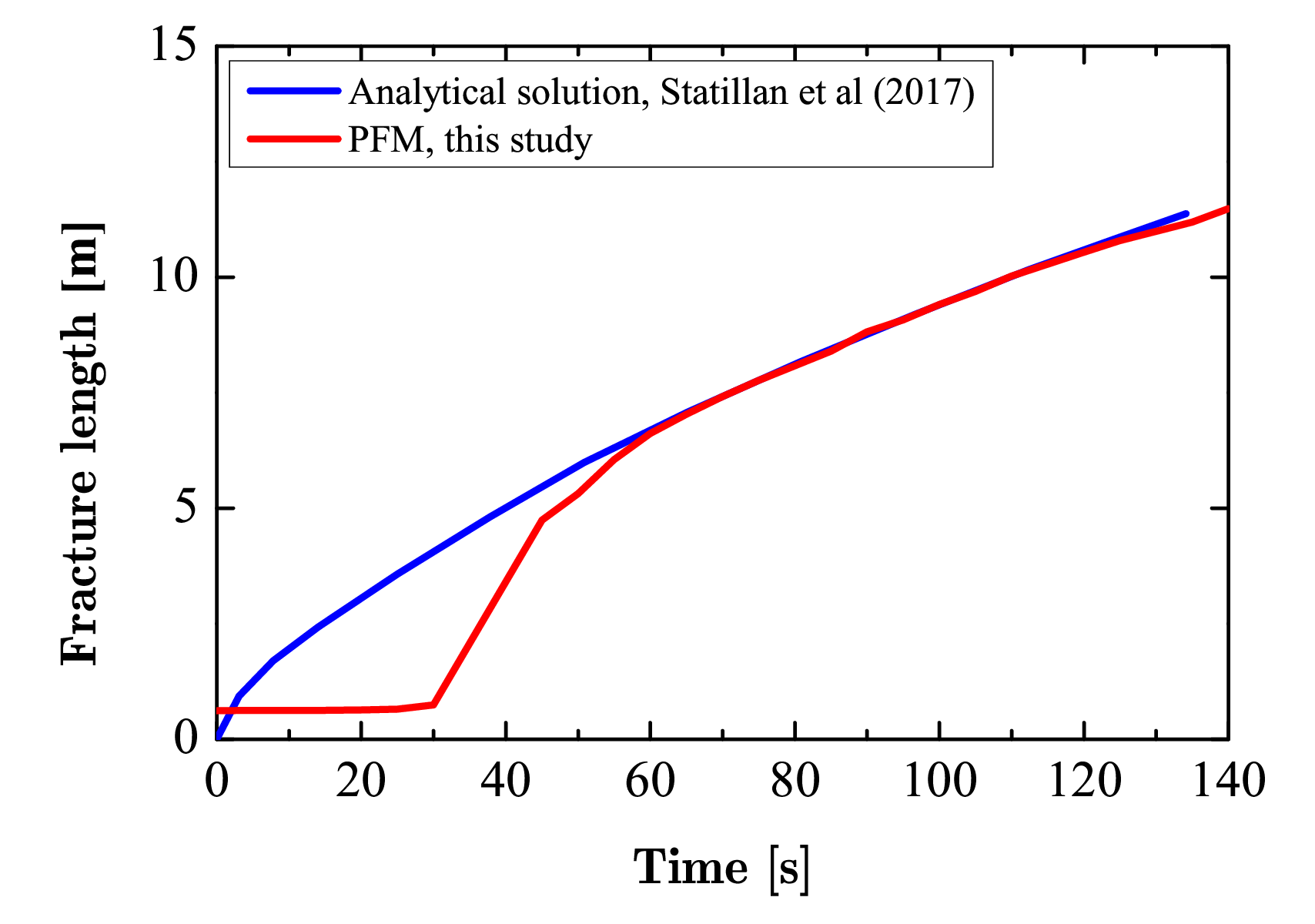}}
	\subfigure[Leak-off dominated]{\includegraphics[width = 5.5cm]{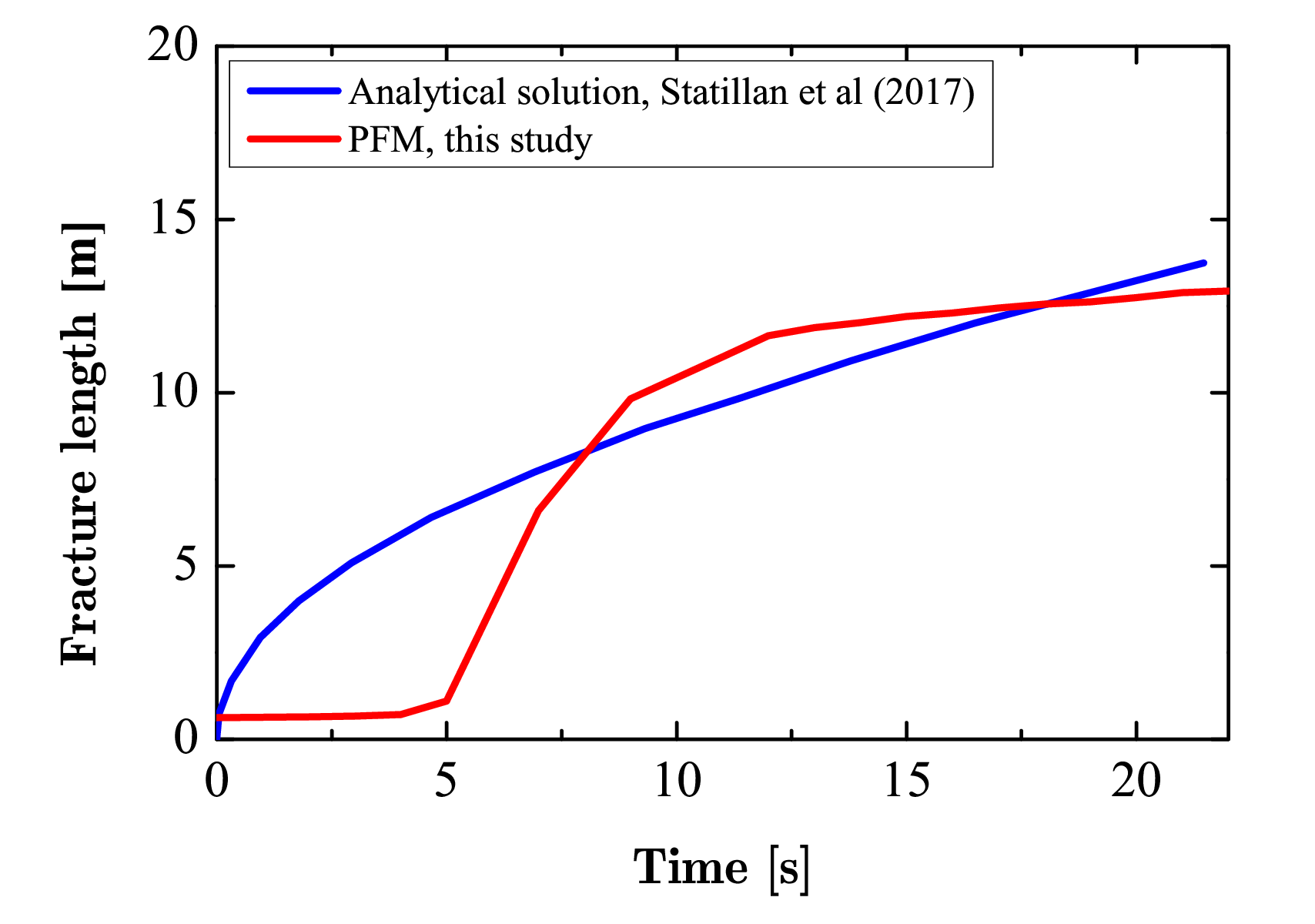}}\\
	\caption{Evolution of half-length of the fracture}
	\label{Evolution of half-length of the fracture}
	\end{figure}

\begin{figure}[htbp]
	\centering
	\subfigure[Toughness dominated]{\includegraphics[width = 5.5cm]{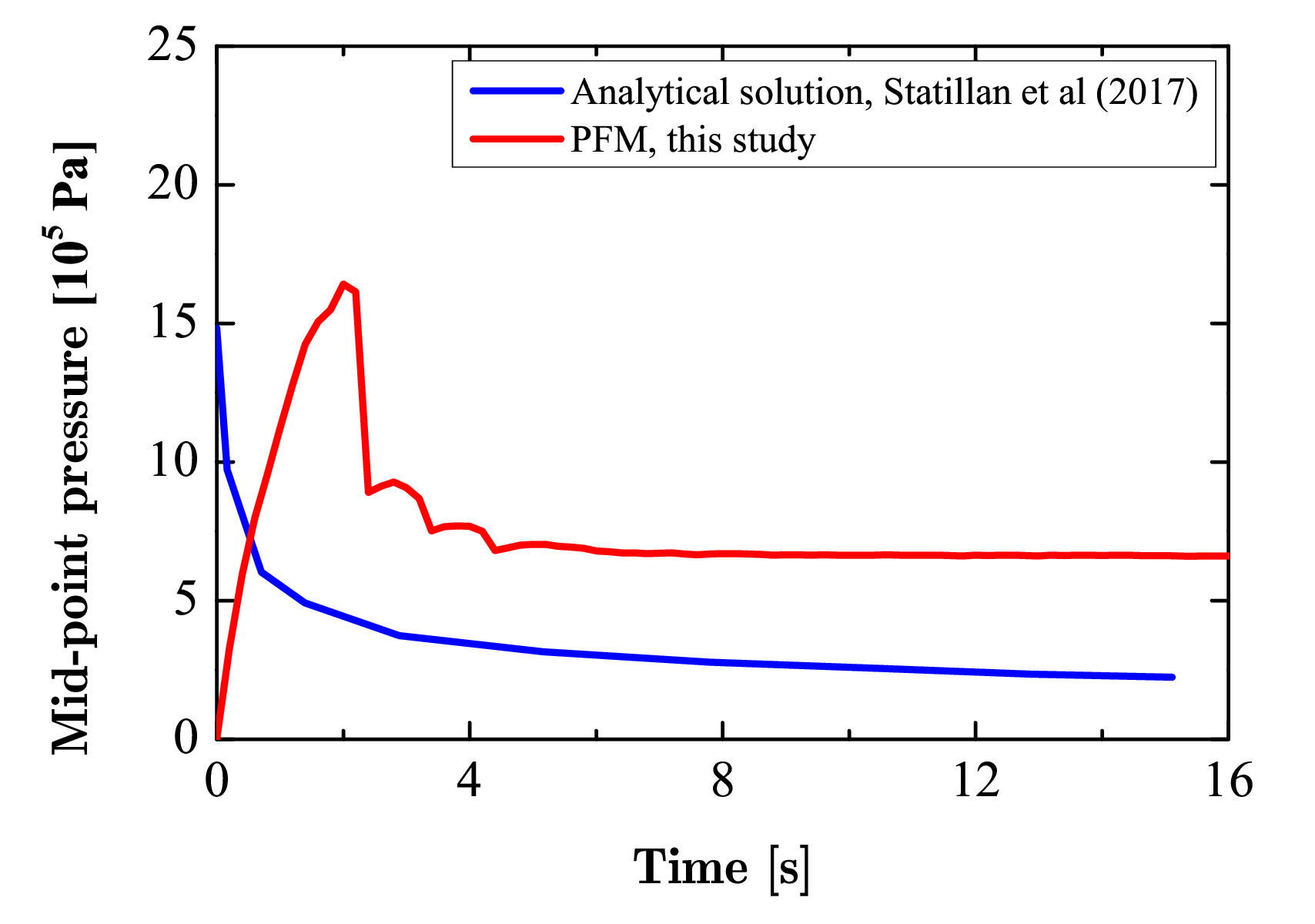}}
	\subfigure[Viscous dominated]{\includegraphics[width = 5.5cm]{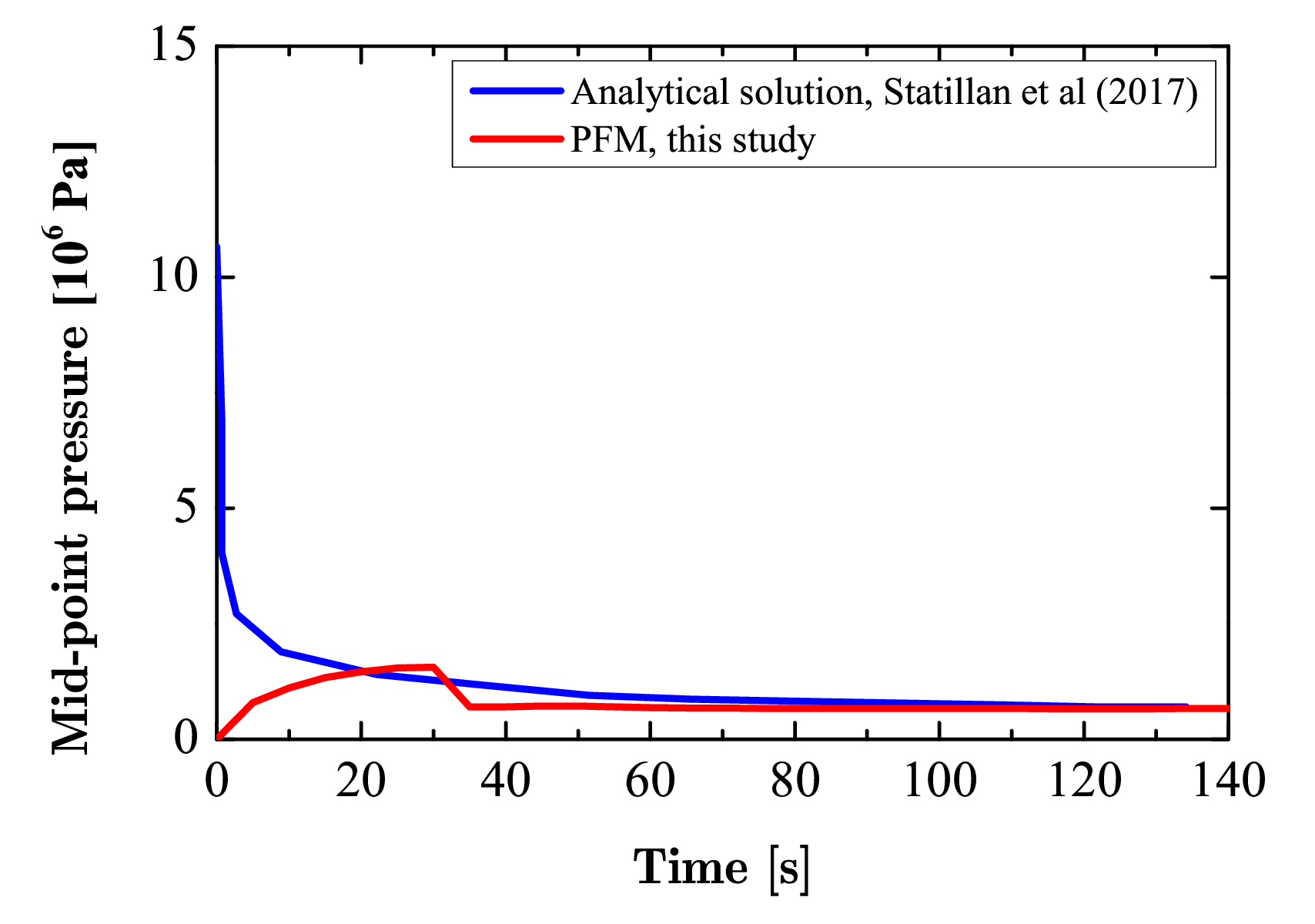}}
	\subfigure[Leak-off dominated]{\includegraphics[width = 5.5cm]{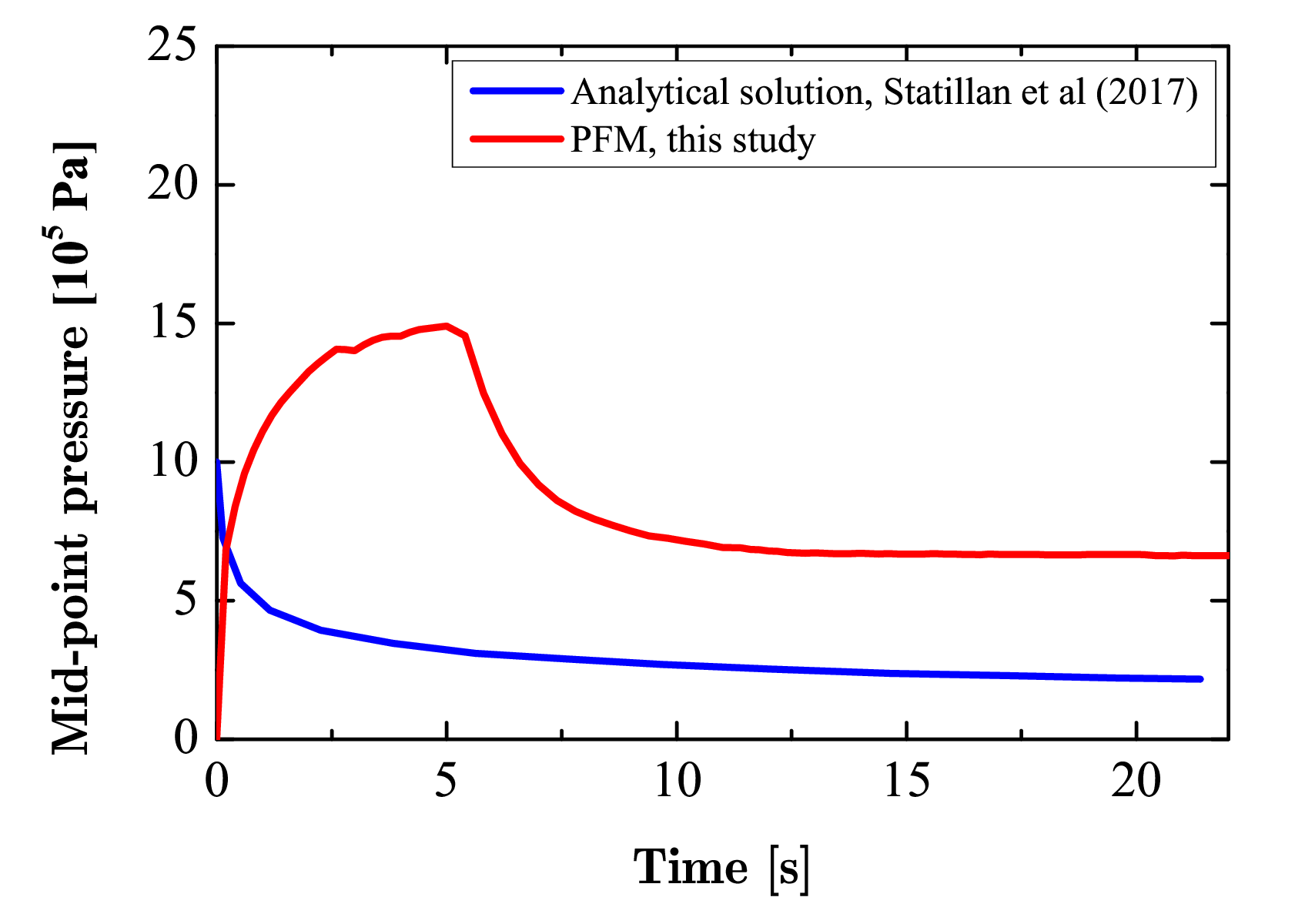}}\\
	\caption{Evolution of fluid pressure at the center of the fracture}
	\label{Evolution of fluid pressure at the center of the fracture}
	\end{figure}

We discretize the domain with uniform quadrilateral elements with the maximum element size of $h=0.2$ m. In all simulations, a time increment $\Delta t= 0.025$ s is used. Figures \ref{Evolution of half-length of the fracture} and \ref{Evolution of fluid pressure at the center of the fracture} show the evolution of half-length of the fracture and fluid pressure at the fracture center for the toughness-dominated, viscous-dominated, and leak-off dominated regimes. The two figures also compare the results obtained by PFM and those by using the analytical approach presented by \citet{santillan2017phase}. Figures. \ref{Evolution of half-length of the fracture} and \ref{Evolution of fluid pressure at the center of the fracture} show that compared with the analytical method, the used PFM reproduces consistent trends in the evolution of the half-length of the fracture and mid-point fluid pressure. 
	
Multiple reasons account for the differences in the actual values obtained by both methods. First, the used PFM aims for a permeable porous medium, whereas the analytical approach of \citet{santillan2017phase} is for an impermeable medium. Despite using parameters that approximate an impermeable solid, this important difference cannot be completely eliminated in our phase field simulation owing to the inevitable fluid penetration. Second,  the fracture in the phase field simulation evolves from an initial notch with a length of 1.2 m. Therefore, the half-length of the fracture starts from 0.6 m, and an evident ascending stage is observed before the fracture initiation in the fluid pressure-time curves. However, the analytical method assumes point-injection fluid and the initial fracture length is 0, while the increase stage before fracture initiation is disregarded. Moreover, the used PFM and analytical method include different flow models. Especially, for the leak-off toughness-dominated regime, the leak-off term in PFM is related to the volumetric strain of the domain while a fixed leak-off coefficient of $5\times10^{-4}$ m/s$^{1/2}$ is used in the analytical method according to \citet{santillan2017phase}. Another reason is that PFM smears the fracture in a finite width, while the analytical method deals with the fracture in a discrete setting.

\section {Soft-to-stiff configuration}\label{The soft to stiff configuration}
\subsection{Geometry and boundary conditions}\label{Geometry and boundary conditions}

\begin{figure}[htbp]
	\centering
	\includegraphics[width = 8cm]{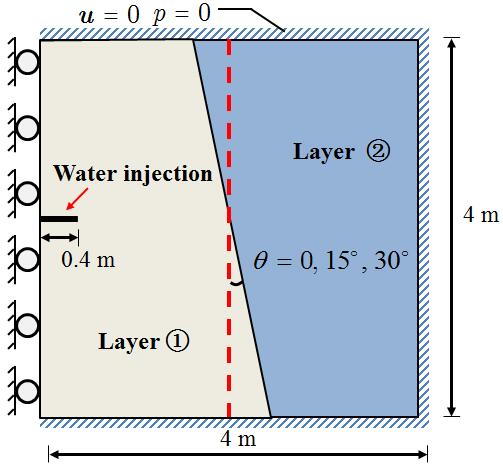}
	\caption{Geometry and boundary conditions of the hydraulic fracture propagation in two porous layers}
	\label{Geometry and boundary conditions of the hydraulic fracture propagation in two porous layers}
	\end{figure}

We investigate the hydraulic fracture pattern in two porous layers, which is a relatively typical setting in engineering geology, rock engineering, and oil/gas exploration. The setup for the two-media problem is described in Fig. \ref{Geometry and boundary conditions of the hydraulic fracture propagation in two porous layers}; the center of the analysis domain is set as the origin of the coordinate system. 2D simulations are performed and we mark the two layers with the notes, $\textcircled{1}$ and $\textcircled{2}$, respectively. A pre-existing notch is set in the layer $\textcircled{1}$, with its position and length also shown in Fig. \ref{Geometry and boundary conditions of the hydraulic fracture propagation in two porous layers}. The width of the pre-existing notch is $l_0$, while the fluid is injected into the notch with the source term $q_f=$ 10 kg/(m$^3\cdot \textrm s$). Except the boundary conditions in Fig. \ref{Geometry and boundary conditions of the hydraulic fracture propagation in two porous layers}, the left boundary of the layer $\textcircled{1}$ is impermeable. The interface of the two layers has an inclination angle of $\theta$. We also test the influence of the inclination angle and choose $\theta = 0^\circ$, $15^\circ$, and $30^\circ$. Note that although our simulations are performed in a relative small scale that corresponds to the XFEM simulations of \citet{vahab2018x}, the phase field simulation can be easily extended to large-scale problem by adjusting the geometry, mesh size, and other relevant parameters.

A total of twelve cases are examined using the various values of Young's modulus and $G_c$ listed in Table \ref{Young's modulus and critical energy release rate of the two layers}. The Young's modulus $E$ and $G_c$ of the two layers are denoted as $E_1$, $E_2$, $G_{c1}$, and $G_{c2}$. Note that a stiffer layer exhibits a larger $E$ and $G_c$ than the softer layer. The Poisson's ratios of the two layers are the same(i.e., $\nu_1=\nu_2=0.3$). The parameter setting in this paper is similar to that of \citet{vahab2018x} where fracture toughness $k_I$ is used but $G_c$ is used in the PFM. Note that the selection of the different values of Young's modulus and critical energy release rate will not affect the fracture patterns in layered domain which are depicted in Fig. \ref{Three fracture patterns when hydraulic fractures reach the layer interface}. The other parameters of the two layers for calculation are listed in Table \ref{Parameters for the two layers}. It is quite often the case in substantial rock formation the contrast of Young's modulus and tensile strength between soft and hard layers can be few times  such as the pseudo fracture studies in \cite{Gu2006SPE}. Therefore, the following cases with various combinations of material parameters for soft and hard layers relatively represent classic situations such as contract about 3 times in Young's modulus were used in \cite{SmithSPE2001}. 

	\begin{table}[htbp]
	\small
	\caption{Young's modulus and $G_c$ of the two layers}
	\label{Young's modulus and critical energy release rate of the two layers}
	\centering
	\begin{tabular}{lllllll}
	\hline
	Case number & Relationship & $E_1$/ GPa &$E_2$ / GPa&$G_{c1}$/ (N/m)&$G_{c2}$/ (N/m)&$\theta$/$^\circ$\\
	\hline
	\makecell{1\\ 2 \\ 3} & $E_2/E_1=2$ & 60 &120 & 2400 & 4800 & \makecell{0\\ 15 \\ 30}\\
	\hline
	\makecell{4\\ 5 \\ 6} & $E_2/E_1=4$ & 30 &120 & 1200 & 4800 & \makecell{0\\ 15 \\ 30}\\
	\hline
	\makecell{7\\ 8 \\ 9} & $E_2/E_1=1/2$ & 120 &60 & 4800 & 2400 & \makecell{0\\ 15 \\ 30}\\
	\hline
	\makecell{10\\ 11 \\ 12} & $E_2/E_1=1/3$ & 120 & 40 & 4800 & 1600 & \makecell{0\\ 15 \\ 30}\\
	\hline
	\normalsize
	\end{tabular}
	\end{table}

	\begin{table}[htbp]
	\small
	\caption{Parameters for the two layers}
	\label{Parameters for the two layers}
	\centering
	\begin{tabular}{llllllllll}
	\hline
	$l_0$ & $4\times10^{-3}$ m & $\alpha_r$ &0.002 & $\rho_{r}$ & $1.0\times10^{3}$ kg/m$^3$ & $k$ &$1\times10^{-9}$ \\
	$c_1$ & 0.4 & $c_2$ & 1.0 & $\varepsilon_{pr}$ & 0.002 & $\rho_{f}$& $1.0\times10^{3}$ kg/m$^3$  \\ $q_r$ & 0 & $q_f$ & 0 & $K_r$ & $1\times10^{-15}$ m$^2$ & $K_f$ & $1.333\times10^{-6}$ m$^2$\\
$c_r$ & $1\times10^{-8}$ 1/Pa & $c_f$ & $1\times10^{-8}$ 1/Pa & $\mu_r$ & $1\times10^{-3}$ Pa$\cdot$s & $\mu_f$ & $1\times10^{-3}$ Pa$\cdot$s\\
	\hline
	\normalsize
	\end{tabular}
	\end{table}

\begin{figure}[htbp]
	\centering
	\subfigure[$\theta = 0^\circ$]{\includegraphics[width = 5cm]{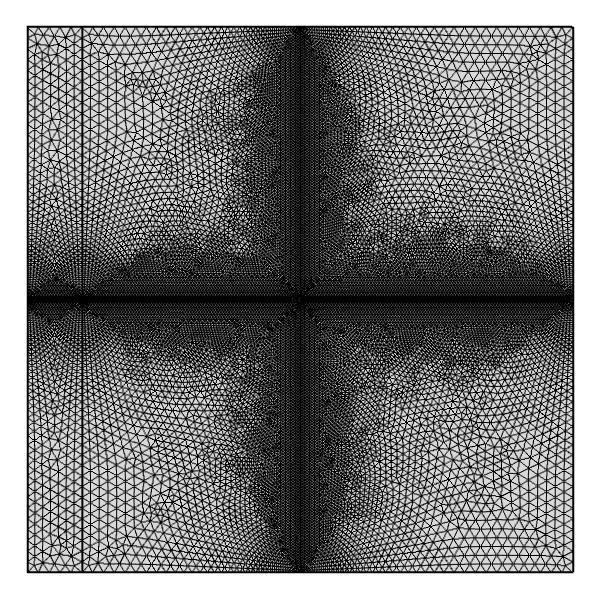}}
	\subfigure[$\theta = 15^\circ$]{\includegraphics[width = 5cm]{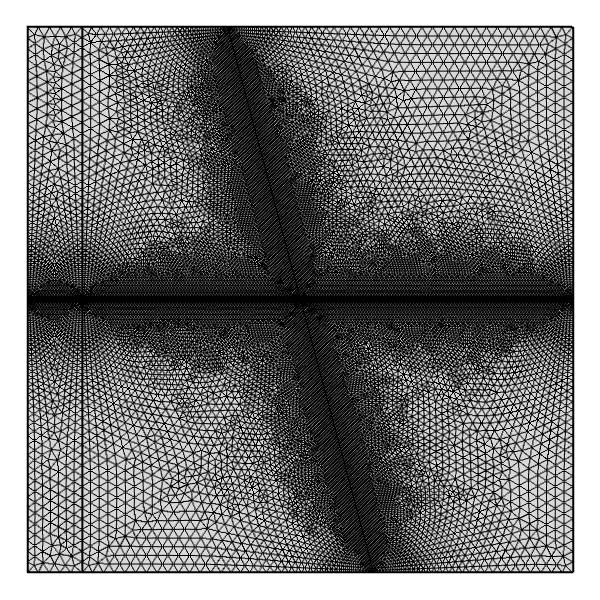}}
	\subfigure[$\theta = 30^\circ$]{\includegraphics[width = 5cm]{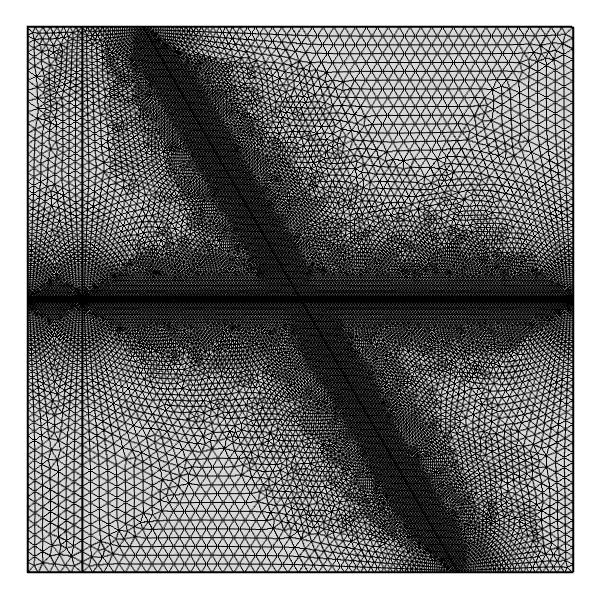}}\\
	\caption{Meshes for calculation}
	\label{Meshes for calculation}
	\end{figure}

The computation is performed based on the finite element meshes shown in Fig. \ref{Meshes for calculation}. In most of the domain we use linear triangular elements that have a maximum size $h=8$ mm. In addition, we restrict the mesh size in the region where fractures may initiate and propagate to $h=$ 2 mm. The pre-existing notch is established through the initial history field $H_0$ and the notch width is fixed as 4 mm. In the simulation, a time step $\Delta t$ = 0.05 s is used.

\subsection{Fracture pattern}\label{Fracture pattern of soft to stiff}

The settings of the numerical model are classified into two types, namely, soft-to-stiff configuration (Cases 1 to 6) and stiff-to-soft configuration (Cases 7 to 12). Note again that only 2D simulations are performed because 3D cases are time-consuming and unsuitable for analysis on multiple influencing factors. In addition, gravity is disregarded for simplicity, while $x$- and $y$- axes correspond to the horizontal and vertical directions, respectively. 

\begin{figure}[htbp]
	\centering
	\subfigure[$\theta = 0^\circ$]{\includegraphics[width = 5cm]{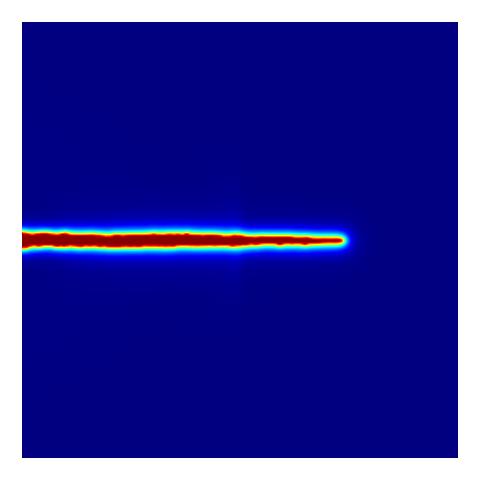}}
	\subfigure[$\theta = 15^\circ$]{\includegraphics[width = 5cm]{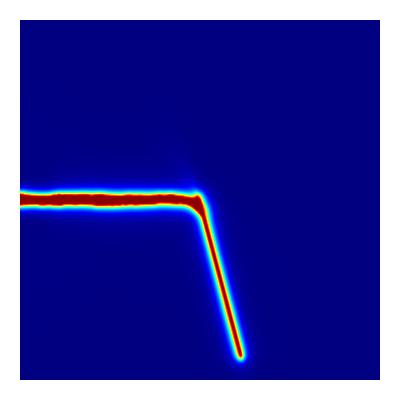}}
	\subfigure[$\theta = 30^\circ$]{\includegraphics[width = 5cm]{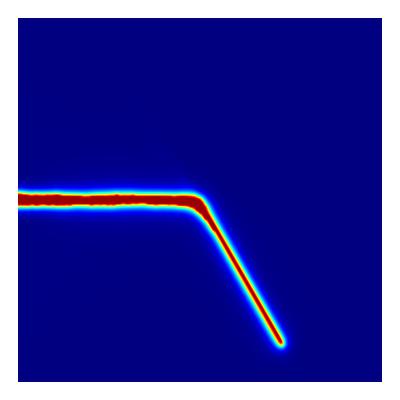}}\\
	\caption{Fracture propagation patterns for $E_2=2E_1$ at t = 200 s}
	\label{Fracture propagation patterns for E_2=2E_1 at t = 200 s}
	\end{figure}

Figure \ref{Fracture propagation patterns for E_2=2E_1 at t = 200 s} depicts the hydraulic fracture patterns for  $E_2=2E_1$ at $t$ = 200 s, in which the fractures completely propagate and their paths are evident. When the inclination angle $\theta = 0^\circ$, the penetration scenario is shown. The fracture propagates perpendicular to the interface between two layers of rock and penetrates deep into the layer $\textcircled{2}$. The reason is that the stiffness of the layer $\textcircled{2}$ is insufficiently large to depress the fracture across the layer interface. The increasing fluid pressure increases the tensile stress around the fracture tip. Thereafter, the effective stress becomes higher than the tensile strength depending on the Young's modulus, fracture toughness, and length scale in the stiffer layer at the layer interface according to \citet{zhou2018phase3}. Therefore, the fracture cannot be prevented.

A singly-deflected scenario is observed when the inclination angle $\theta= 15^\circ$ and $30^\circ$. At an inclination angle, the fracture propagation is blocked when it reaches the layer interface. The fracture is arrested by the interface and then propagates along the interface towards the bottom boundary of the domain because of the increasing fluid injection. The singly-deflected scenario results from the fact that the fluid pressure-induced stress rotates at the layer interface in a complex manner. In this case, the decomposed effective stress exceeds the tensile strength along the interface more easily than that in the second layer.

\begin{figure}[htbp]
	\centering
	\subfigure[$\theta = 0^\circ$, $t$ = 100 s]{\includegraphics[width = 5cm]{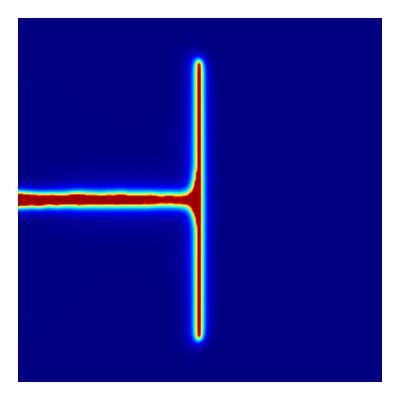}}
	\subfigure[$\theta = 15^\circ$, $t$ = 100 s]{\includegraphics[width = 5cm]{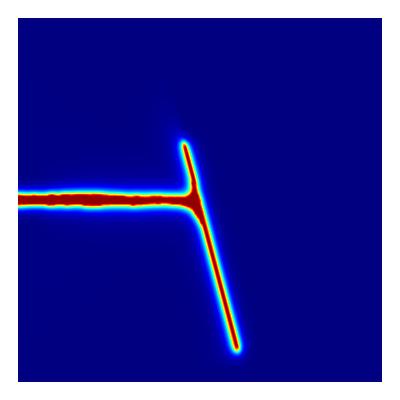}}
	\subfigure[$\theta = 30^\circ$, $t$ = 63.02 s]{\includegraphics[width = 5cm]{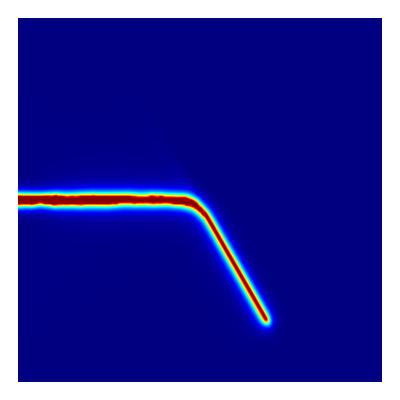}}\\
	\caption{Fracture propagation patterns for $E_2=4E_1$}
	\label{Fracture propagation patterns for E_2=4E_1}
	\end{figure}

Figure \ref{Fracture propagation patterns for E_2=4E_1} presents the hydraulic fracture patterns for $E_2=4E_1$ at different times. When $E_2=4E_1$, the fracture patterns are not the same as those for $E_2=2E_1$. The doubly-deflected scenario is observed when $\theta = 0^\circ$ and $15^\circ$. The stiffness of the layer $\textcircled{2}$ is sufficiently large and the increasing elastic energy cannot support a fracture penetration into the second layer. The tensile strength of the layer $\textcircled{2}$ is approximately four times that of the layer $\textcircled{1}$ according to \citet{zhou2018phase3}. The fracture branching is a natural result of the evolution equation of the phase field after the fracture reaches the layer interface. Given different inclination angles, the time for a fracture propagates to the same distance is relatively different. Therefore, the time point displayed is 100 s in Fig. \ref{Fracture propagation patterns for E_2=4E_1}a and \ref{Fracture propagation patterns for E_2=4E_1}b and 63.02 s in Fig. \ref{Fracture propagation patterns for E_2=4E_1}c.

The branched fractures are symmetrical for $\theta = 0^\circ$ but asymmetric for $\theta = 15^\circ$. The lower fracture is longer than the upper fracture because of the asymmetry of the layer interface along the vertical direction. For the inclination angle $\theta =30^\circ$, the fracture pattern is similar to that of $E_2=2E_1$ in Fig. \ref{Fracture propagation patterns for E_2=2E_1 at t = 200 s}c. Only the singly-deflected scenario is shown and the propagating fracture moves to the bottom boundary while the case of $E_2=4E_1$ produces a relatively large fracture propagation velocity. In addition, Figs. \ref{Fracture propagation patterns for E_2=2E_1 at t = 200 s} and \ref{Fracture propagation patterns for E_2=4E_1} show a narrower fracture after the fracture penetrates or deflects because of unavoidable fluid penetration in the fractures formed at a relatively small fluid pressure level with normal extension of damage.

\subsection{Effective maximum stress}\label{Effective maximum stress of soft to stiff}
 
\begin{figure}[htbp]
	\centering
	\subfigure[$\theta = 0^\circ$, $t$ = 55 s]{\includegraphics[width = 7.5cm]{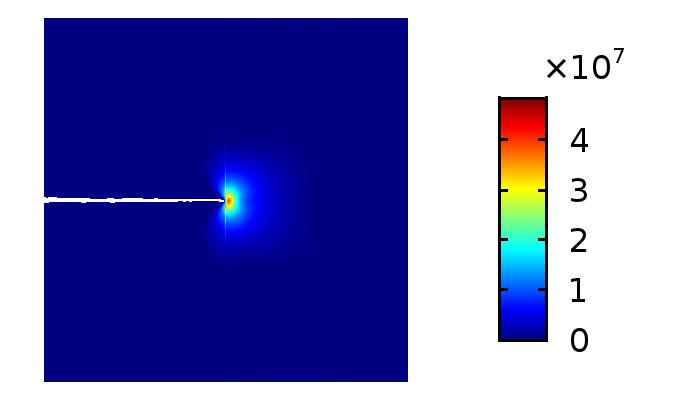}}
	\subfigure[$\theta = 0^\circ$, $t$ = 200 s]{\includegraphics[width = 7.5cm]{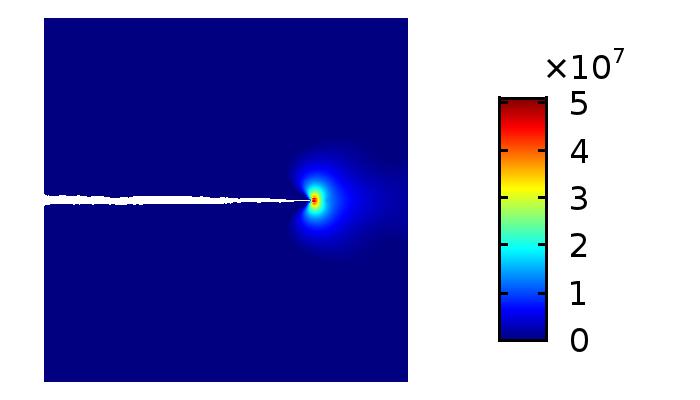}}\\
	\subfigure[$\theta = 15^\circ$, $t$ = 55 s]{\includegraphics[width = 7.5cm]{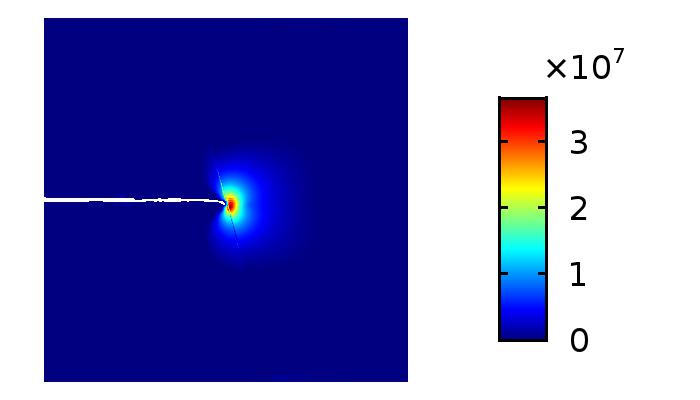}}
	\subfigure[$\theta = 15^\circ$, $t$ = 200 s]{\includegraphics[width = 7.5cm]{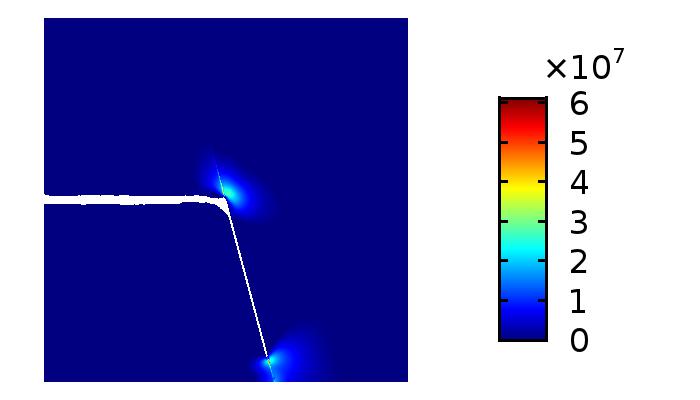}}\\
	\subfigure[$\theta = 30^\circ$, $t$ = 55 s]{\includegraphics[width = 7.5cm]{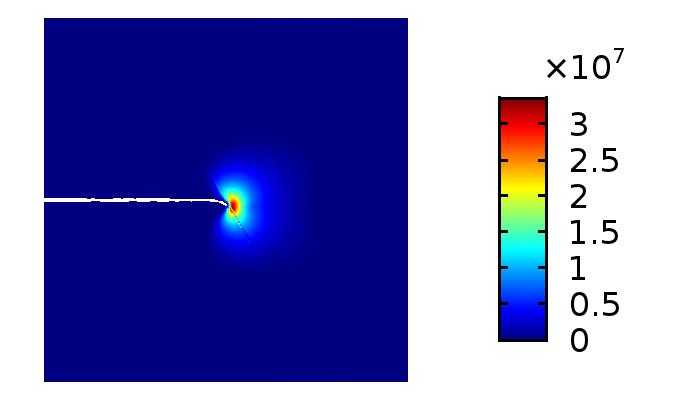}}
	\subfigure[$\theta = 30^\circ$, $t$ = 200 s]{\includegraphics[width = 7.5cm]{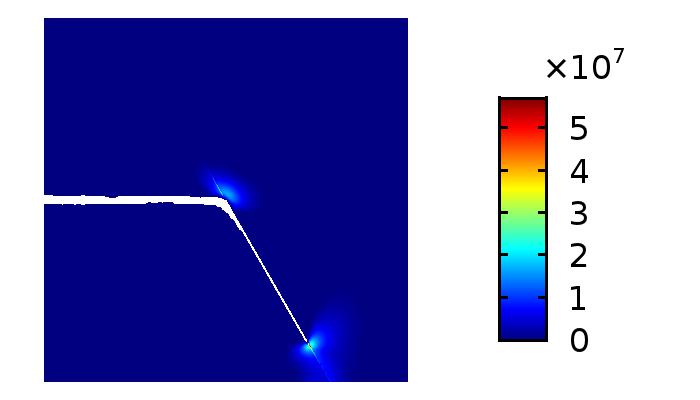}}\\
	\caption{Effective maximum stress distributions for $E_2=2E_1$ (Unit: Pa)}
	\label{Effective maximum stress distributions for E_2=2E_1}
	\end{figure}

\begin{figure}[htbp]
	\centering
	\subfigure[$\theta = 0^\circ$, $t$ = 34 s]{\includegraphics[width = 7.5cm]{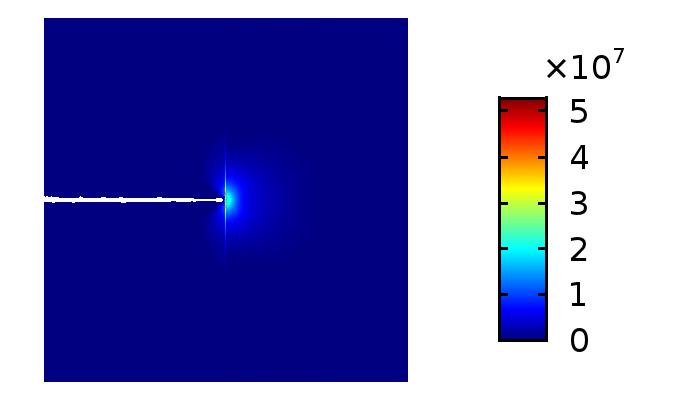}}
	\subfigure[$\theta = 0^\circ$, $t$ = 100 s]{\includegraphics[width = 7.5cm]{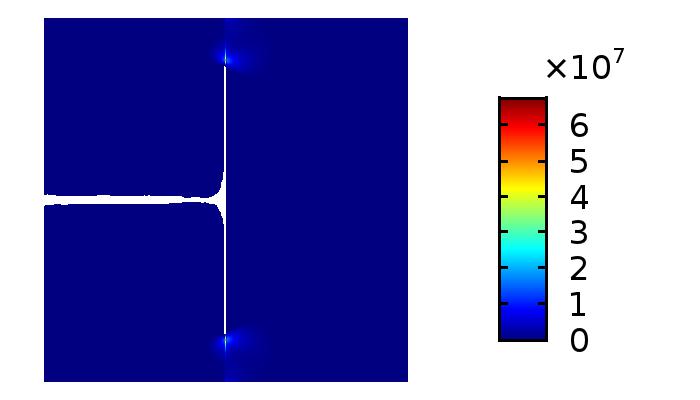}}\\
	\subfigure[$\theta = 15^\circ$, $t$ = 34 s]{\includegraphics[width = 7.5cm]{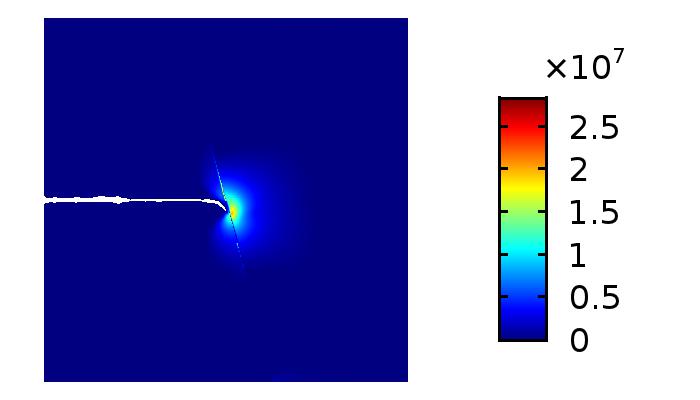}}
	\subfigure[$\theta = 15^\circ$, $t$ = 100 s]{\includegraphics[width = 7.5cm]{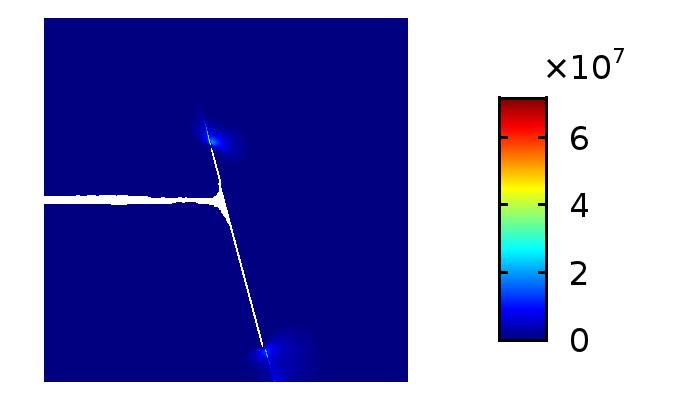}}\\
	\subfigure[$\theta = 30^\circ$, $t$ = 34 s]{\includegraphics[width = 7.5cm]{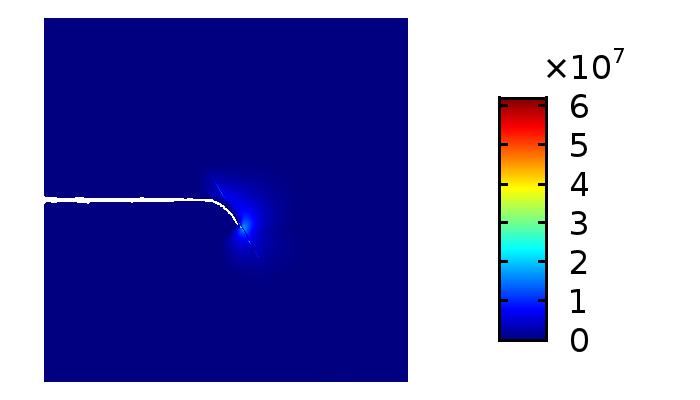}}
	\subfigure[$\theta = 30^\circ$, $t$ = 63.02 s]{\includegraphics[width = 7.5cm]{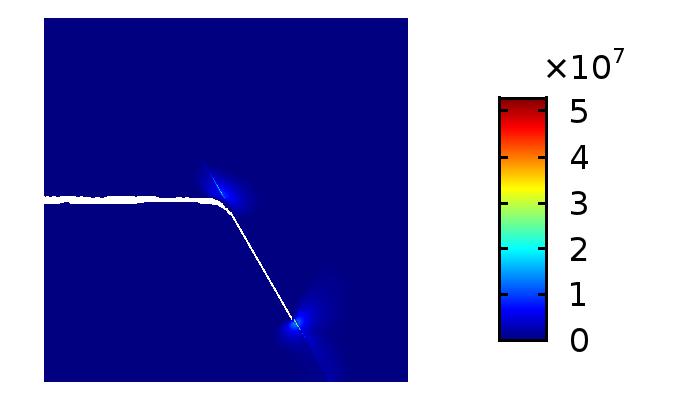}}\\
	\caption{Effective maximum stress distributions for $E_2=4E_1$ (Unit: Pa)}
	\label{Effective maximum stress distributions for E_2=4E_1}
	\end{figure}

In the soft-to-stiff configuration, the effective maximum stress distributions for $E_2=2E_1$ and $E_2=4E_1$ are displayed in Figs. \ref{Effective maximum stress distributions for E_2=2E_1} and \ref{Effective maximum stress distributions for E_2=4E_1}, respectively. Note that the effective maximum stress in this study is defined as max$(\sigma_1,0)$, with $\sigma_1$ being the effective first principal stress. In addition, the region with $\phi\ge0.95$ is removed from Figs. \ref{Effective maximum stress distributions for E_2=2E_1} and \ref{Effective maximum stress distributions for E_2=4E_1} to show an improved fracture shape. The stress distributions coincide with the fracture patterns. The stress concentration is observed only around the fracture tip for the penetration and doubly-deflected scenarios. However, stress for the singly-deflected scenario also concentrates in the area where the fracture deflects, except the fracture tip. Note that the stress values in Figs. \ref{Effective maximum stress distributions for E_2=2E_1} and \ref{Effective maximum stress distributions for E_2=4E_1} are realistic because stress singularity exists at a fracture tip and the stress value theoretically approaches infinity based on linear elastic fracture mechanics. However, the phase field in a PFM smears the sharp fracture. Therefore, the tensile stress value at the fracture tip cannot be infinity. By contrast, this stress decreases as the length scale $l_0$ increases.
 
\subsection{Displacement field}\label{Displacement field of soft to stiff}

\begin{figure}[htbp]
	\centering
	\subfigure[$E_2=2E_1$]{\includegraphics[width = 7.5cm]{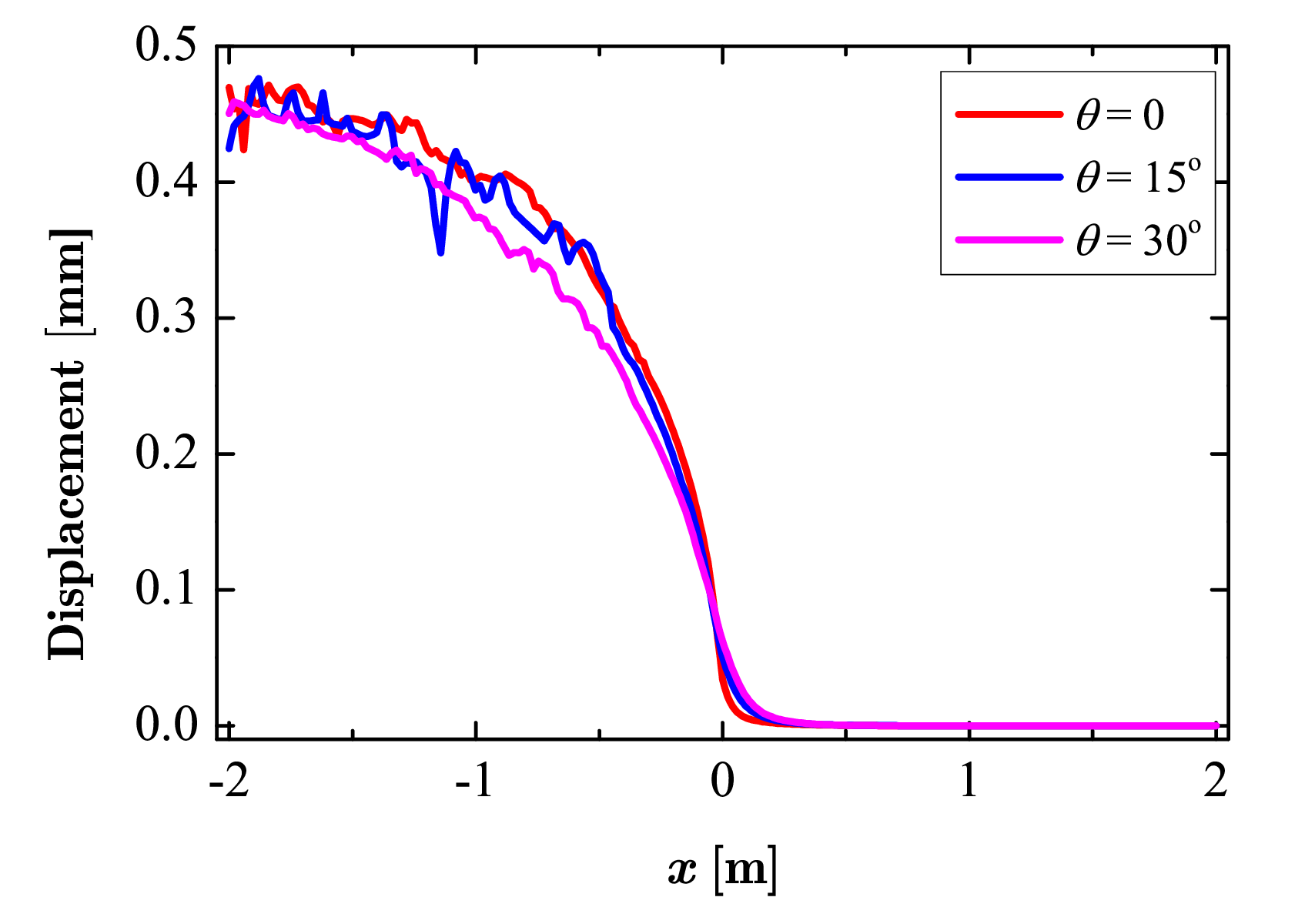}}
	\subfigure[$E_2=4E_1$]{\includegraphics[width = 7.5cm]{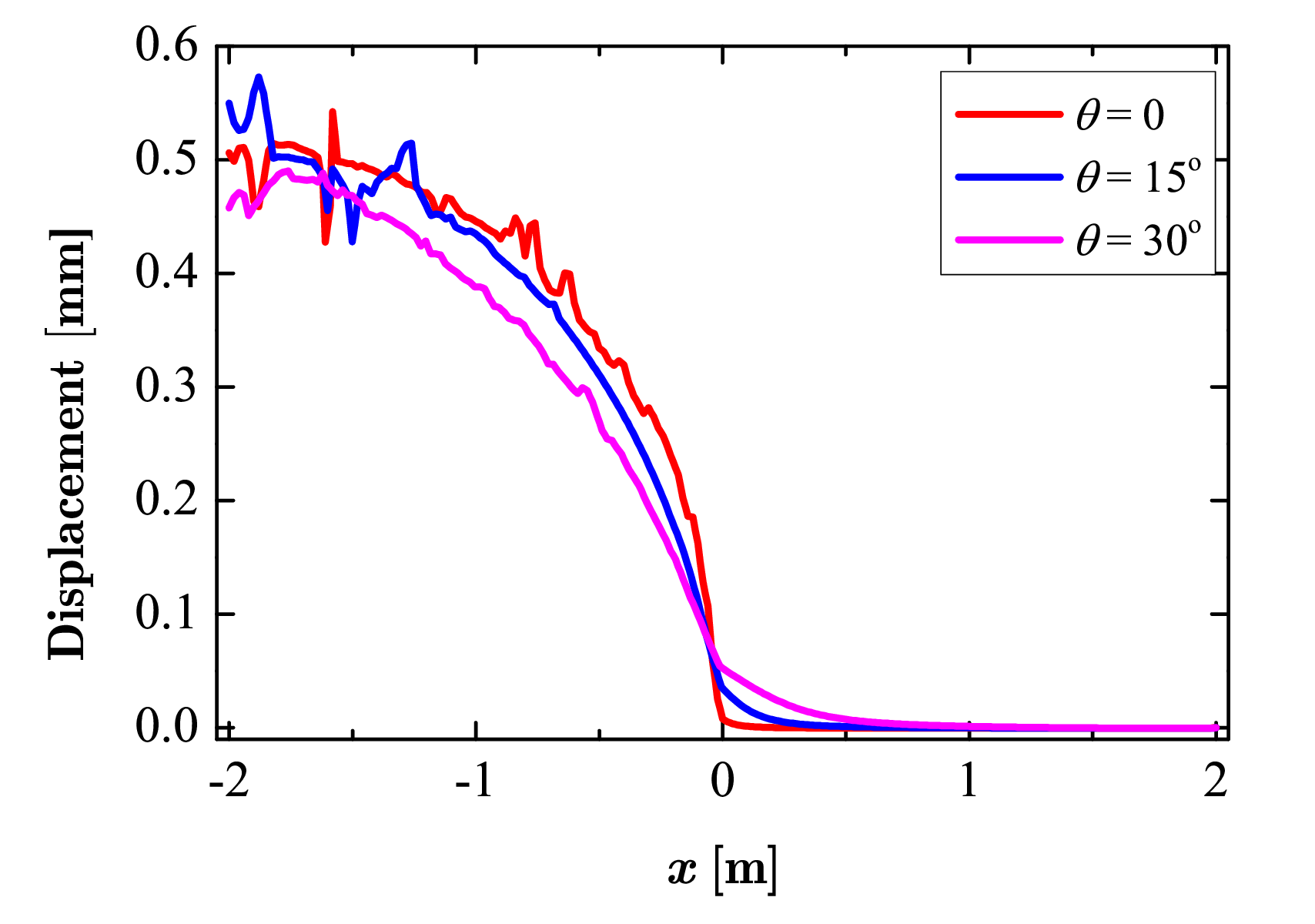}}\\
	\caption{Vertical displacement along the path L1 when the fracture reaches the layer interface in the soft-to-stiff configuration}
	\label{Vertical displacement along the path L1 when the fracture reaches the layer interface in the soft to stiff configuration}
	\end{figure}

We also investigate the vertical displacement distributions of the calculation domain for $E_2=2E_1$ and $E_2=4E_1$ at different time $t$. In the simulations, the vertical displacement is relatively large around the fracture region and the displacement increases with fracture propagation. We set a straight monitoring path L1 with its starting point (-2 m, 0.02 m) and ending point (2 m, 0.02 m). That is, the vertical coordinate of the path L1 is identical to that of the upper edge of the pre-existing notch. Figure \ref{Vertical displacement along the path L1 when the fracture reaches the layer interface in the soft to stiff configuration} shows the vertical displacement along path L1 for $E_2=2E_1$ and $E_2=4E_1$ when the fracture reaches the layer interface. The displacement pattern is similar to those in \citet{mikelic2013phase, zhou2018phase2}. The vertical displacement decreases along the direction of the fracture propagation, which is consistent with the analytical solution of \citet{sneddon1969crack}. In addition, the increase in the inclination angle $\theta$ slightly decreases the vertical displacement. The reason is that a larger inclination angle results in a larger stiffer layer volume in the direction normal to the propagating fracture, thereby restricting the upward deformation of the softer layer.

\begin{figure}[htbp]
	\centering
	\subfigure[$E_2=2E_1$]{\includegraphics[width = 7.5cm]{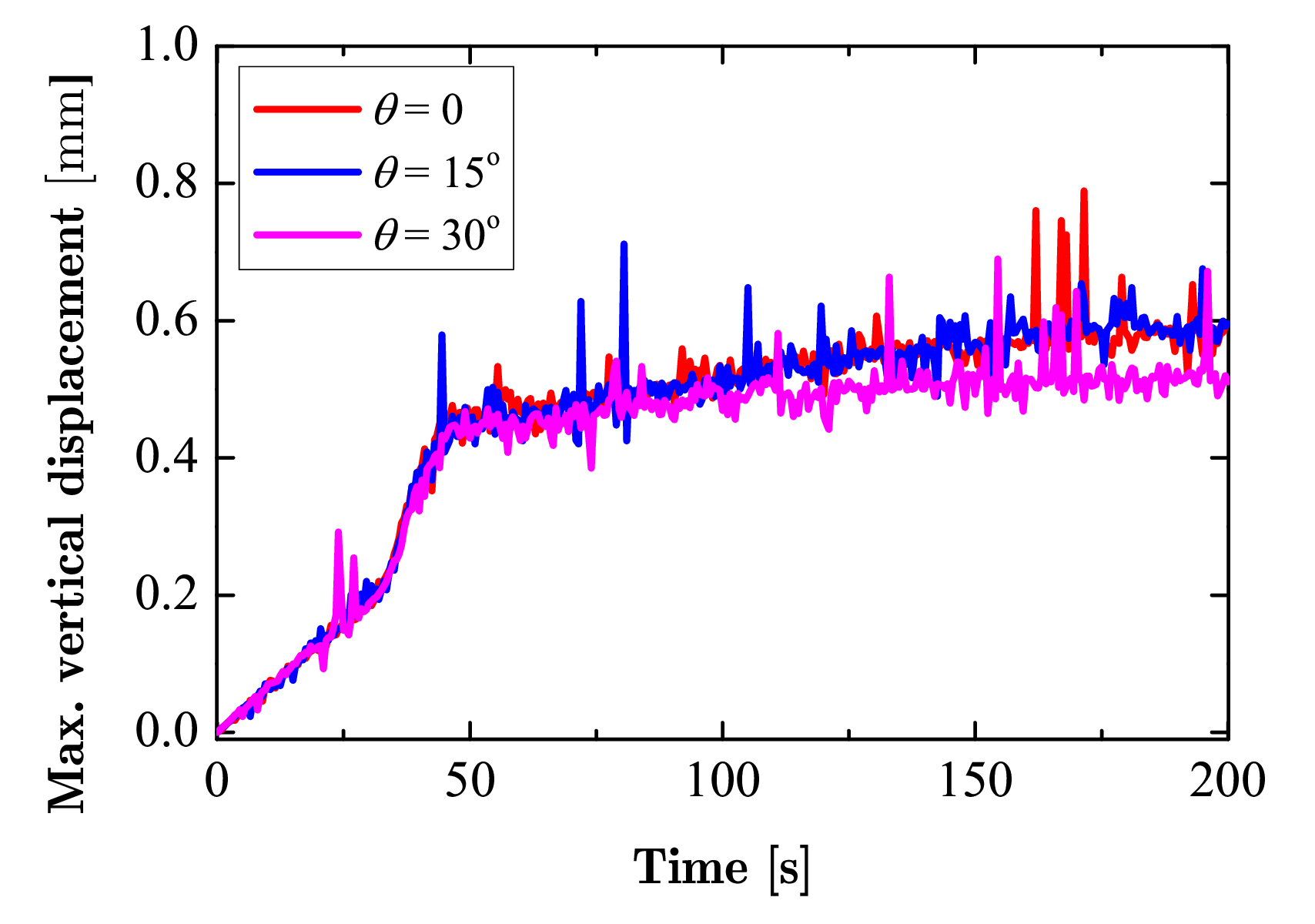}}
	\subfigure[$E_2=4E_1$]{\includegraphics[width = 7.5cm]{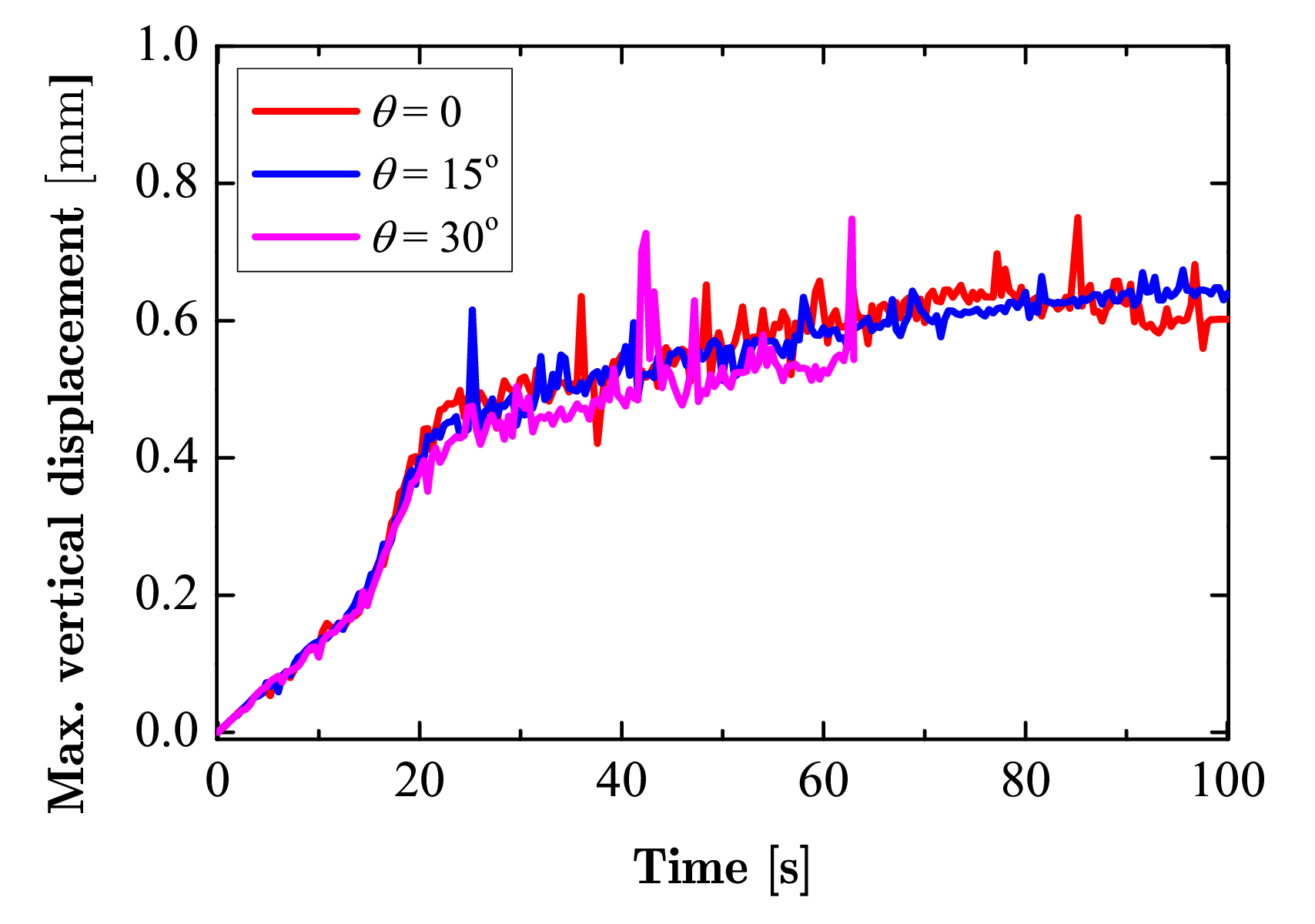}}\\
	\caption{Maximum vertical displacement along the upper left boundary in the soft-to-stiff configuration}
	\label{Maximum vertical displacement along the upper left boundary in the soft to stiff configuration}
	\end{figure}

Figure \ref{Maximum vertical displacement along the upper left boundary in the soft to stiff configuration} shows the maximum vertical displacement on the upper left edge of the domain when the time increases. In the soft-to-stiff configuration, the maximum displacement increases rapidly in the first 50 s ($E_2=2E_1$) and 25 s ($E_2=4E_1$). Thereafter, the maximum displacement increases at a low rate because the fracture propagation is depressed by the stiffer layer $\textcircled{2}$. In particular, the inclination angle $\theta=30^\circ$ achieves a relatively small maximum vertical displacement. Another reason for the second stage with a small increase rate in the vertical displacement is fracture deflection at the layer interface. In addition, slight displacement fluctuation is observed in the XFEM simulation of \citet{vahab2018x}.

\subsection{Fluid pressure}\label{Fluid pressure of soft to stiff}
\begin{figure}[htbp]
	\centering
	\subfigure[$\theta = 0^\circ$, $t$ = 55 s]{\includegraphics[width = 6cm]{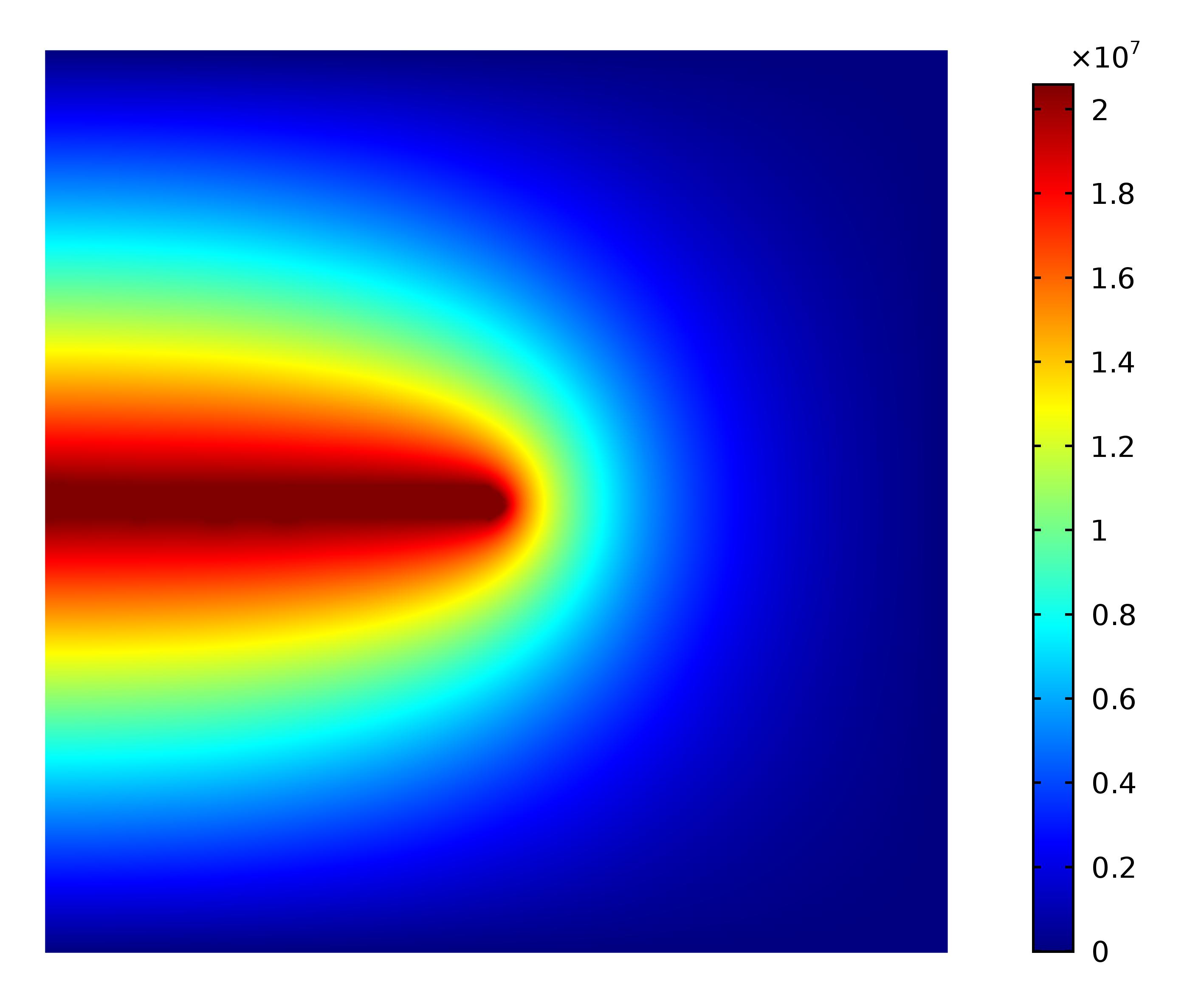}}
	\subfigure[$\theta = 0^\circ$, $t$ = 200 s]{\includegraphics[width = 6cm]{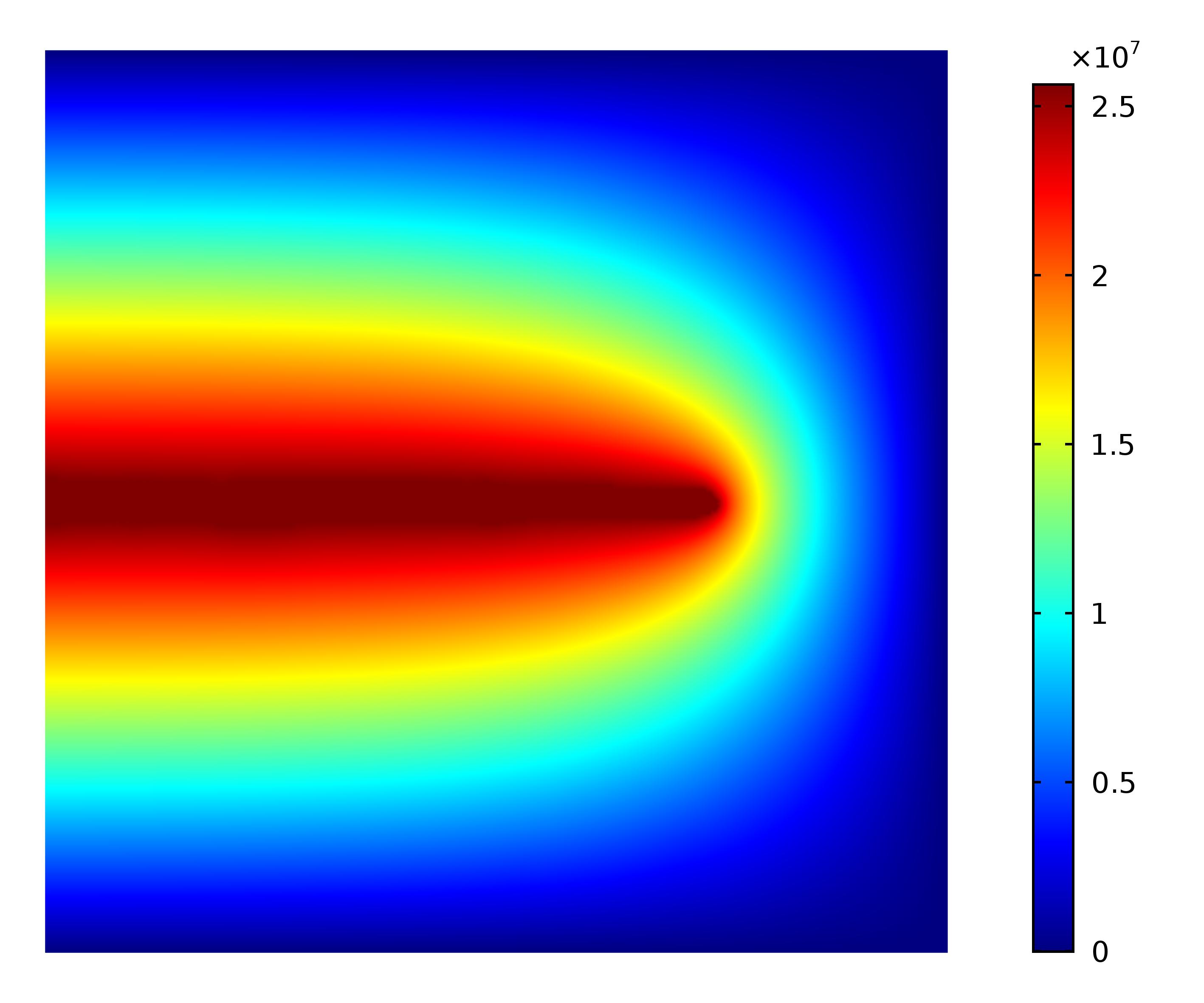}}\\
	\subfigure[$\theta = 15^\circ$, $t$ = 55 s]{\includegraphics[width = 6cm]{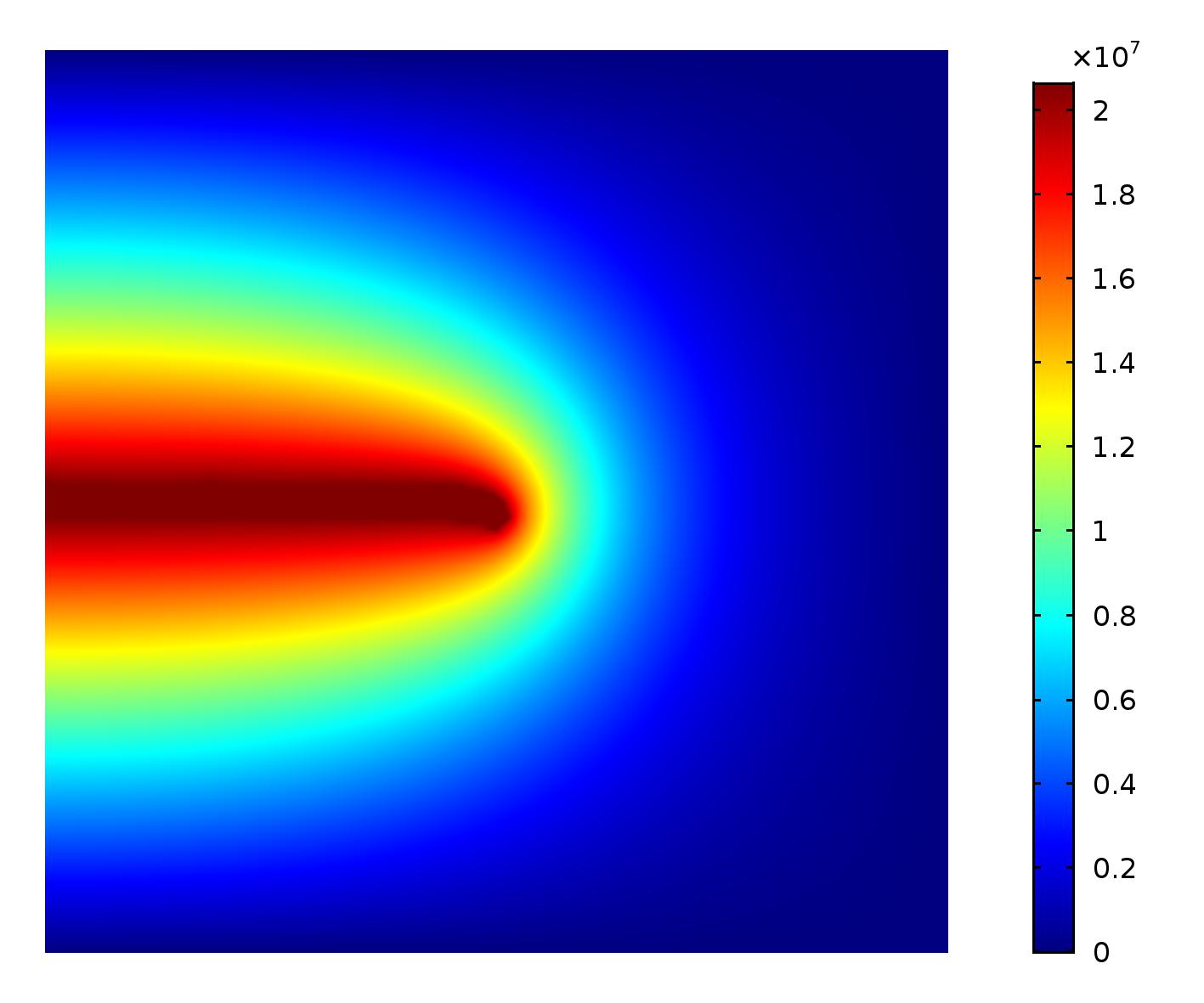}}
	\subfigure[$\theta = 15^\circ$, $t$ = 200 s]{\includegraphics[width = 6cm]{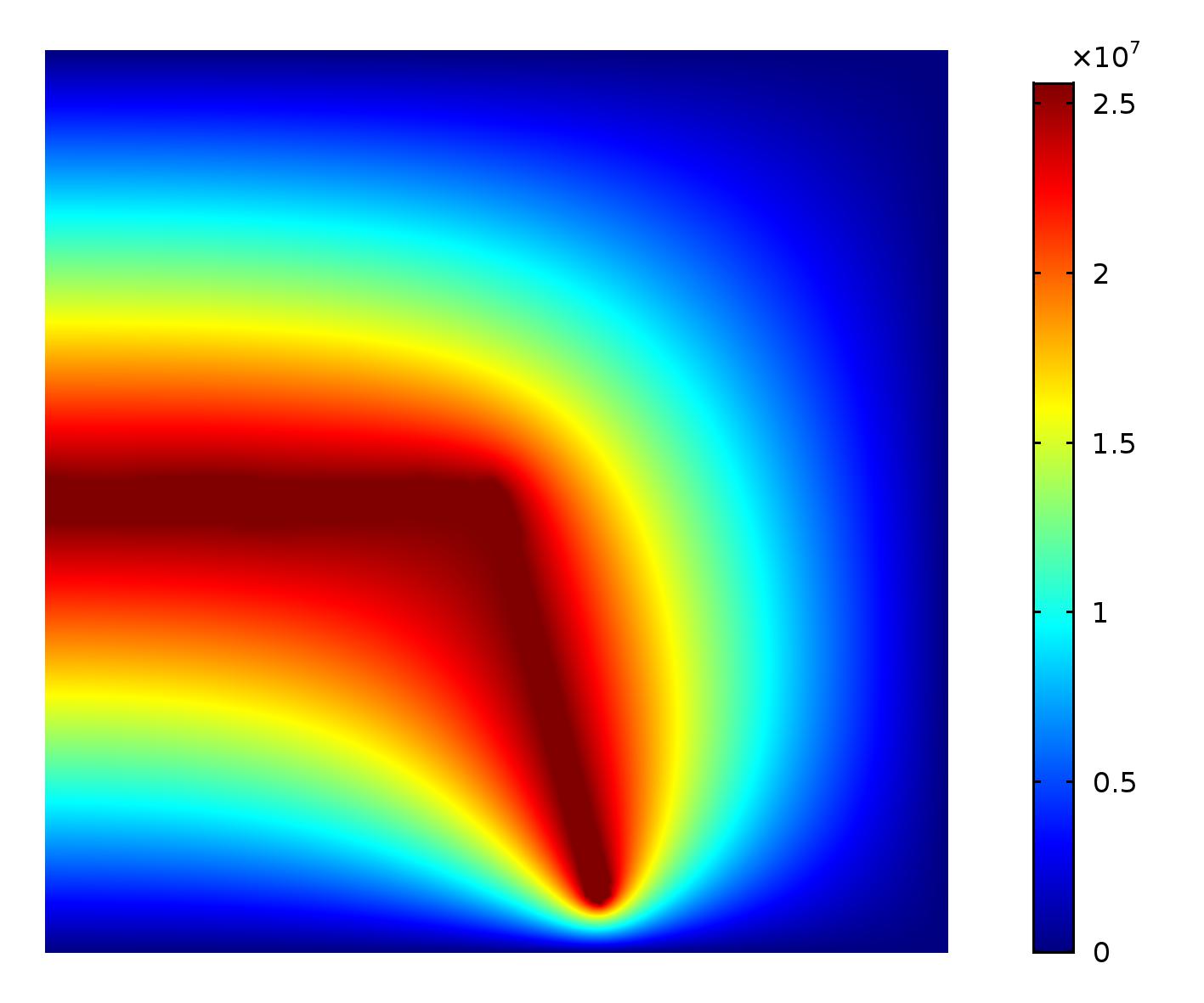}}\\
	\subfigure[$\theta = 30^\circ$, $t$ = 55 s]{\includegraphics[width = 6cm]{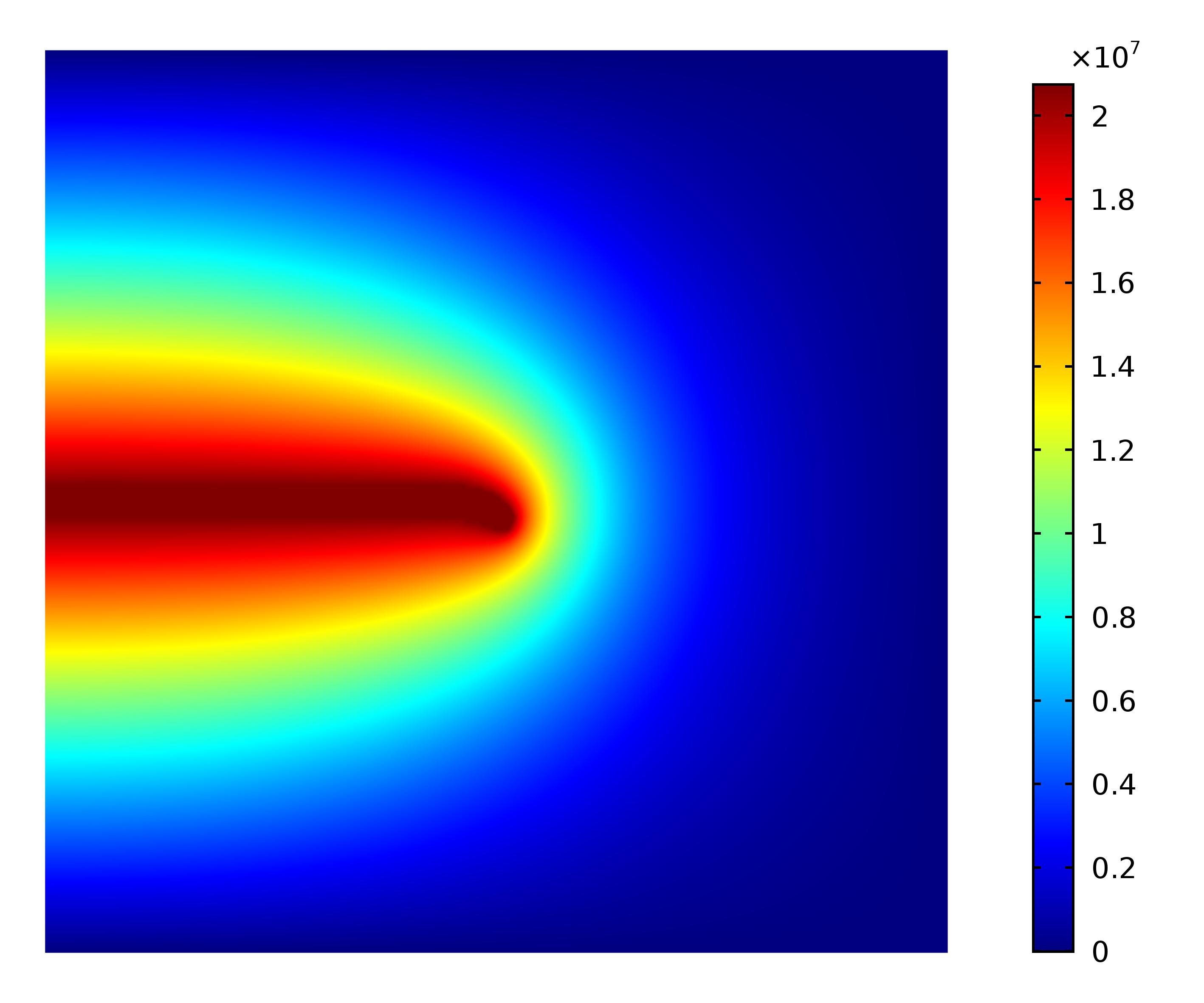}}
	\subfigure[$\theta = 30^\circ$, $t$ = 200 s]{\includegraphics[width = 6cm]{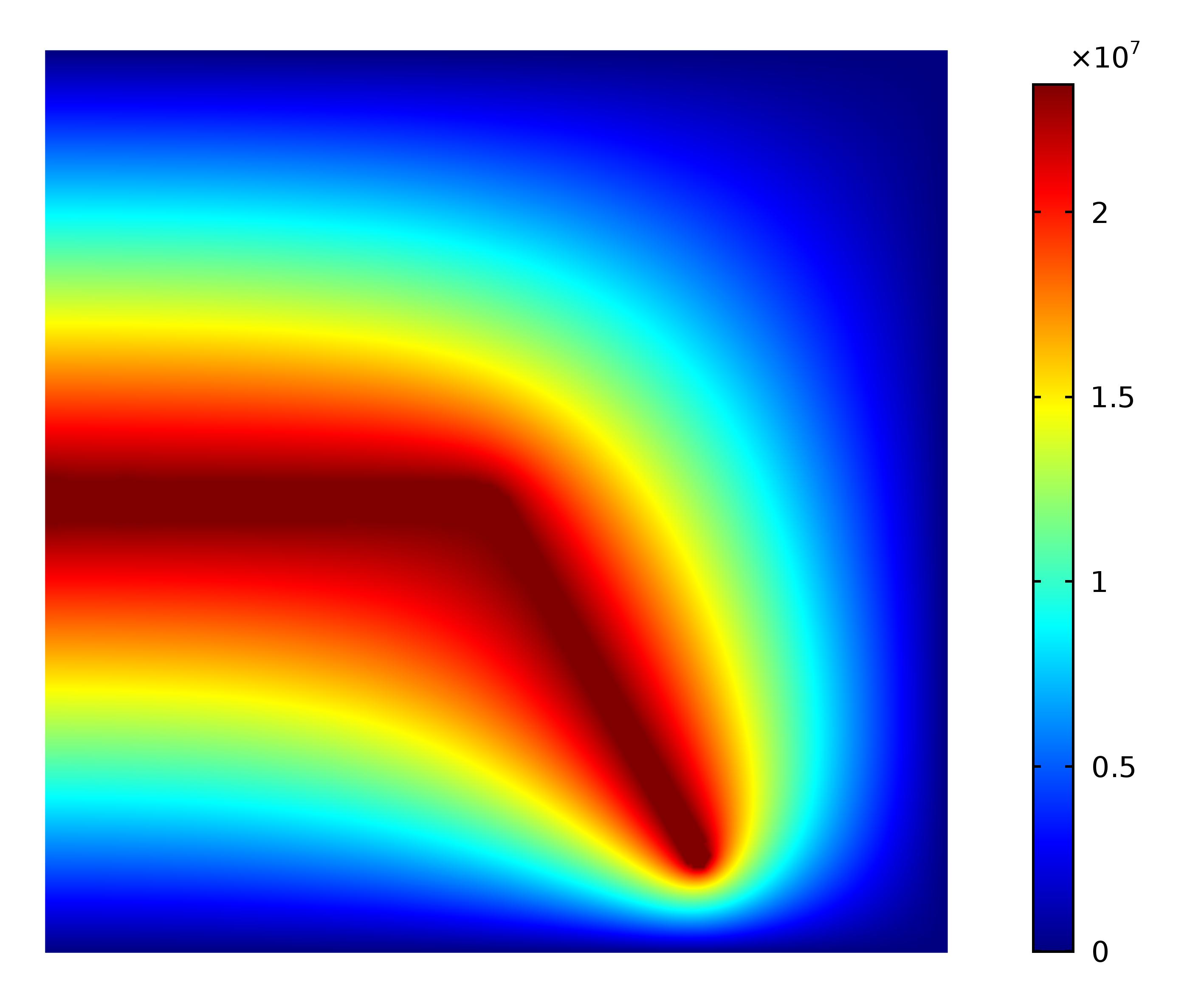}}\\
	\caption{Fluid pressure distributions for $E_2=2E_1$ (Unit: Pa)}
	\label{Fluid pressure distributions for E_2=2E_1}
	\end{figure}

\begin{figure}[htbp]
	\centering
	\subfigure[$\theta = 0^\circ$, $t$ = 34 s]{\includegraphics[width = 6cm]{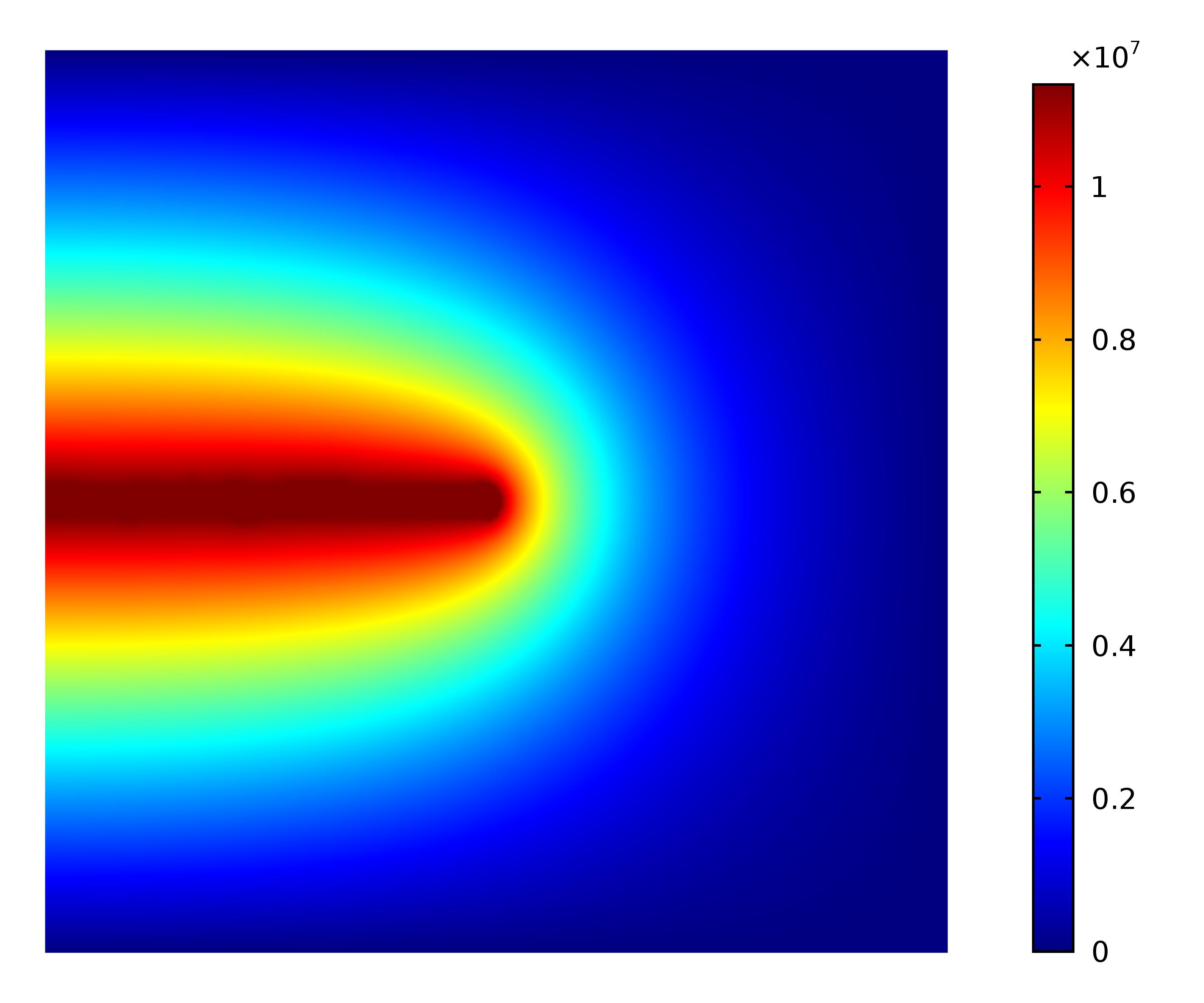}}
	\subfigure[$\theta = 0^\circ$, $t$ = 100 s]{\includegraphics[width = 6cm]{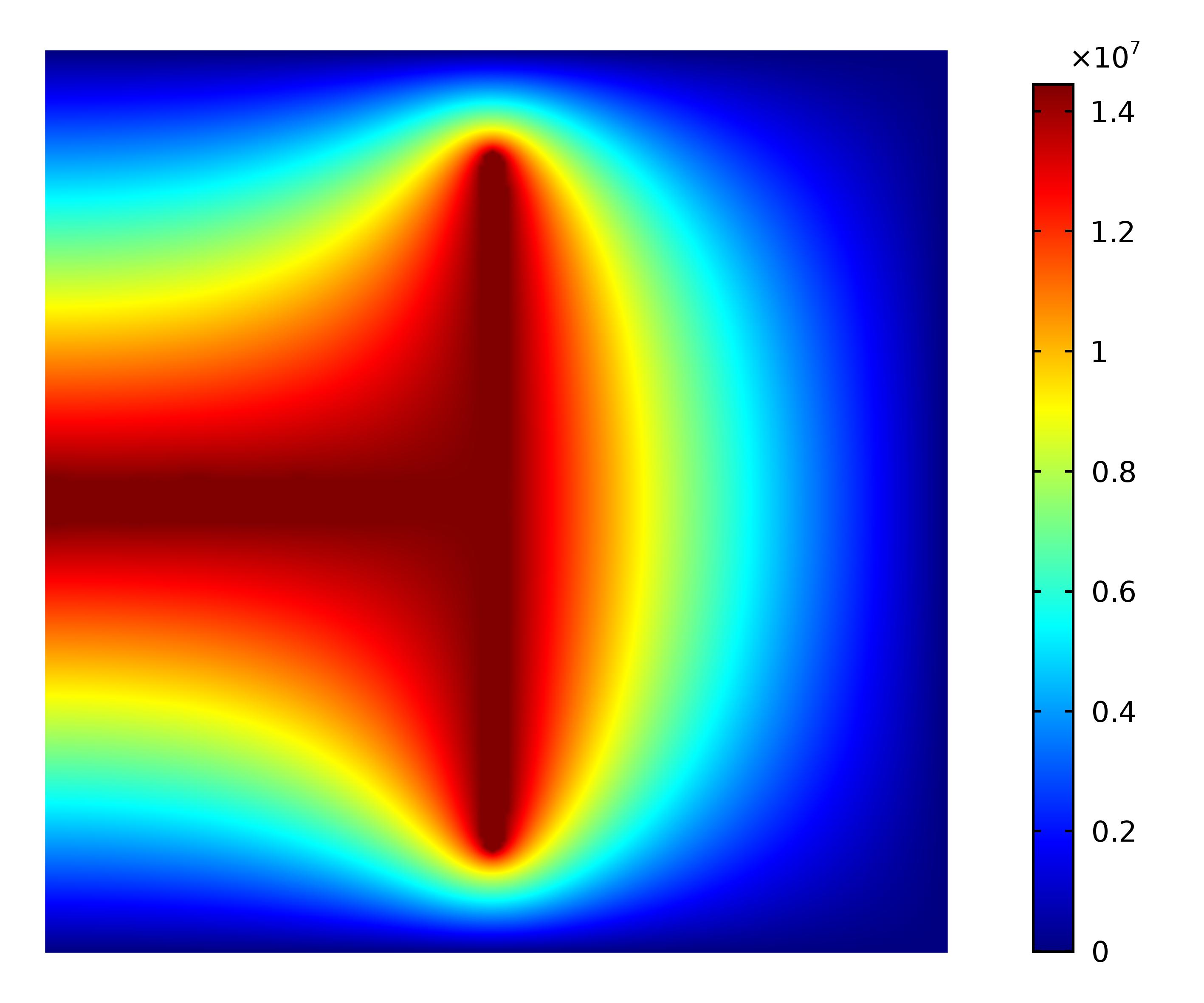}}\\
	\subfigure[$\theta = 15^\circ$, $t$ = 34 s]{\includegraphics[width = 6cm]{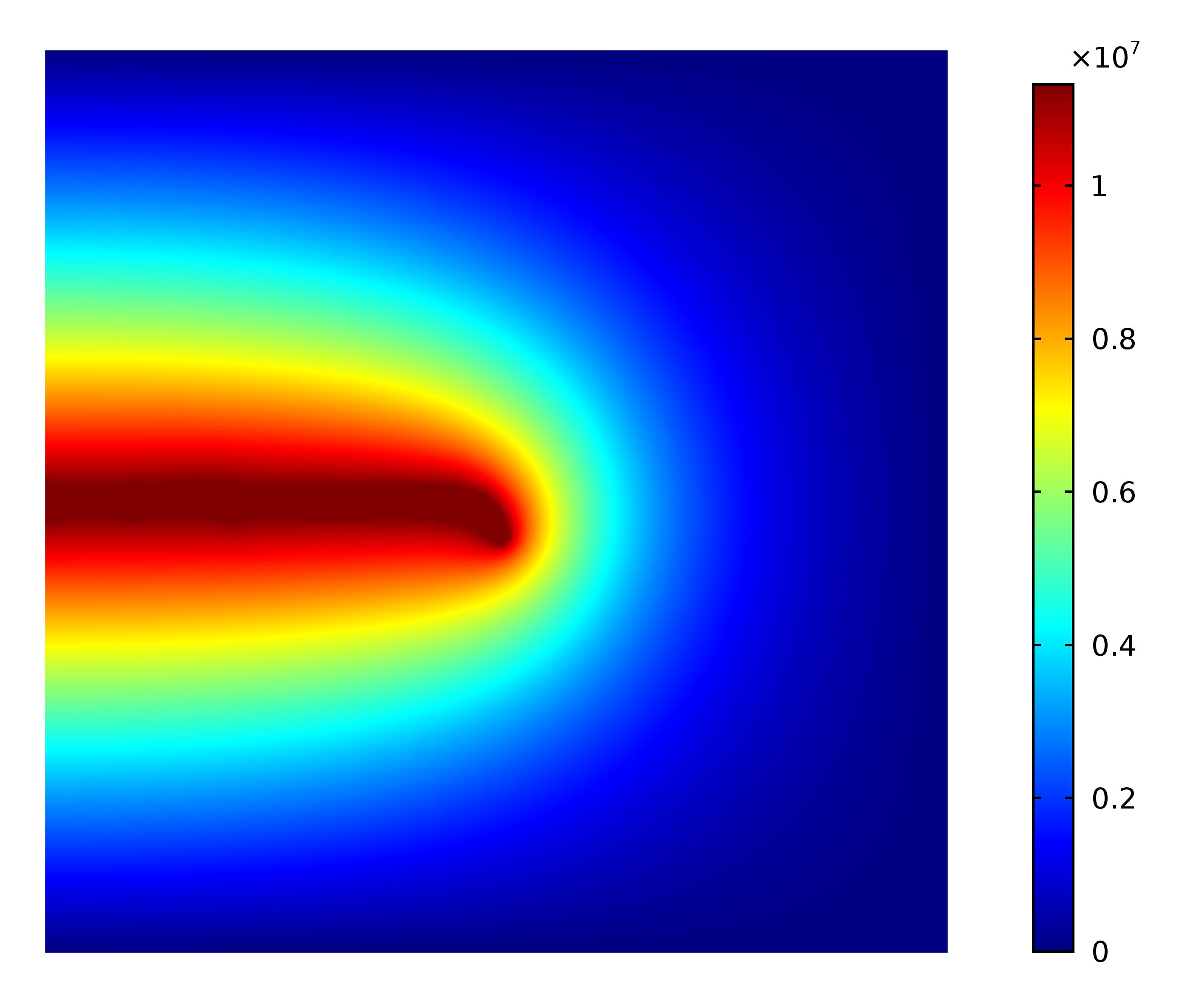}}
	\subfigure[$\theta = 15^\circ$, $t$ = 100 s]{\includegraphics[width = 6cm]{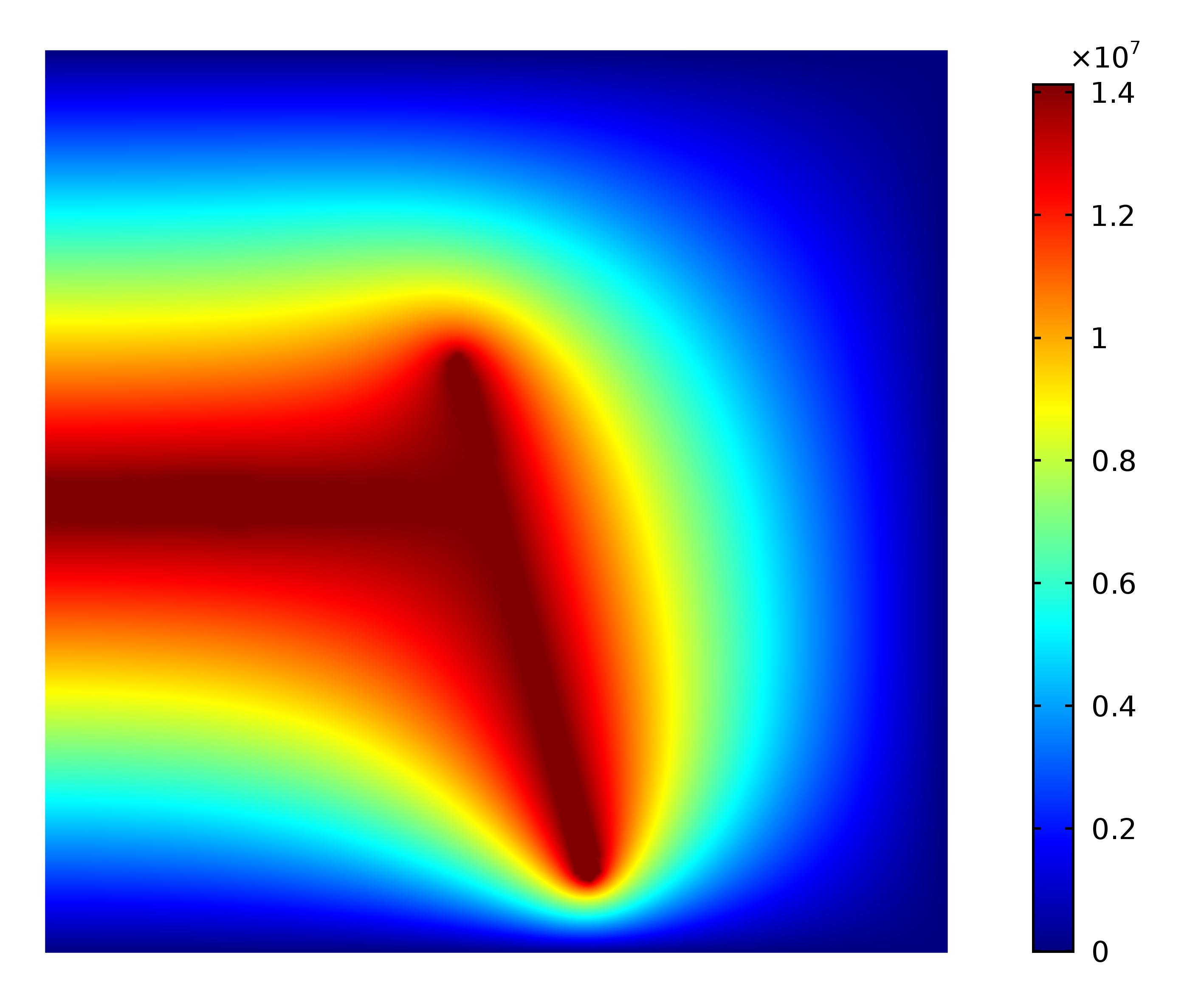}}\\
	\subfigure[$\theta = 30^\circ$, $t$ = 34 s]{\includegraphics[width = 6cm]{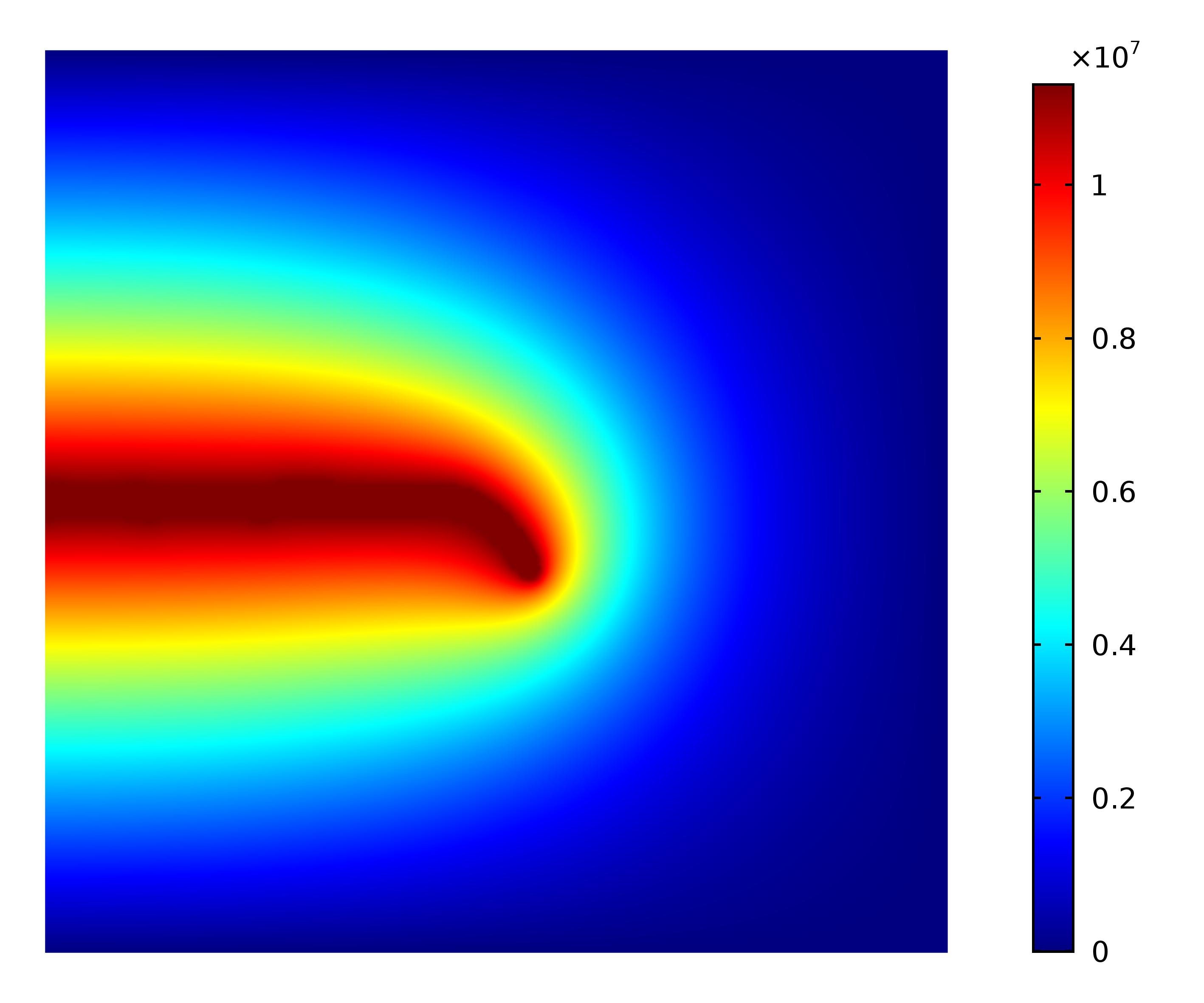}}
	\subfigure[$\theta = 30^\circ$, $t$ = 63.02 s]{\includegraphics[width = 6cm]{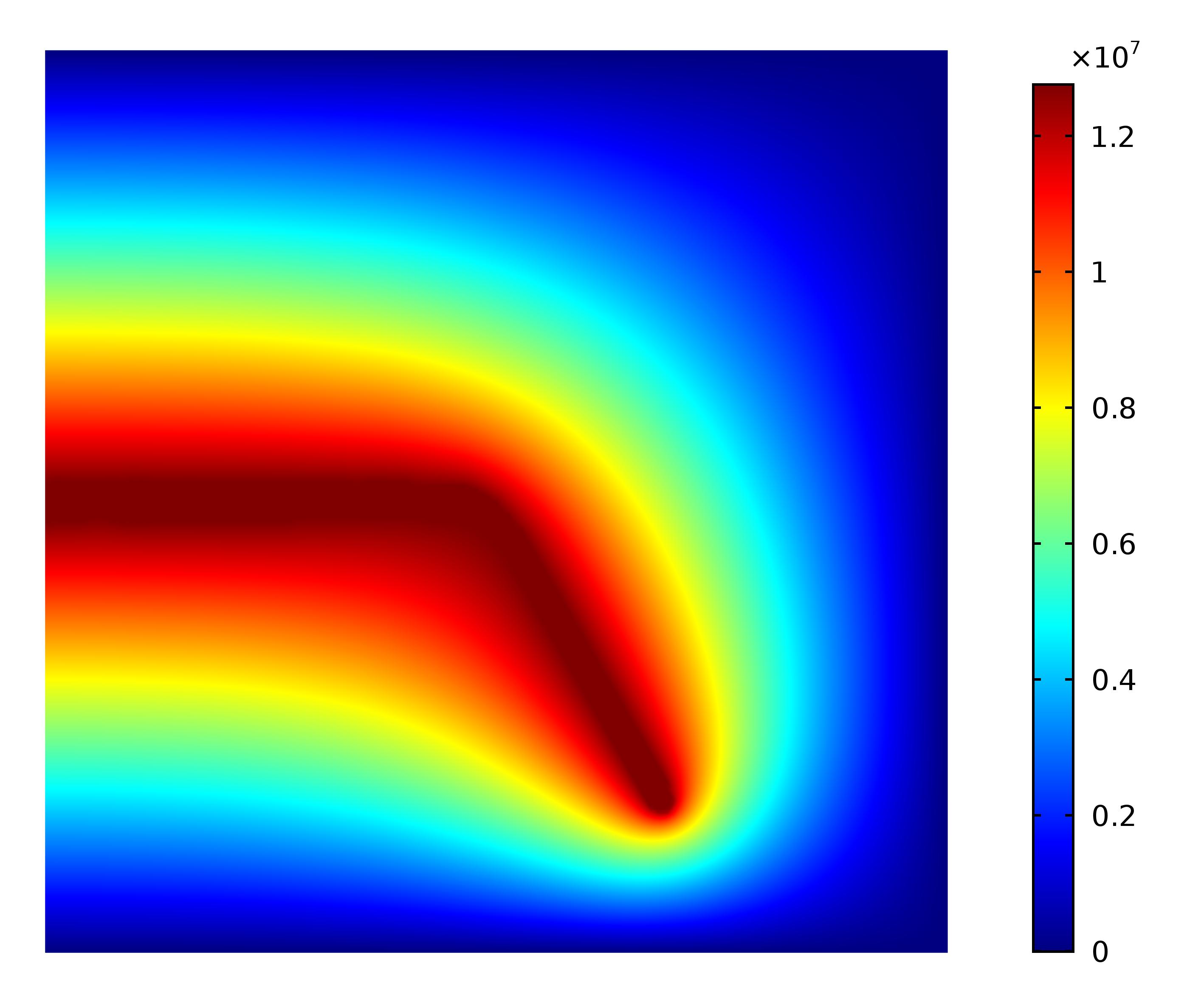}}\\
	\caption{Fluid pressure distributions for $E_2=4E_1$ (Unit: Pa)}
	\label{Fluid pressure distributions for E_2=4E_1}
	\end{figure}

\begin{figure}[htbp]
	\centering
	\subfigure[$E_2=2E_1$]{\includegraphics[width = 7.5cm]{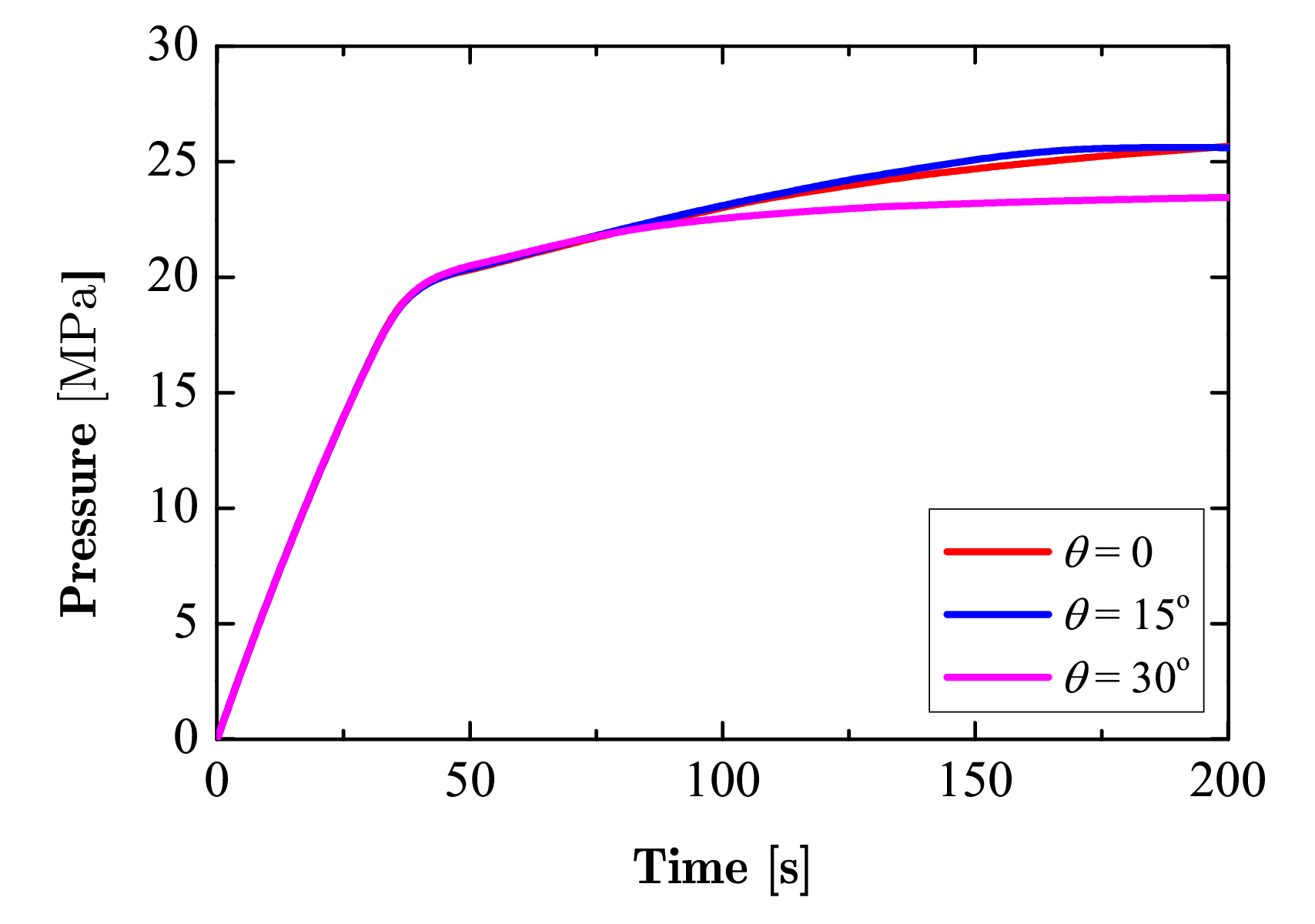}}
	\subfigure[$E_2=4E_1$]{\includegraphics[width = 7.5cm]{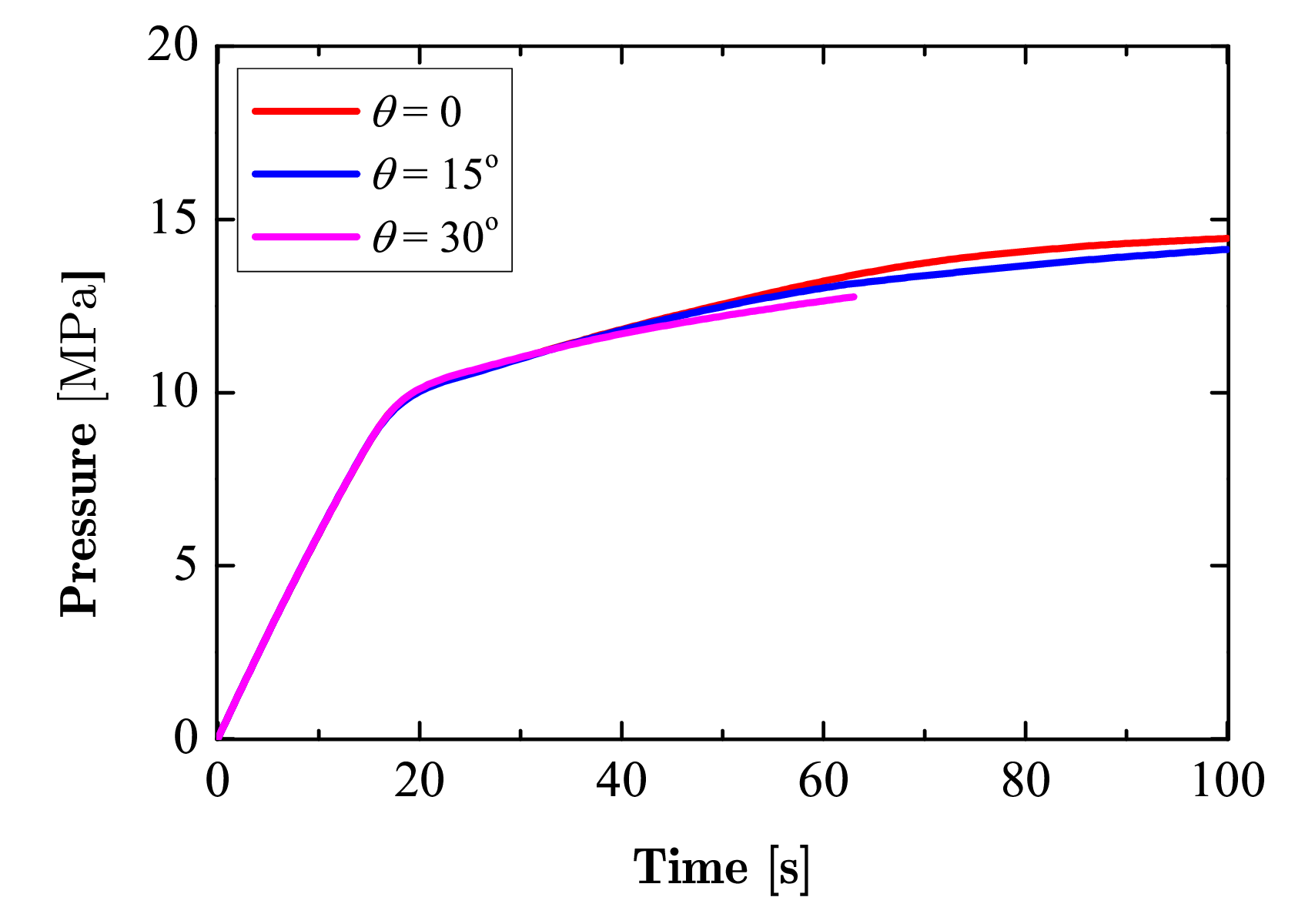}}\\
	\caption{Fluid pressure in the pre-existing notch in the soft-to-stiff configuration}
	\label{Fluid pressure in the pre-existing notch in the soft to stiff configuration}
	\end{figure}

Figures \ref{Fluid pressure distributions for E_2=2E_1} and \ref{Fluid pressure distributions for E_2=4E_1} show the fluid pressure field for $E_2=2E_1$ and $E_2=4E_1$. The fluid pressure field is consistent with the phase field while the maximum pressure appears in the hydraulic fractures. Figure \ref{Fluid pressure in the pre-existing notch in the soft to stiff configuration} shows the fluid pressure with the increasing time. The data is picked at the point (-2 m, 0). The fluid pressure has a similar trend with the maximum vertical displacement in Fig. \ref{Maximum vertical displacement along the upper left boundary in the soft to stiff configuration}. Note that the fluid pressure increases rapidly in the first 40 s ($E_2=2E_1$) and 20 s ($E_2=4E_1$). Thereafter, the increasing rate of the fluid pressure decreases. When $\theta=0^\circ$ and $30^\circ$, the fluid pressure-time curves are similar. However, the inclination angle $\theta=30^\circ$ obtains a relatively low fluid pressure. The pattern of fluid pressure mainly results from the fracture pattern, soft-to-stiff setting, and relatively low porosity used in the simulation.

\section{Stiff-to-soft configuration}\label{The stiff to soft configuration}
\subsection{Fracture pattern}\label{Fracture pattern of stiff to soft}

\begin{figure}[htbp]
	\centering
	\subfigure[$\theta = 0^\circ$, $t=$ 83 s]{\includegraphics[width = 5cm]{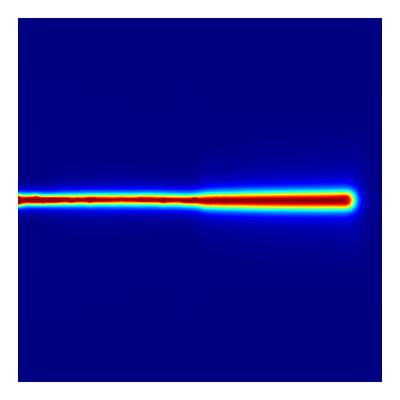}}
	\subfigure[$\theta = 15^\circ$, $t=$ 84.4 s]{\includegraphics[width = 5cm]{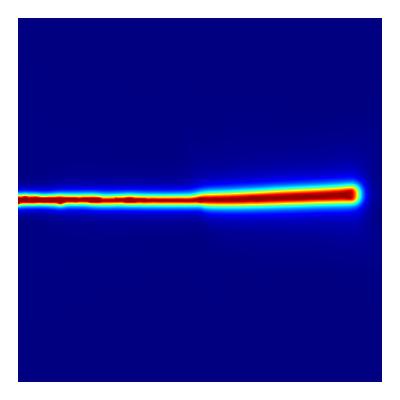}}
	\subfigure[$\theta = 30^\circ$, $t=$ 82.05 s]{\includegraphics[width = 5cm]{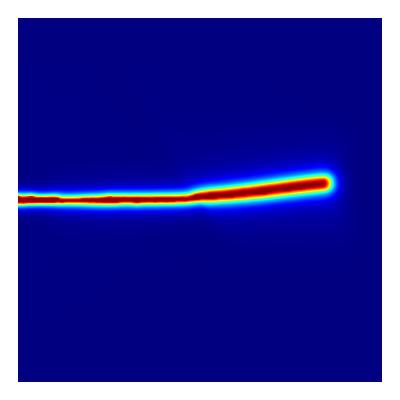}}\\
	\caption{Fracture propagation patterns for $E_1=2E_2$}
	\label{Fracture propagation patterns for E_1=2E_2}
	\end{figure}

\begin{figure}[htbp]
	\centering
	\subfigure[$\theta = 0^\circ$, $t=$ 80.54 s]{\includegraphics[width = 5cm]{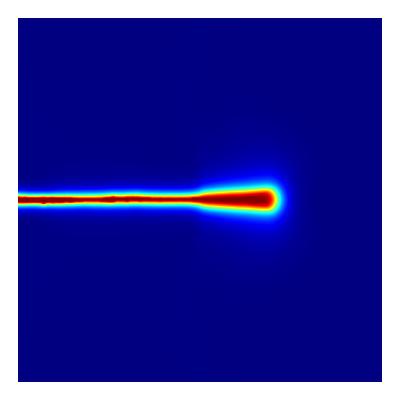}}
	\subfigure[$\theta = 15^\circ$, $t$ = 80.11 s]{\includegraphics[width = 5cm]{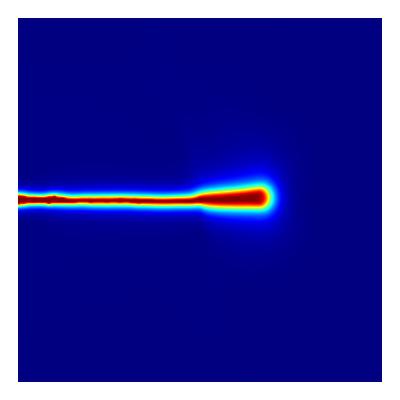}}
	\subfigure[$\theta = 30^\circ$, $t$ = 79.78 s]{\includegraphics[width = 5cm]{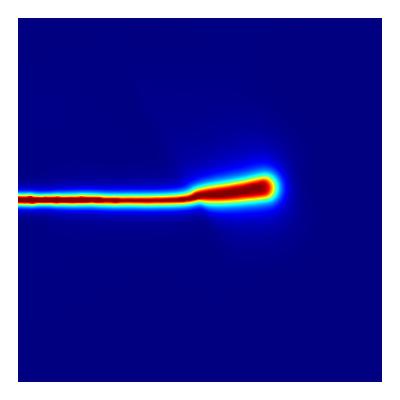}}\\
	\caption{Fracture propagation patterns for $E_1=3E_1$}
	\label{Fracture propagation patterns for E_1=3E_1}
	\end{figure}

This section presents the numerical results for the stiff-to-soft configuration ($E_1=2E_2$ and $E_1=3E_2$). Figures \ref{Fracture propagation patterns for E_1=2E_2} and \ref{Fracture propagation patterns for E_1=3E_1} represent the hydraulic fracture patterns for $E_1=2E_2$ and $E_1=3E_2$ at different time $t$, respectively. Only the penetration scenario is obtained in the stiff-to-soft configuration, which is different from the observations in the soft-to-stiff configuration. Singly-deflected, doubly-deflected, and penetration scenarios can be all simulated in the soft-to-stiff configuration. The fracture penetration scenario in the stiff-to-soft configuration is a natural result calculated from the evolution equation of the phase field, and the fracture penetration pattern is formed because the softer layer $\textcircled{2}$ has a lower tensile strength than the stiffer layer $\textcircled{1}$ according to \citet{zhou2018phase3}.

Figures \ref{Fracture propagation patterns for E_1=2E_2} and \ref{Fracture propagation patterns for E_1=3E_1} also show that the fracture width increases when the fracture penetrates into the softer layer $\textcircled{2}$. By comparing Figs. \ref{Fracture propagation patterns for E_1=2E_2} and \ref{Fracture propagation patterns for E_1=3E_1}, it is observed that the fracture in the layer $\textcircled{2}$ has a larger width when the stiffness of the layer $\textcircled{2}$ is smaller. The inclination angle also influences the fracture pattern. When $\theta=0$, the fractures in the layers $\textcircled{1}$ and $\textcircled{2}$ propagate horizontally. However, when $\theta=15^\circ$ and $30^\circ$, the hydraulic fracture deflects slightly after it crosses the layer interface. The hydraulic fracture in the layer $\textcircled{2}$ intersects the fracture in the layer $\textcircled{1}$ at a small angle, which increases slightly as $\theta$ increases.

\subsection{Effective maximum stress}\label{Effective maximum stress of stiff to soft}

\begin{figure}[htbp]
	\centering
	\subfigure[$\theta = 0^\circ$, $t$ = 80 s]{\includegraphics[width = 7.5cm]{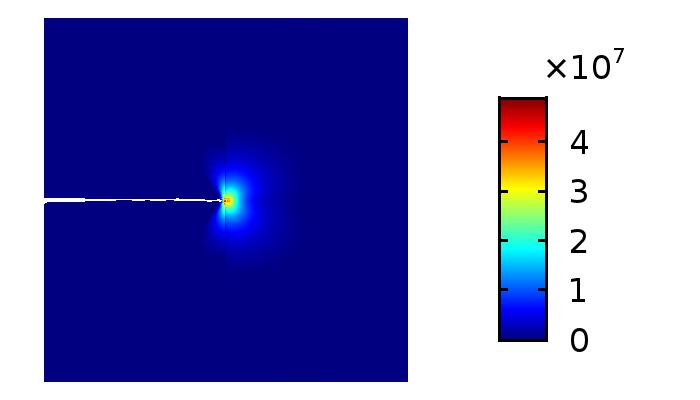}}
	\subfigure[$\theta = 0^\circ$, $t$ = 83 s]{\includegraphics[width = 7.5cm]{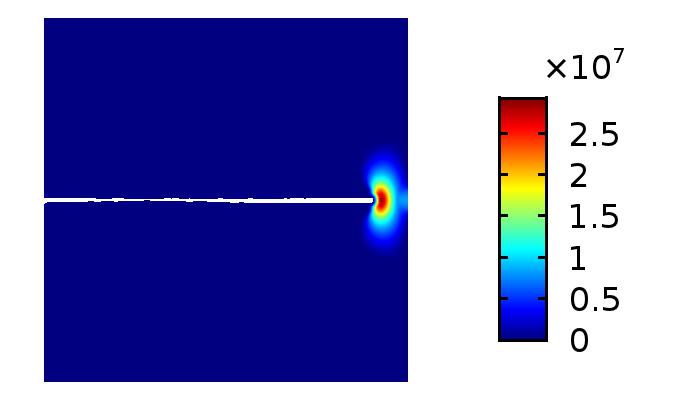}}\\
	\subfigure[$\theta = 15^\circ$, $t$ = 81.2 s]{\includegraphics[width = 7.5cm]{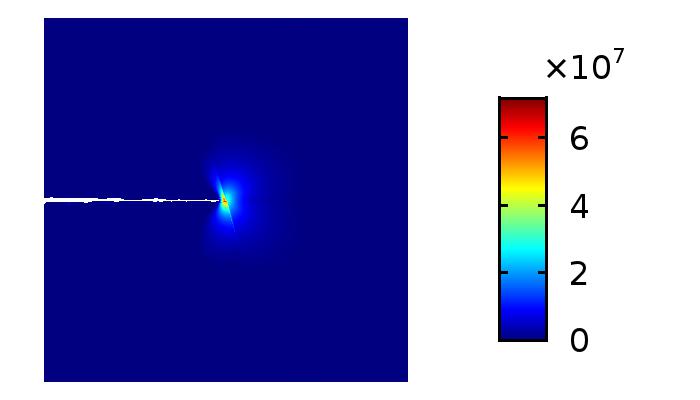}}
	\subfigure[$\theta = 15^\circ$, $t$ = 84.4 s]{\includegraphics[width = 7.5cm]{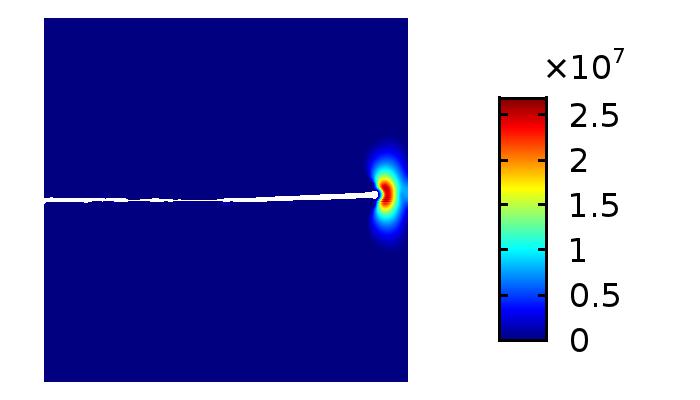}}\\
	\subfigure[$\theta = 30^\circ$, $t$ = 79.4 s]{\includegraphics[width = 7.5cm]{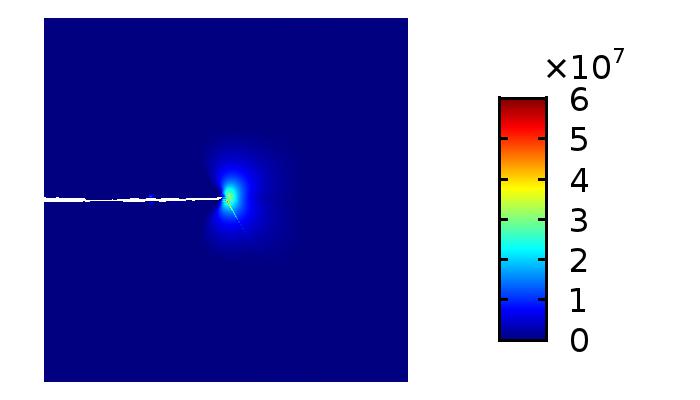}}
	\subfigure[$\theta = 30^\circ$, $t$ = 82.05 s]{\includegraphics[width = 7.5cm]{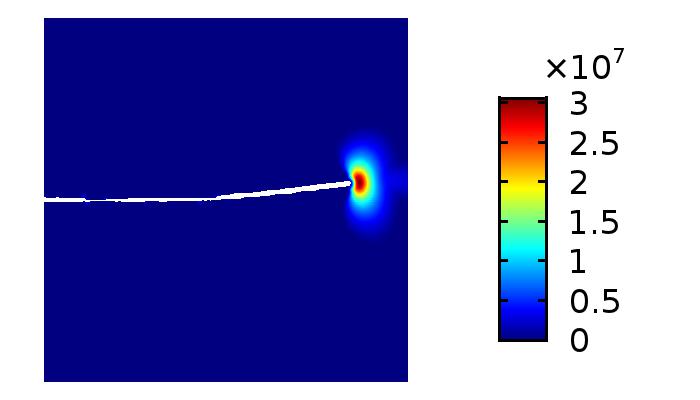}}\\
	\caption{Effective maximum stress distributions for $E_1=2E_2$ (Unit: Pa)}
	\label{Effective maximum stress distributions for E_1=2E_2}
	\end{figure}

\begin{figure}[htbp]
	\centering
	\subfigure[$\theta = 0^\circ$, $t$ = 79.4 s]{\includegraphics[width = 7.5cm]{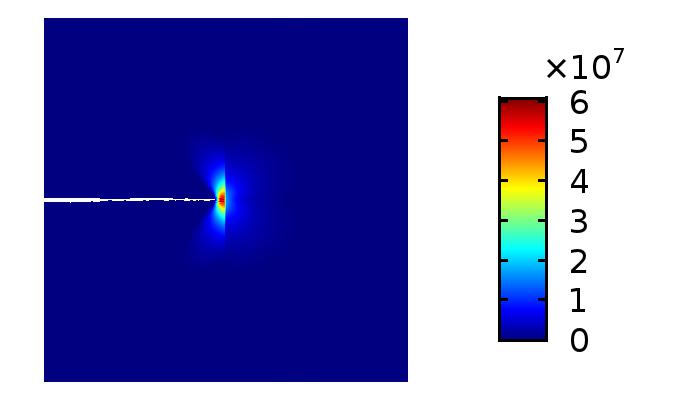}}
	\subfigure[$\theta = 0^\circ$, $t$ = 80.54 s]{\includegraphics[width = 7.5cm]{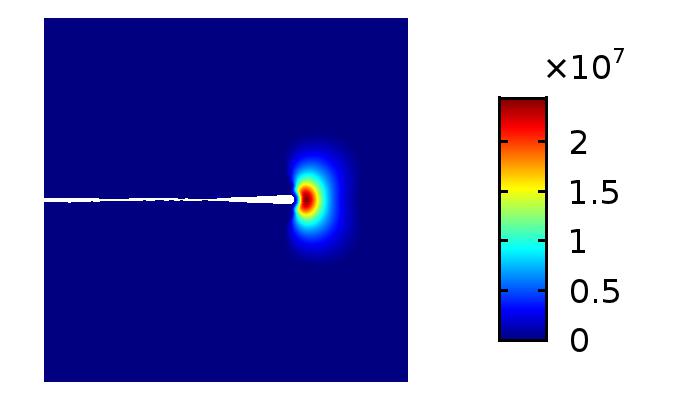}}\\
	\subfigure[$\theta = 15^\circ$, $t$ = 79.4 s]{\includegraphics[width = 7.5cm]{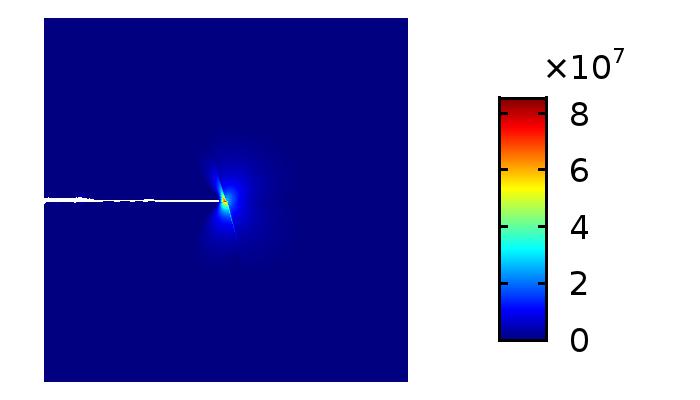}}
	\subfigure[$\theta = 15^\circ$, $t$ = 80.11 s]{\includegraphics[width = 7.5cm]{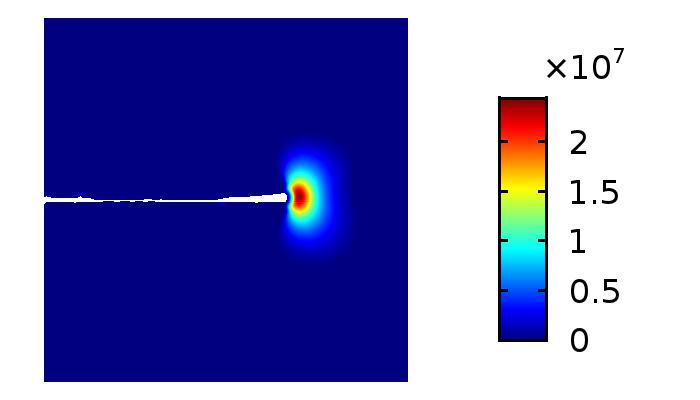}}\\
	\subfigure[$\theta = 30^\circ$, $t$ = 79 s]{\includegraphics[width = 7.5cm]{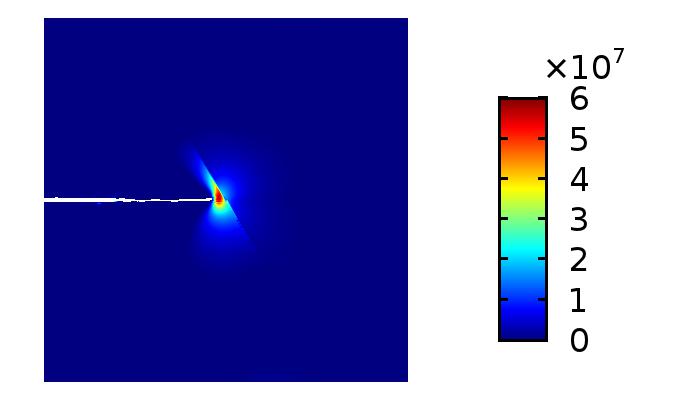}}
	\subfigure[$\theta = 30^\circ$, $t$ = 79.78 s]{\includegraphics[width = 7.5cm]{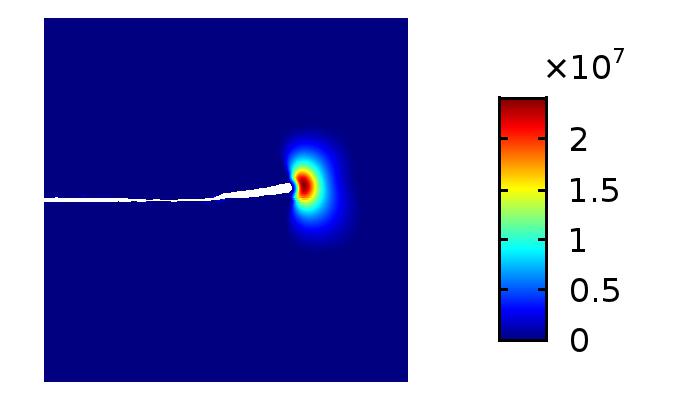}}\\
	\caption{Effective maximum stress distributions for $E_1=3E_2$ (Unit: Pa)}
	\label{Effective maximum stress distributions for E_1=3E_2}
	\end{figure}

In the stiff to soft configuration, the effective maximum stress distributions for $E_1=2E_2$ and $E_1=3E_2$ at different time $t$ are shown in Figs. \ref{Effective maximum stress distributions for E_1=2E_2} and \ref{Effective maximum stress distributions for E_1=3E_2}, respectively. The stress distributions coincide with the fracture patterns. The stress concentration is observed only around the fracture tip owing to the penetration scenario. In addition, the effective maximum stress around the fracture tip decreases when the hydraulic fracture propagates into the layer $\textcircled{2}$. The difference between the stress in the two layers $\textcircled{1}$ and $\textcircled{2}$ reflects again the difference in the tensile strength for resisting fracturing.

\subsection{Displacement field}\label{Displacement field of stiff to soft}

The vertical displacement distributions of the calculation domain for $E_1=2E_1$ and $E_1=3E_2$ at different time $t$ is investigated. When hydraulic fracture propagates in the stiffer layer $\textcircled{1}$, the maximum vertical displacement occurs around the fluid injection region and the displacement increases along with the fracture propagation. However, increasing vertical displacement exists around the fracture domain in the softer layer $\textcircled{2}$ when the hydraulic fracture penetrates deep into the layer $\textcircled{2}$.

\begin{figure}[htbp]
	\centering
	\subfigure[$E_1=2E_2$]{\includegraphics[width = 7.5cm]{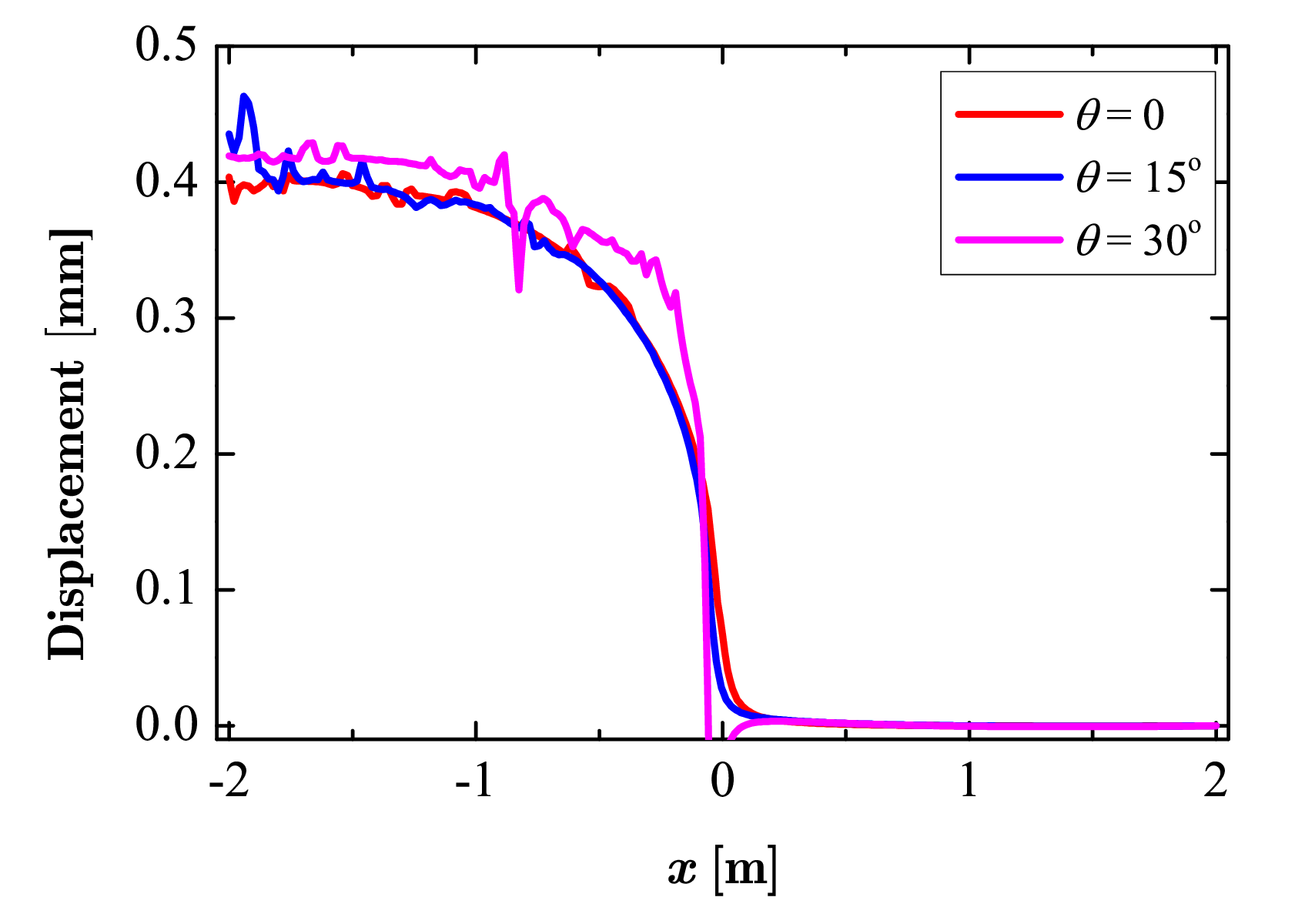}}
	\subfigure[$E_1=3E_2$]{\includegraphics[width = 7.5cm]{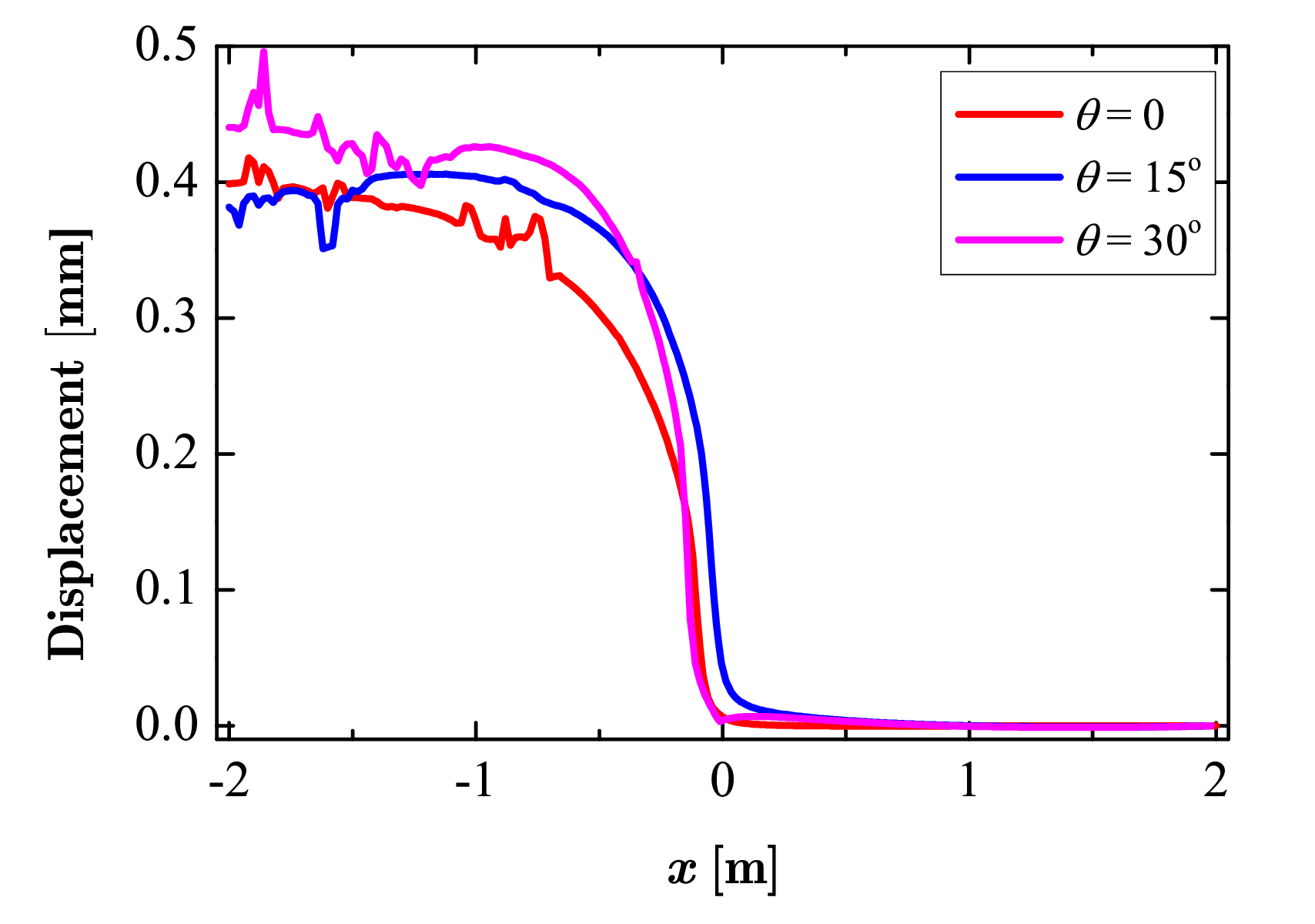}}\\
	\caption{Vertical displacement along the path L1 when the fracture reaches the layer interface in the stiff-to-soft configuration}
	\label{Vertical displacement along the path L1 when the fracture reaches the layer interface in the stiff to soft configuration}
	\end{figure}

Figure \ref{Vertical displacement along the path L1 when the fracture reaches the layer interface in the stiff to soft configuration} shows the vertical displacement along the path L1 for $E_1=2E_2$ and $E_1=3E_2$ when the fracture reaches the layer interface. The displacement pattern is similar to those in the soft to stiff configuration. The vertical displacement decreases along the direction of fracture propagation while the displacement increases slightly as the inclination angle $\theta$ increases. 

\begin{figure}[htbp]
	\centering
	\subfigure[$E_1=2E_2$]{\includegraphics[width = 7.5cm]{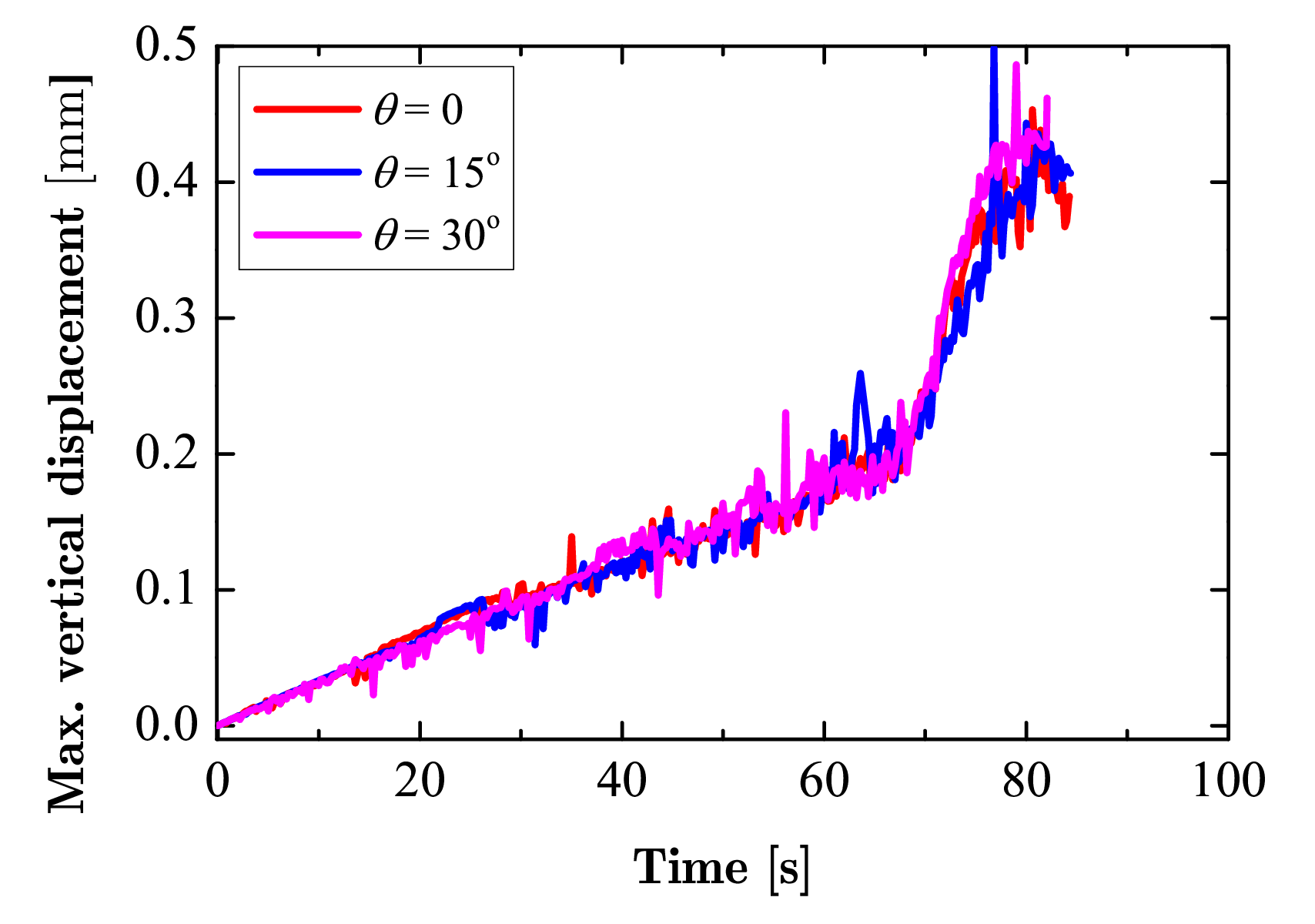}}
	\subfigure[$E_1=3E_2$]{\includegraphics[width = 7.5cm]{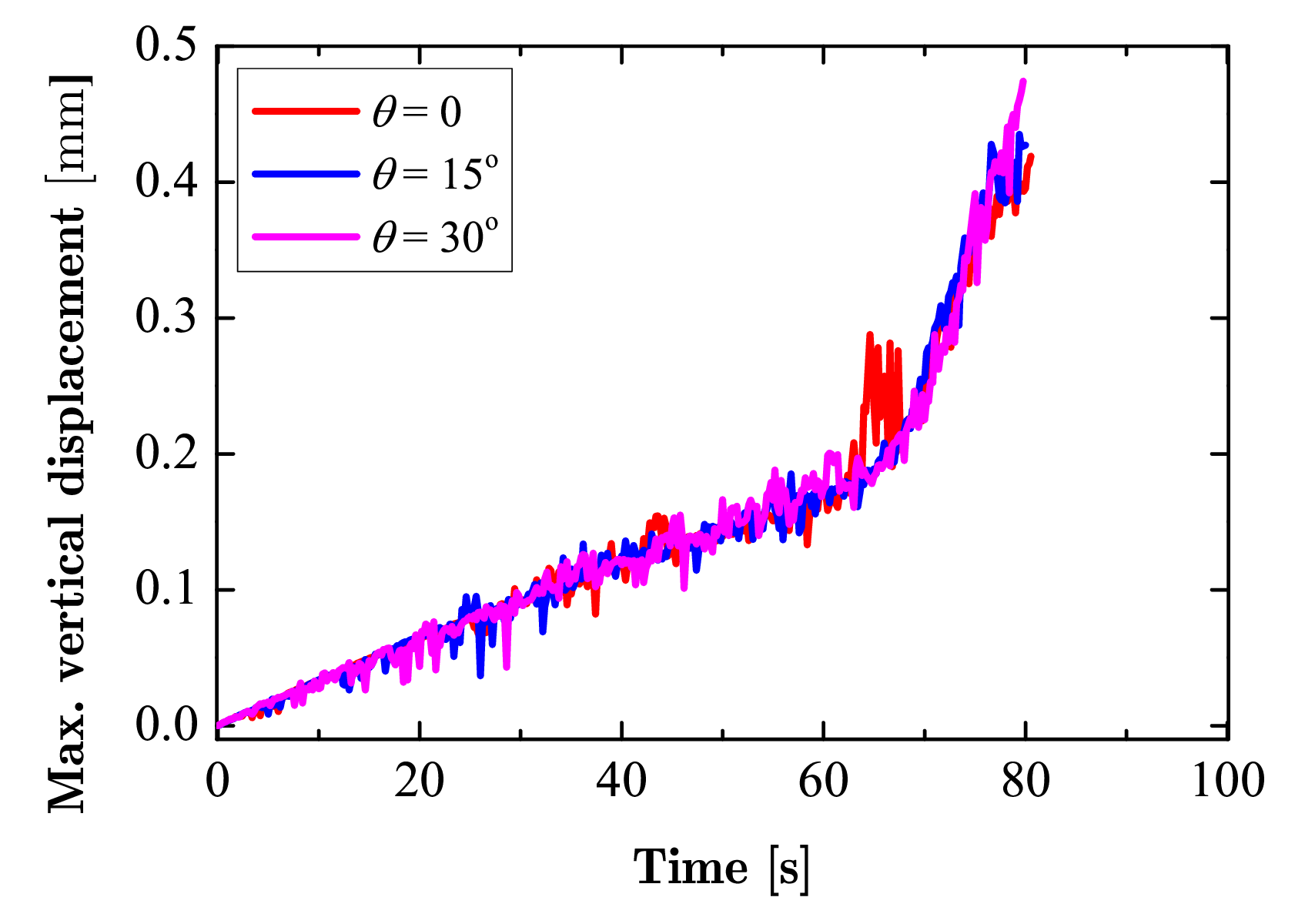}}\\
	\caption{Maximum vertical displacement along the upper left boundary in the stiff-to-soft configuration}
	\label{Maximum vertical displacement along the upper left boundary in the stiff to soft configuration}
	\end{figure}

Figure \ref{Maximum vertical displacement along the upper left boundary in the stiff to soft configuration} shows the maximum vertical displacement on the upper left edge when time increases in the stiff-to-soft configuration. As shown in Fig. \ref{Maximum vertical displacement along the upper left boundary in the stiff to soft configuration}, the maximum displacement increases at a relative small rate in the first 65 s. Thereafter, when the hydraulic fracture reaches the layer interface and propagates deep into the layer $\textcircled{2}$, the maximum displacement increases at a relatively large rate. In addition, the curves of the maximum displacement versus time are minimally affected by the inclination angle $\theta$.

\subsection{Fluid pressure}\label{Fluid pressure of stiff to soft}

\begin{figure}[htbp]
	\centering
	\subfigure[$E_1=2E_2$]{\includegraphics[width = 7.5cm]{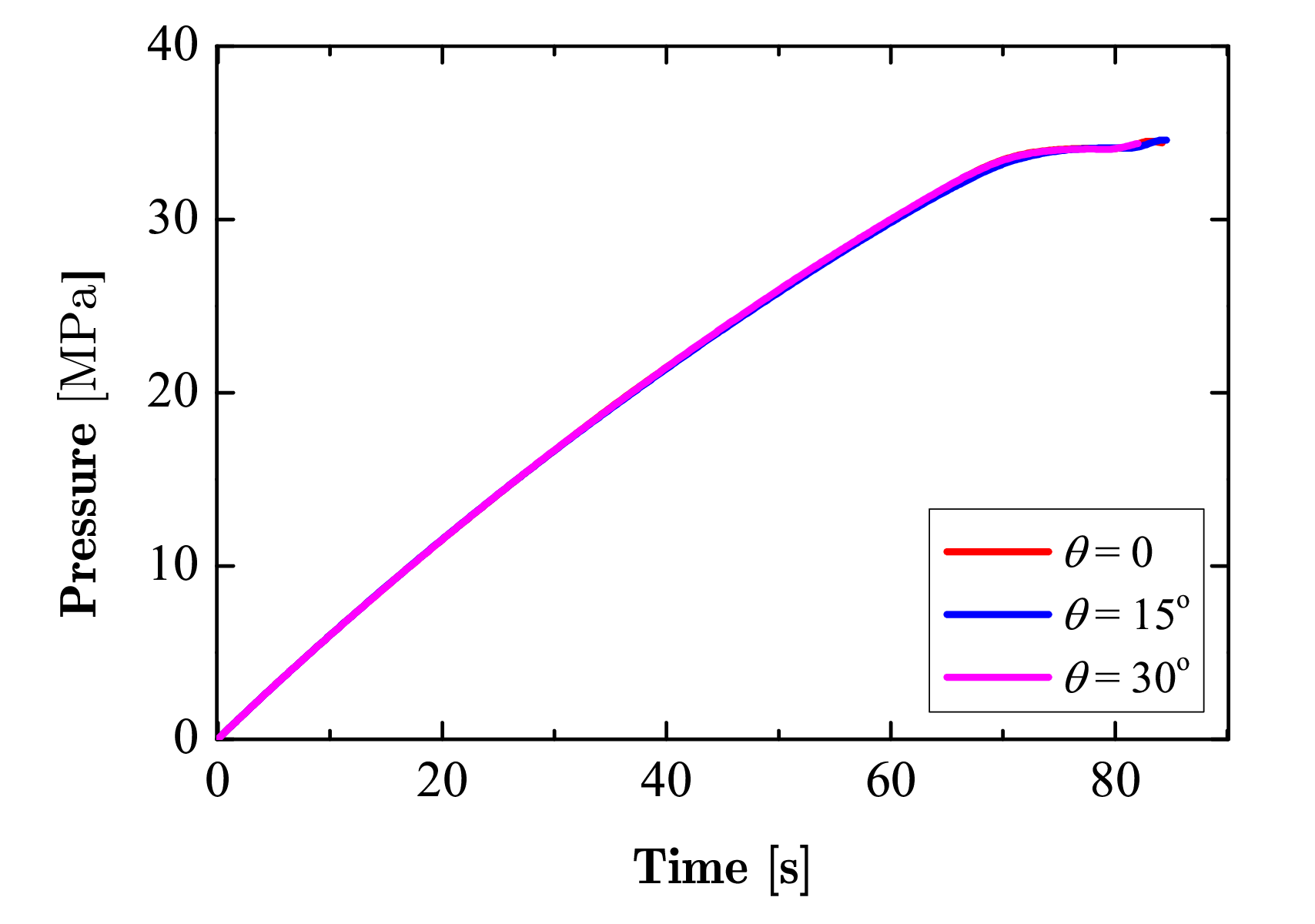}}
	\subfigure[$E_1=3E_2$]{\includegraphics[width = 7.5cm]{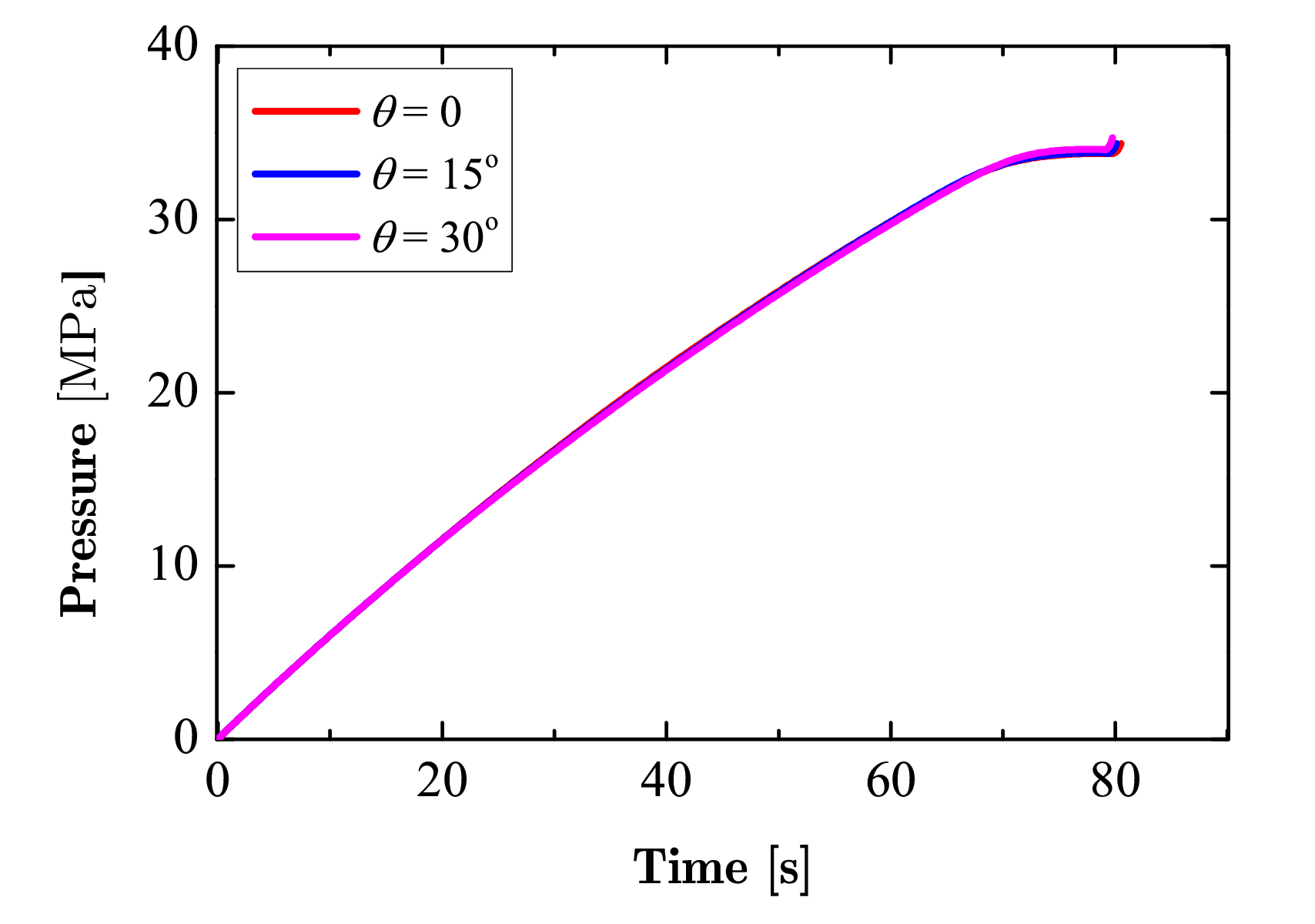}}\\
	\caption{Fluid pressure in the pre-existing notch in the stiff-to-soft configuration}
	\label{Fluid pressure in the pre-existing notch in the stiff to soft configuration}
	\end{figure}

Figure \ref{Fluid pressure in the pre-existing notch in the stiff to soft configuration} shows the fluid pressure when the time increases in the stiff-to-soft configuration. The data is also picked at the point (-2 m, 0). The fluid pressure has a similar trend to that in the soft-to-stiff configuration in Fig. \ref{Fluid pressure in the pre-existing notch in the soft to stiff configuration}. The fluid pressure increases rapidly in the first 65 s. Thereafter, the increasing rate of the fluid pressure decreases and the pressure is nearly stable. Furthermore, for the stiff-to-soft configuration, the second stage is observed to be considerably shorter than the soft-to-stiff configuration. The inclination angle $\theta$ has minimal effect on the fluid pressure-time curve.

\subsection{Guidance for HF practice}

\begin{figure}[htbp]
	\centering
	\includegraphics[width = 7.5cm]{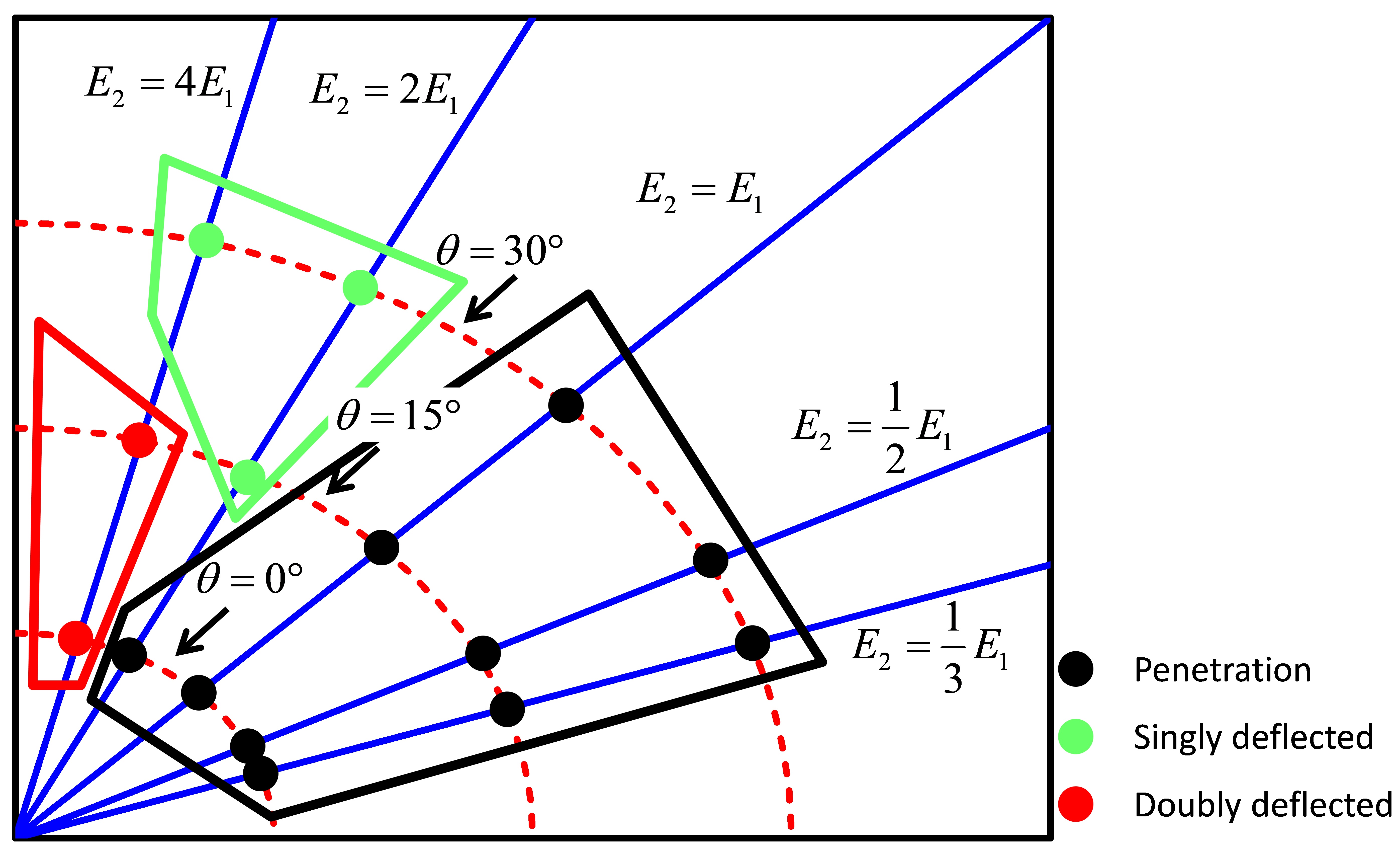}\\
	\caption{Summation of fracture patterns for different $E_2/E_1$ and interface angle $\theta$}
	\label{Fracture patterns for different E_2/E_1 and interface angle theta}
\end{figure}

Figure \ref{Fracture patterns for different E_2/E_1 and interface angle theta} summarizes all the fracture patterns at the layer interface for different $E_2/E_1$, inclination angle $\theta$, and the data in a single porous layer \citep{zhou2018phase2} ($E_2=E_1$). On the basis of the relationship between $E_2/E_1$ and $\theta$, three regions formed in this figure represent the three fracture patterns--penetration, singly-deflected, and doubly-deflected scenarios. Note that a low $E_2/E_1$ or a medium $E_2/E_1$ with a low $\theta$ produces the penetration scenario. A high $\theta$ with a medium or high $E_2/E_1$ corresponds to the singly-deflected scenario, while only a high $E_2/E_1$ with a low $\theta$ will form the doubly-deflected scenario. Therefore, the phase field model and numerical investigation in this research can easily reflect the influence of layer stiffness and interface angle on fracture patterns. In addition, the summary figure, which will be further improved by using PFM in future studies, can be applied for unconventional HF in shale gas development and to study safety and environmental concerns and efficiency issues in current HF practices. For example, if the elastic parameters of two neighboring layers are known, then the perforation direction in HF can be optimized according to Fig. \ref{Fracture patterns for different E_2/E_1 and interface angle theta} to avoid fracture penetration into the neighboring layer due to the potential risks for water contamination and stability of the geological system.

\section {Conclusions}\label{Conclusions}

This study applies a phase field framework to examine hydraulic fracture growth in naturally layered geological formations. The total energy functional used fully includes the influence of fluid pressure, and the fracture pattern is the natural result of minimization of the energy functional. In addition, we consider the soft-to-stiff and the stiff-to-soft configurations where the layer interface exhibits different inclination angles. Therefore, the relationship between the phase field (fracture pattern) and the stiffness contrast and inclination angle of the geological formations, is revealed. The numerical investigation in this study supports the following:
 
(1) The huge advantage of the phase field framework over the X-FEM framework lies in the fact that penetration criteria are also not required when hydraulic fractures reach the material interfaces. In addition, the phase field implementation does not require tracking the fracture paths algorithmically.

(2) Penetration, singly-deflected, and doubly-deflected fracture scenarios can be predicted using PFM. In the soft-to-stiff configuration, the simulations exhibit penetration or symmetrical doubly-deflected scenarios when the pre-existing fracture is perpendicular to the layer interface. When the interface angle is $15^\circ$, singly-deflected or asymmetric doubly-deflected scenarios are obtained. Only the singly-deflected scenario is obtained for the interface angle of $30^\circ$. 
	
(3) In the stiff-to-soft configuration, only the penetration scenario is obtained with widening fractures when the hydraulic fractures penetrate into the softer layer. The fracture in the softer layer deflects at a small angle with the fracture in the stiffer layer and the angle increases as the layer interface angle increases.
 
Note that our study includes a perfect bonding at the layer interface, which is not always the case in geological settings \citep{cooke2001fracture, xing2018laboratory}. Therefore, in future studies, layer interfaces with weak bonding and methods on fixing the interface parameters from the neighboring layers should be involved in the PFM on fracture propagation in layered formations. Laboratory experiments on fractures along weakly-bonded interfaces \citep{xing2018laboratory} will also be used for further validation of PFM. Another limitation of PFM is that the direct coupling of the permeability and crack opening is difficult to apply because of the smeared representation of fracture. In this sense, future PFMs should introduce substantially accurate permeability models for fractured porous domains.

\section*{Acknowledgment}
The authors gratefully acknowledge the financial support provided by the Natural Science Foundation of China (51474157), and the RISE-project BESTOFRAC (734370).

\bibliography{references}

\end{document}